\documentclass[manuscript,screen]{acmart}

\AtBeginDocument{%
  \providecommand\BibTeX{{%
    \normalfont B\kern-0.5em{\scshape i\kern-0.25em b}\kern-0.8em\TeX}}}

\setcopyright{acmcopyright}
\copyrightyear{2018}
\acmYear{2018}
\acmDOI{XXXXXXX.XXXXXXX}

\acmConference[Conference acronym 'XX]{Make sure to enter the correct
  conference title from your rights confirmation emai}{June 03--05,
  2018}{Woodstock, NY}
\acmPrice{15.00}
\acmISBN{978-1-4503-XXXX-X/18/06}

\usepackage{natbib}
\usepackage{verbatim}
\usepackage{caption} 
\usepackage{subcaption}
\usepackage[]{algorithm2e}
\usepackage{graphicx}
\usepackage{booktabs}
\usepackage{relsize}
\usepackage{colortbl}
\usepackage{amsmath}
\usepackage{tikz}
\usepackage{pgf}
\usepackage{pbox}
\usepackage{rotating}
\usepackage{multirow}
\usepackage{balance}
\usepackage{tablefootnote}
\usepackage{tcolorbox}
\usepackage{tabularx}
\usepackage{pdfpages}
\usepackage{enumitem} 
\usepackage{longtable}
\usepackage{pdflscape}

\usepackage{hanging}
	\newcolumntype{L}{>{\raggedright\arraybackslash}X}
\usepackage{xcolor}[table,xcdraw]

\usepackage{cleveref}[2012/02/15]
\usepackage{arydshln}
\crefformat{footnote}{#2\footnotemark[#1]#3}

\usepackage{dblfloatfix}

\usepackage{soul}

\renewcommand{\labelitemi}{--}
\newcommand{\startlist}{\begin{list}{\labelitemi}{\leftmargin=1em}\setlength{\itemsep}{-1mm}}
\newcommand{\stoplist}{\end{list}}

\newcommand{\resultBox}[1]{\begin{tcolorbox}[boxsep=1pt,left=4pt,right=4pt,top=4pt,bottom=4pt]#1\end{tcolorbox}}

\newcommand{\smallheading}[1]{{\bf #1:}\hspace{1mm}}

\let\oldFootnote\footnote
\newcommand\nextToken\relax

\renewcommand\footnote[1]{%
    \oldFootnote{#1}\futurelet\nextToken\isFootnote}

\newcommand\isFootnote{%
    \ifx\footnote\nextToken\textsuperscript{,}\fi}

\newcommand{\spfull}{Story Points}

\newcommand{\devtime}{development time}

\newcommand{\wi}{work item}

\newcommand{\wis}{work items}

\newcommand{\makeitred}[1]{#1}
\newcommand{\revised}[2]{#2}

\newcommand{\revisedTextOnly}[1]{{#1}}

\newcommand{\reason}{reason for inaccurate estimations}
\newcommand{\reasons}{reasons for inaccurate estimations}
\newcommand{\approaches}{approaches to improve effort estimation}
\newcommand{\approach}{approach to improve effort estimation}
\newcommand{\litrevrqone}{What are the discovered \reasons{} in Agile iterative development?}
\newcommand{\litrevrqtwo}{What are the approaches proposed to improve effort estimation in Agile iterative development?}

\newcommand{\purposeestimate}{estimate the effort}
\newcommand{\purposeestimatelarge}{Estimate the effort}
\newcommand{\purposesupport}{support the effort estimation process}
\newcommand{\purposesupportlarge}{Support the effort estimation process}

\newcommand{\purposeestimatepredictinglarge}{Predicting the effort}
\newcommand{\purposeestimatepredict}{predict the effort}

\newcommand{\purposeestimatepracticelarge}{Improving the estimation technique}

\newcommand{\litrevreasonqualitylarge}{Quality issues of the available information}

\newcommand{\litrevreasonqualitysmall}{\litrevreasonquality{}}
\newcommand{\litrevreasonquality}{quality issues of the available information}
\newcommand{\litrevreasonteamlarge}{Team-related}
\newcommand{\litrevreasonteam}{team-related}
\newcommand{\litrevreasonestimationlarge}{Estimation Practices}
\newcommand{\litrevreasonestimation}{estimation practices}
\newcommand{\litrevreasonprojectlarge}{Project management}
\newcommand{\litrevreasonproject}{project management}
\newcommand{\litrevreasonbusinesslarge}{Business influence}
\newcommand{\litrevreasonbusiness}{business influence}

\newcommand{\litrevsearchresultunique}{519}
\newcommand{\litrevtotalpaper}{82}
\newcommand{\litrevreasonspaper}{eight}
\newcommand{\litrevapproachpaper}{75}
\newcommand{\litrevestimatepaper}{66}
\newcommand{\litrevestimatepaperpredict}{56}
\newcommand{\litrevestimatepaperpractice}{10}
\newcommand{\litrevestimatepaperwithplanninglevel}{33}
\newcommand{\litrevestimatepredictpapersprintplanning}{13}
\newcommand{\litrevestimatepapersprintplanning}{13}
\newcommand{\litrevestimatepaperprojectplanningtextbig}{Nine}
\newcommand{\litrevexcludeicfour}{150}
\newcommand{\litrevsupportpaper}{9}
\newcommand{\litrevsupportpapertext}{nine}
\newcommand{\litrevphaseascreening}{163}
\newcommand{\litrevphasebscreening}{109}
\newcommand{\litrevsupportforsprintplanning}{five} 
\newcommand{\litrevsupportnotspecified}{four} 
\newcommand{\litrevestimateforotherlevels}{18}

\newcommand{\litrevestimatenotspecified}{36}
\newcommand{\litrevsearchresultpaper}{578} 

\newcommand{\litrevselectioncriteria}{13}

\newcommand{\litrevsupportcatone}{Improving the available information quality}
\newcommand{\litrevsupportcattwo}{Identifying additional information}
\newcommand{\litrevsupportcatthree}{Identifying estimation uncertainty}
\newcommand{\litrevsupportcatonesmall}{improving the available information quality}
\newcommand{\litrevsupportcattwosmall}{identifying additional information}
\newcommand{\litrevsupportcatthreesmall}{identifying estimation uncertainty}
\newcommand{\litrevreasonnumber}{the number of answered participants}
\newcommand{\litrevreasonagreement}{the agreement scores provided by the participants}
\newcommand{\litrevreasonobserve}{the occurrences reported by the participants}

\begin{document}

\title{A Systematic Literature Review on Reasons and Approaches for Accurate Effort Estimations in Agile}

\author{Jirat Pasuksmit}
\email{jpasuksmit@student.unimelb.edu.au}
\orcid{0000-0003-4059-757X}
\authornotemark[1]
\authornote{jpasuksmit@student.unimelb.edu.au}
\affiliation{%
  \institution{The University of Melbourne}
  \country{Australia}
}

\author{Patanamon Thongtanunam}
\email{patanamon.t@unimelb.edu.au}
\orcid{0000-0001-6328-8839}
\authornote{patanamon.t@unimelb.edu.au}
\affiliation{%
  \institution{The University of Melbourne}
  \country{Australia}
}

\author{Shanika Karuasekera}
\email{karus@unimelb.edu.au}
\orcid{0000-0001-7080-5064}
\authornote{karus@unimelb.edu.au}
\affiliation{%
  \institution{The University of Melbourne}
  \country{Australia}
}

\renewcommand{\shortauthors}{Pasuksmit et al.}

\begin{abstract}
\textbf{Background}:
Accurate effort estimation is crucial for planning in Agile iterative development.
Agile estimation generally relies on consensus-based methods like planning poker, which require less time and information\revised{R1-5}{ than other formal methods (e.g., COSMIC)} but are prone to inaccuracies. 
Understanding the common reasons for inaccurate estimations and how proposed approaches can assist practitioners is essential.
However, prior systematic literature reviews (SLR) only focus on the estimation practices (e.g.,~\cite{conoscenti2019combining, usman2015effort}) and the effort estimation approaches (e.g.,~\cite{alsaadi2022data}).
\textbf{Aim}: We aim to identify themes of \reasons{} and classify \approaches{}.
\textbf{Method}: We conducted an SLR and identified the key themes and a taxonomy.
\textbf{Results}: \revised{R2-1}{The reasons for inaccurate estimation are related to information quality, team, estimation practice, project management, and business influences.
The effort estimation approaches were the most investigated in the literature, while only a few aim to \purposesupport{}.
Yet, few automated approaches are at risk of data leakage and indirect validation scenarios.
\textbf{Recommendations}: Practitioners should enhance the quality of information for effort estimation, potentially by adopting an automated approach. 
Future research should aim to improve the information quality, while avoiding data leakage and indirect validation scenarios.
}

\end{abstract}




\maketitle

\section{Introduction}\label{sec:introduction}

Effort estimation is an important process in Agile iterative development.
When planning an iteration (i.e., a sprint in Scrum), an Agile software development team estimates the effort of a \wi{} (i.e., a task or a story to develop the software).
Based on the estimated effort, the team then selects a set of \wis{} to be included in the sprint, while ensuring that the accumulated effort of the selected \wis{} fits within the sprint capacity.
The sprint capacity (or team capacity) is the available capacity of the team effort to work in a sprint~\cite[p.340]{rubin2012essential}, which is derived from the estimated effort of the delivered \wis{} in the past sprints.

To achieve reliable sprint planning, the estimated effort should accurately reflect the size (or \devtime{}) of the \wi{}.
However, prior work pointed out that lightweight methods of Agile effort estimation (e.g., Planning poker) are prone to be inaccurate~\cite{dantas2018effort}.
Several studies reported that the team re-estimates the effort after sprint planning is finished (or even during the implementation) to maintain the estimation accuracy~\cite{hoda2016multilevel, jpsurvey, masood2020real}.
Such late re-estimation may invalidate the original sprint plan and may cost additional effort for re-planning~\cite{rubin2012essential}. 

\begin{table}[]
\caption{Overview of review studies on effort estimation.}
\label{tab:litrev_of_litrev}
\centering
\resizebox{1\linewidth}{!}{
\setlength\tabcolsep{2pt}
\begin{tabular}{p{3.7cm}|p{0.7cm}|p{1cm}|p{1.9cm}|p{1.7cm}|p{2cm}|p{0.7cm}|p{0.7cm}|p{3cm}}
\hline
\textbf{Authors} & \textbf{Year} & \textbf{Agile Context} & \textbf{Reasons for Inaccurate Estimations} & \textbf{Approaches to Estimate the Effort} & \textbf{Approaches to Support Effort Estimation} & \textbf{Size Metrics} & \textbf{Cost Drivers} & \textbf{Others} \\
\hline \hline
Jorgensen \cite{jorgensen2004review} &  2004 &    &      &   \checkmark{}&       &     &     &       \\
Grimstad et al. \cite{grimstad2006software} &  2006 &    &      &     &       &     &     & Terminology \\
Trendowicz \cite{trendowicz2011state} &  2011 &    &      &     &       &     &     & Industrial practices \\
Wen et al. \cite{wen2012systematic} &  2012 &    &      &   \checkmark{}&       &     &     &       \\
Andrew and Selamat \cite{andrew2012systematic} &  2012 &    &      &     &       &     &     & Data imputation \\
Dave and Dutta \cite{dave2014neural} &  2014 &    &      &   \checkmark{}&       &   \checkmark{}&   \checkmark{}&       \\
Usman et al. \cite{Usman2014} &  2014 &  \checkmark{}&      &   \checkmark{}&       &   \checkmark{}&   \checkmark{}&       \\
Idri et al. \cite{idri2015analogy} &  2015 &    &      &   \checkmark{}&       &     &     &       \\
Idri et al. \cite{idri2016systematic} &  2016 &    &      &   \checkmark{}&       &     &     &       \\
Sharma and Singh\cite{sharma2017systematic} &  2017 &    &      &   \checkmark{}&       &   \checkmark{}&   \checkmark{}&       \\
Gautem and Singh \cite{gautam2018state} &  2018 &    &      &   \checkmark{}&       &     &     &       \\
Dantas et al. \cite{dantas2018effort} &  2018 &  \checkmark{}&      &   \checkmark{}&       &   \checkmark{}&   \checkmark{}&       \\
Hacaloglu and Demirors \cite{hacalouglu2018challenges} &  2018 &  \checkmark{}&      &     &       &     &     & Challenges \\ 
Fernandez et al. \cite{fernandez2020update} &  2020 &  \checkmark{}&      &   \checkmark{}&       &   \checkmark{}&   \checkmark{}&       \\
Perkusich et al. \cite{perkusich2020intelligent} &  2020 &  \checkmark{}&      &   \checkmark{}&       &     &     & Automated approaches \\
Alsaadi and Saeedi \cite{alsaadi2022data} &  2022 &  \checkmark{}&      &   \checkmark{}&       &   \checkmark{}&   \checkmark{}&       \\
\hline
Ours &  2024 &  \checkmark{}&    \checkmark{}&   \checkmark{}&     \checkmark{}&     &     &       \\
\hline 
\end{tabular}
}
\end{table}

Many studies were conducted to understand and improve the effort estimation in Agile.
While several systematic literature reviews (SLRs) aggregated these Agile studies, the focuses of these SLRs were limited to the estimation practices or effort prediction approaches.
\revised{R2-2}{Table~\ref{tab:litrev_of_litrev} shows the overview of the existing SLRs in the effort estimation context.
While several SLRs were conducted in the Agile context, most of them focused on effort estimation approaches, size metrics, and cost drivers.}
For example, Usman et al.\cite{Usman2014}, Dantas et al.\cite{dantas2018effort}, and Fernandez et al.~\cite{fernandez2020update} reviewed and aggregated the estimation practices in Agile.
Hacaloglu and Demirors~\cite{hacalouglu2018challenges} reviewed the challenges of using estimation units in Agile.
However, none has reviewed the \reasons{}.
Understanding the common \reasons{} will allow one to address the core common problems in effort estimation, leading to an improvement in the estimation accuracy.
Table~\ref{tab:litrev_of_litrev} also shows that many studies reviewed approaches to estimate the effort. 
For example, Alsaadi et al.~\cite{alsaadi2022data} focused on reviewing machine learning-based effort prediction approaches.
Indeed, there could be other approaches than effort prediction that can \purposesupport{} in Agile.
Yet, such \approaches{} have not been comprehensively reviewed in the existing literature. 
These gaps highlight the need for a SLR in these areas, which forms the basis of our research.



Therefore, this paper presents an SLR on the \reasons{} (RQ1) and the \approaches{} (RQ2) in the Agile context.
We searched for the studies on five digital libraries, i.e., ACM Digital Library, IEEE Xplore, Scopus, Web of Science, and Wiley Online Library.
We obtained \litrevsearchresultpaper{} studies from the search.
\revised{R1-1}{We conducted the systematic literature review (SLR) following a well-established guideline, i.e., SEGRESS (Software Engineering Guidelines for REporting Secondary Studies) guideline of Kitchenham et al.~\cite{kitchenham2022segress} (see checklists in Table~\ref{tab:table_litrev_segress}).}
We applied a total of \litrevselectioncriteria{} inclusion, exclusion, and quality criteria to select the studies that fit in the scope of our RQs and are of quality.
In total, \litrevtotalpaper{} studies passed all the criteria.
To answer our RQs, we performed card sorting to derive the themes of the \reasons{} and the taxonomy of the \approaches{}.


In RQ1, we identified five categories of the \reasons{}: (1) \litrevreasonquality{}, (2) \litrevreasonteam{}, (3) \litrevreasonestimation{}, (4) \litrevreasonproject{}, and (5) \litrevreasonbusiness{}.
We found that the \litrevreasonquality{} are the commonly reported \reason{}.
In RQ2, we categorized the \approaches{} based on their purposes (i.e., to \purposeestimate{} and to \purposesupport{}). 
We found that \litrevestimatepaper{} out of \litrevapproachpaper{} proposed approaches aimed to \purposeestimate{}.
\makeitred{However, only \litrevsupportpapertext{} approaches were proposed to address the \litrevreasonquality{}, which is the commonly reported \reason{}.
While many of these approaches were proposed to be used for sprint planning, we observed that it is unclear whether they used only the information available during sprint planning or used the latest information version (which is considered as using the future data).}
Our findings suggest that the quality of the available information should be improved for effort estimation.
We found that there is a lack of approaches to help the team improve the information quality.
In addition, the prior effort prediction approaches may need to be revisited in a realistic usage scenario (i.e., using only the available information in training and validation).


\textbf{Paper organization.}
The remainder of this paper is organized as follows.
Section~\ref{sec:litrevrq} motivates for our two research questions.
Section~\ref{sec:litrevmethod} presents our research methodology for the systematic literature review and subsequent analyses. Section~\ref{sec:litrevresult} presents the study results.
Section~\ref{sec:litrevdiscussion} provides broader implications and recommendations based on the results.
Section~\ref{sec:litrevthreats} discloses the threats to validity.
Section~\ref{sec:litrevconclusions} draws the conclusion.
Finally, Section~\ref{sec:litrevappendix} (Appendix) lists the SEGRESS checklist and details of the selected studies.

\revised{SEGRESS 25, 26}{
\textbf{Declaration.} This research was conducted without external funding. Jirat Pasuksmit was associated with the University of Melbourne during the research phase of this work and is presently employed by Atlassian Pty Ltd. at the time of revision and publication.
}
\section{Research Questions}\label{sec:litrevrq}

\begin{table}[]
\caption{Our PICOC criteria based on our research questions.}
\label{tab:table_litrev_PICOC}
\centering

\begin{tabular}{l|p{6.5cm}|p{6cm}}
\hline 
\textbf{Criteria} & \textbf{Description} & \textbf{Our PICOC} \\
\hline \hline
P (Population)    & The target population of the study.                                                                & Academic studies in software engineering                        \\
I (intervention)  & A methodology, tool, technology, procedure to address a specific issue or perform a specific task. & Effort estimation                                               \\
C (Comparison)    & The methodology, tool, technology, procedure that the intervention is being compared.              & N/A (No comparison between different interventions)             \\
O (Outcome)       & The relevant outcomes of the study.                                                                & Reasons for inaccurate estimation (RQ1) and \approaches{} (RQ2) \\
C (Context)       & The context in which the study take place.                                                         & Agile Iterative Development                                     \\
\hline 
\end{tabular}
\end{table}

Similar to Usman et al.~\cite{Usman2014} and Fernandez et al.~\cite{fernandez2020update}, we framed our research questions based on the PICOC criteria.
As suggested by Petticrew and Roberts~\cite{petticrew2008systematic}, PICOC is used to frame a research question for a systematic literature review.
Table~\ref{tab:table_litrev_PICOC} provides the description of each criterion and our PICOC for this study.
Below, we describe the motivation for the two expected outcomes (see Table~\ref{tab:table_litrev_PICOC}), which framed the two research questions of this study.

Since Agile teams plan the sprint based on the estimated effort~\cite{coelho2012effort, rubin2012essential}, inaccurate estimations may cause the sprint plan to become inaccurate.
Identifying the common \reasons{} would help the teams and researchers pinpoint the possible improvements in the effort estimation process.
Yet, to the best of our knowledge, there is no SLR that focuses on the \reasons{} in Agile iterative development.
Therefore, we aim to address RQ1:

\resultBox{(RQ1) \litrevrqone{}}

Prior SLRs show that many approaches were proposed to improve the effort estimation in Agile~\cite{alsaadi2022data, dantas2018effort, fernandez2020update}.
However, these aggregated approaches are limited to effort estimation practices and effort prediction models.
On the other hand, there could be other approaches proposed to improve the effort estimation.
Therefore, we propose this RQ2: 

\resultBox{(RQ2) \litrevrqtwo{}}

\section{Methodology: Systematic Literature Review}\label{sec:litrevmethod}

\makeitred{
To ensure the quality of our work and to align with other SLRs, we conducted this systematic literature review (SLR) following a well-established guideline, i.e., SEGRESS (Software Engineering Guidelines for REporting Secondary Studies) guideline of Kitchenham et al.~\cite{kitchenham2022segress}.
SEGRESS is a revision of the PRISMA 2020 checklist~\cite{page2021prisma} that explains how to apply PRISMA 2020 in the software engineering context.
We provided the \textbf{SEGRESS checklist~\cite{kitchenham2022segress}} with notes on how we comply with each item in the appendix (Section~\ref{sec:litrevappendix}
, Table~\ref{tab:table_litrev_segress}).
In addition, we also follow the another guideline by Kitchenham and Charter~\cite{kitchenham2007guidelines} in evidence collection to avoid bias and provide reproducibility.
In this section, we describe our search strategy, study selection, and data analysis processes.}

\subsection{Search Strategy}\label{sec:searchstrategy}

\begin{figure}
    \centering
    \includegraphics[width=1\linewidth]{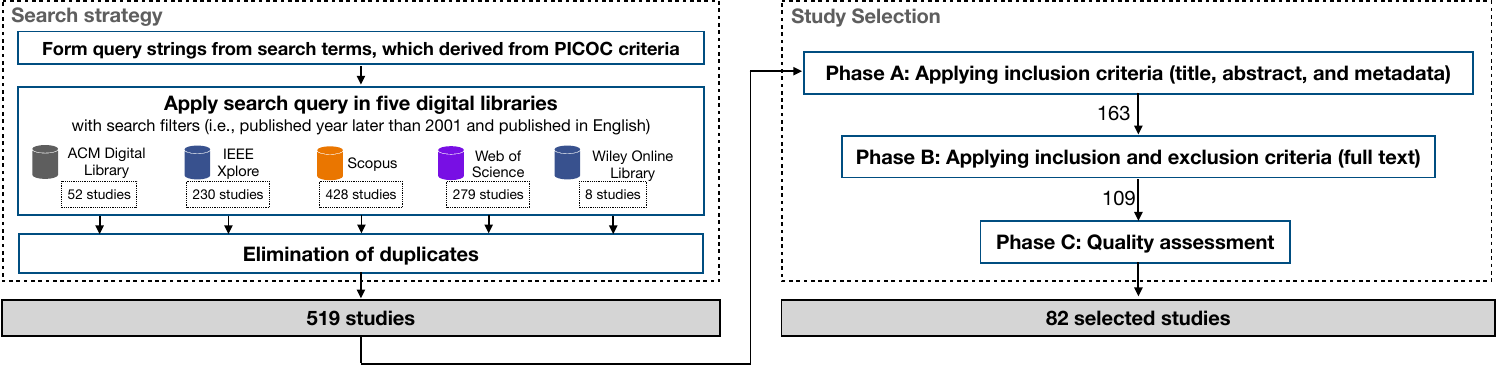}
    \caption{The search strategy (left) and study selection (part) approaches we used in the systematic literature review.}
    \label{fig:litrevapproach}
\end{figure}

To search for the studies, we formed the query string based on the PICOC criteria~\cite{kitchenham2007guidelines} and applied it to five digital libraries in June 2023.
Figure~\ref{fig:litrevapproach} (left) outlines the search strategy we used to collect the studies.

\subsubsection{\textbf{Query String}}

\begin{table}
    \centering
    \caption{Keywords used in search queries.}
    \label{tab:searchterm}
    \begin{tabular}{l|r|l}
    \hline
        \textbf{Criteria} & \textbf{Main Search terms} & \textbf{Alternatives or Synonyms} \\ \hline \hline
        Population    & software & \revisedTextOnly{code} \\ 
        Intervention (1)  & estimat* & predict, forecast, calculat*, assess*, xmeasur* \\
        Intervention (2)       & effort & size, story point \\ 
        Outcome (RQ1) & reason & cause, impact, factor \\
        Outcome (RQ2) & *accura* & *stable, *stabili*, *certain*, *reliab*, error\revisedTextOnly{, *precise*} \\
        Context & agile & extreme programming, scrum\revisedTextOnly{, kanban, scrumban, lean, crystal} \\ \hline
    \end{tabular}
\end{table}

We designed our query string to search for academic studies in software engineering (i.e., P in PICOC) that cover two research areas, i.e., investigating the \reasons{} (RQ1) and proposing the \approaches{} (RQ2).
To do so, we determined our search terms and aggregated them to form our query string.
Table~\ref{tab:searchterm} lists our search terms.
Aligning with our PICOC, we selected the following main search terms based on each PICOC criteria (see Table~\ref{tab:searchterm}).
For each main search term, we added alternative search terms derived from prior work~\cite{Usman2014, fernandez2020update, dantas2018effort} and their synonyms to broaden our search.
Then, we formed our query string using AND/OR operations as suggested by the guideline~\cite{kitchenham2007guidelines}.
In particular, we first used the OR operator to incorporate all alternatives and synonyms to each main search term to form a set of search terms.
After that, we used the AND operator to combine all sets of search terms to form one search query. 
In conclusion, our query string is:

\revised{R1-2}{}
\resultBox{
\revisedTextOnly{
(\textbf{software} OR code) AND (\textbf{estimat*} OR predict* OR forecast OR calculat* OR assess* OR measur*)
AND (\textbf{effort} OR size OR "story point")
AND ((\textbf{reason} OR cause OR impact OR factor) OR (\textbf{*accura*} OR *stable OR *stabili* OR *certain* OR *reliab* OR *precise OR error))
AND (\textbf{agile} OR "extreme programming" OR scrum OR kanban OR scrumban OR "agile lean" OR "lean development" OR "lean methodology" OR "crystal method" OR "crystal agile")
}
}

\subsubsection{\textbf{Searching Methodology}}\label{sec:searchmethodology}


We applied our query string to five academic digital libraries that were used by the prior studies~\cite{alsaadi2022data, Usman2014}, i.e., ACM Digital Library, IEEE Xplore, Scopus, Web of Science, and Wiley Online Library.
We applied filters to focus on the studies that were written in English and were published after 2001 (i.e., the year that the Agile Manifesto was published~\cite{fowler2001agile}).
Then, we downloaded the search results from all digital libraries in Bibtex, combined the results from five digital libraries together, and used a simple Python program to remove duplicate studies based on their title.
Figure~\ref{fig:litrevapproach} shows the number of studies in the search results.

\subsection{Study selection}\label{sec:studyselection}

To ensure that all studies fit in the scope of our RQs and of the quality, we applied \litrevselectioncriteria{} inclusion, exclusion, and quality criteria on the key information manually extracted from the paper in each phase.
Similar to prior work~\cite{Usman2014, alsaadi2022data}, we conducted a multiple-phase study selection.
Figure~\ref{fig:litrevapproach} (right) outlines our study selection process.
We first applied the inclusion criteria to the title, abstract, and metadata of each study (Phase A).
Then, we applied the exclusion criteria to the full text (Phase B).
Lastly, we applied the quality criteria to the full text (Phase C).
\revised{SEGRESS 8}{Note that the first author performed these processes manually without using any review automation tools.
The results along with justifications for the deviant cases were then reviewed by the second author.
When disagreement arose, the first and second authors discussed until they reached a consensus then the first author applied the criteria to the whole collection again to ensure consistency. 
}
We describe how we applied the criteria below.

\begin{table}[]
\caption{The study inclusion criteria.}
\label{tab:table_litrev_inclusion_criteria}
\centering
\begin{tabular}{r|l}
\hline
\multicolumn{2}{l}{\textbf{Inclusion criteria}}                                \\
\hline \hline
IC-1     & Must be peer-reviewed and published at a journal, conference,         \\
         & or workshop                                                           \\
IC-2     & Focusing on any of the Agile Iterative Development                    \\
IC-3     & Investigating the reasons for inaccurate effort estimation (IC-3A)    \\
         & OR proposing an approach to improve the effort estimation accuracy (IC-3B) \\
\hline
\end{tabular}
\end{table}

\begin{table}[]
\caption{The study exclusion criteria.}
\label{tab:table_litrev_exclusion_criteria}
\centering
\begin{tabular}{r|l}
\hline
\multicolumn{2}{l}{\textbf{Exclusion criteria}}                                \\
\hline \hline
EC-1     & Provide no access to the full paper                                   \\
EC-2     & Is duplicate or continuation of another included study/approach       \\
EC-3     & Is a literature review study                                          \\
EC-4     & Is an experience report or replication study                          \\
EC-5     & Is an evaluation of the existing effort estimation technique          \\
\hline
\end{tabular}
\end{table}

\subsubsection{\textbf{Phase A: Applying the inclusion criteria to title, abstract, and metadata}}
Table~\ref{tab:table_litrev_inclusion_criteria} lists our inclusion criteria.
To only include the studies that fit our scope, we applied the inclusion criteria (IC) to the title, abstract, and metadata of each study in the search results.
We applied the IC-1 to IC-3 in order.
For each study, we checked the official website of the publication venue to investigate whether a published article needs to be peer-reviewed (IC-1).
Then, we examined whether the study context is related to Agile iterative development (IC-2).
Lastly, we examined whether the study's objective is to investigate the \reasons{} (IC-3A) or propose an \approach{} accuracy (IC-3B).
For some studies that we could not clearly understand their objective based on their title, abstract, and metadata, we further considered the introduction, background, and conclusion of the study.
A study passed phase A only if it satisfied IC-1, IC-2, and either IC-3A or IC-3B.

\subsubsection{\textbf{Phase B: Applying the exclusion criteria on full text}}
Table~\ref{tab:table_litrev_exclusion_criteria} lists our exclusion criteria.
To fully understand a study in the review, we need to analyze the study based on the full text.
First, we attempted to download the published article of each study and excluded the studies that we could not download the full-text articles (EC-1).
Then, we collected the data from each study to have a better understanding of the context.
After that, we applied other exclusion criteria to each study based on the downloaded full-text article.
Since duplication of findings may mislead our review, we excluded the study that is a duplicate (i.e., having similar authors and content) or a continuation (i.e., extended analyses) of another included study (EC-2).
Note that we only excluded the older study from a duplicate pair and excluded a newer study from a continuation pair.
Finally, we applied EC-3 to EC-5 in order.
We excluded literature review studies (EC-3) as they reported their results based on other studies.
We then excluded experience reports or replication studies (EC-4) and the studies that only evaluate the performance of the other effort estimation techniques (EC-5) as they conducted their studies based on existing approaches or findings.
During this phase, we again applied the inclusion criteria based on the full text to validate the results from Phase A.
A study passed phase B only if it was not excluded by any of the exclusion criteria.
\revised{R2-4}{We provided a list of all included and excluded studies in supplementary material~\cite{supplementary}}.

\begin{table}[]
\centering
\caption{The study quality assessment criteria.}
\label{tab:litrev_quality_criteria}
\begin{tabular}{r|l}
\hline 
\multicolumn{2}{l}{\textbf{Quality criteria}}                                  \\
\hline
QC-1     & The research objective is described                                  \\
QC-2     & The techniques or methodologies are described                        \\
QC-3     & The dataset, participants, or case studies of the study are described               \\
QC-4     & The evaluation or validation methods are described                   \\
QC-5     & The results of the study are described                               \\
\hline

\end{tabular}
\end{table}

\subsubsection{\textbf{Phase C: Applying the quality criteria}}\label{sec:litrevphasec}
Table~\ref{tab:litrev_quality_criteria} lists our quality criteria.
\revised{SEGRESS 10b}{To ensure the quality of the studies, we first extracted the \textbf{objectives}, \textbf{techniques}, \textbf{approaches}, and \textbf{results} of each study. 
We then excluded the studies that did not describe this information.}
After that, we applied the quality assessment checklist of Alsaadi et al.~\cite{alsaadi2022data} to the full-text of the remaining studies.
We extracted and determined whether the study describes the important data, i.e., research objective (QC-1), techniques or methodologies (QC-2), dataset, participants, or case studies (QC-3), evaluation or validation methods (QC-4), and results (QC-5).
We selected only the studies that satisfy all the quality criteria into our literature review.

\subsubsection{\textbf{Phase D: Uncertainty and risk of bias assessment}}\label{sec:litrevphased}

\revised{SEGRESS 10a, 11, 14}{
As suggested by Kitchenham et al.~\cite{kitchenham2022segress}, we assessed the uncertainty and risk of bias of our selected studies and reported it along with our results.
To do so, we applied GRADE approach~\cite{guyatt2008grade} (Grading of Recommendations Assessment, Development, and Evaluation) to assess whether there is a high uncertainty (or risk) in any of the five following domains.
\textbf{Risk of bias of individual studies} refers to methodological biases reducing the certainty of the finding.
In our context, we assessed if a study is relied on a students experiment (high risk), used a weak methodology or lack of detailed explanation (high risk), used an old dataset (high risk), conducted on a small scale (moderate risk), or reported only relative performance metric like MMRE (moderate risk).
\textbf{Imprecision} refers to the ambiguity and vagueness in the data or results reported in a study.
We assess the imprecision of the selected studies based on the extracted results (e.g., confidence interval).
\textbf{Inconcsistency} refers to whether there are strong disagreements among one and other studies in a similar context.
We assess the inconsistency among studies of the same category in the taxonomies in Section~\ref{sec:litrevdataanalysis}.
\textbf{Indirectness} refers to when a study is conducted with subjects that are not representative of the target of interest, e.g., conducting Scrum experiment with students or analyzing non-Agile projects.
We assess the indirectness based on the extracted datasets or participants.
\textbf{Publication bias} in our context refers to the soundness of the search process of our study and whether the majority of the studies we found were conducted on small scales.
To avoid publication bias, we clearly explained our search process in Section~\ref{sec:searchstrategy} and recheck whether the majority of the studies included in each taxonomy were not conducted on a small scale.
}

\subsection{Data Analysis}\label{sec:litrevdataanalysis}

In this section, we described the process of discovering the thematic taxonomies of the \reasons{} (RQ1) and the \approaches{} (RQ2).
\revised{SEGRESS 10a}{Broadly speaking, we extracted the reported \reasons{} with their rankings and the key purpose of the \approaches{} with their planning level desired to operate.
The extracted data was used for discovering the themes and taxonomy of search results (Section~\ref{sec:litrevtaxonomy}), ranking the \reasons{} (Section~\ref{sec:litrevrankingmethod}), and identifying the planning level of the \approaches{} (Section~\ref{sec:litrevplanninglevel}).}
\revised{SEGRESS 9}{In general, the first author performed these processes manually without using any automation tools.
The second author then reviewed the discovered themes, taxonomies, and other results.
When disagreed, the first and the second authors discussed until reached a consensus.
Then, the first author conducted another round of analysis again to ensure consistency.}
\revised{SEGRESS 13c}{To report the results, we used R, Microsoft PowerPoint, and Latex to create charts, diagrams, and tables, respectively.}

\subsubsection{\textbf{Discovering the themes and taxonomy}}\label{sec:litrevtaxonomy}

For both RQ1 and RQ2, we applied an open card sorting technique to categorize and extract the themes and taxonomy.
We analyzed the selected studies that passed the three selection phases (see Section~\ref{sec:litrevphasec}).
The card sorting was performed in multiple iterations.

\revised{SEGRESS 13b}{In the first iteration, we classified the studies into groups based on their scope (i.e., whether the scope fits into RQ1, RQ2, or both).
After that, we extracted the key information to discover the theme of the \reasons{} (RQ1) and discover the taxonomy of the \approaches{} (RQ2).
For the studies that fit RQ1, we extracted the \textbf{\reasons{}} from the result section of each study. 
We then sorted all the reasons into groups based on their thematic similarities and defined a theme for each group.
For the studies that fit RQ2, we extracted the \textbf{key purpose} based on the key question: \textit{``How does the approach aim to help the team in effort estimation?.''}
We sorted the proposed approaches into groups based on their thematic similarities of the purposes and defined a theme for each group.}
\revised{SEGRESS 13f}{When an approach was sorted into a group, we did not exclude it from the pool to look for the possible heterogeneity among the approaches.}
\revised{SEGRESS 13e}{In addition, we also mark (if any) the studies with a high risk of bias and describe the impact if we excluded them from our study in the results section.}
Note that the first author conducted the analyses in the first iteration.

\revised{SEGRESS 13d}{To validate the results, the second author, with different expertise and backgrounds, reviewed and discussed the results with the first author until they reached a consensus.
Then, the first author conducted card sorting again to ensure consistency}
This was to ensure that the themes and taxonomies were not only subjective to the first author.

\subsubsection{\textbf{Extracting the ranking of the \reasons{}}}\label{sec:litrevrankingmethod}
In addition to the reasons extracted from the selected studies in RQ1, we further obtained the \textbf{ranking} provided by those studies.
This insight will help us understand which reason the practitioners or researchers should pay attention to address in order to improve the estimation accuracy.
To extract the ranking, we directly obtained the ranking of each reason (or category of the reasons) as reported in the results section of each study.
We found that these rankings were reported based on different methods, i.e., the number of reported participants, the occurrences reported by the participants, and the agreement scores rated by the participants.
Therefore, we only presented them as an additional insight into the common \reasons{}.

\subsubsection{\textbf{Identifying the planning level of the \approaches{}}}\label{sec:litrevplanninglevel}

For RQ2, we further investigated the \textbf{planning levels} that the approaches were proposed to be used for.
This is because the proposed approaches can be designed to fit their usage scenarios of different planning levels.
For example, a prediction model may be designed to predict the total development cost, which will be used for an early planning level (e.g., project bidding).
Knowing the planning level would allow us to better understand the applicability of the proposed approach.
To achieve this, we categorized the proposed approaches into groups based on their planning levels.
We first obtained the list of planning levels that effort estimation is performed from prior studies~\cite{fernandez2020update, usman2017effort, usman2015effort}, i.e., \textit{daily planning}, \textit{sprint planning}, \textit{release planning}, \textit{project planning}, and \textit{project bidding}.
We read the full text to identify the planning level that the proposed approaches were designed to be used for.
In this process, we identified the planning level when the paper indicates the planning level to be used (e.g., specified in the abstract, objective, motivation, background, or an example usage scenario).
\makeitred{If the planning level was not explicitly indicated in the paper, we used information described in the approach, the outcome of the approach, or the benefit to the practitioners to infer the planning level.
For example, Choetiertikul et al.~\cite{Choetkiertikul2019} situated their Story Points prediction outcome to be used in Scrum sprint-based development: \textit{``Story point sizes are used for [...], planning and scheduling for future iterations and releases, [...].''}
With sufficient information provided, we categorized this approach as suitable to be used for sprint planning and release planning.}
Otherwise, we marked the planning level as ``not specified'' when the planning level is not indicated or the information in the paper is insufficient to infer one of the five planning levels.
Lastly, the first author revisited the categorized results of the planning levels.
To mitigate the risk that the results are subjective to the first author, the second authors revisited the categorization results and discussed them with the first author until both reached a consensus.
Then, the first author conducted another round of categorization to ensure consistency.

\section{Results}\label{sec:litrevresult}

This section presents the search results and describes our findings for each research question.

\subsection{Search Results}\label{sec:litrevsearchresult}

\begin{figure}
\centering
\begin{subfigure}{.3\textwidth}
    \includegraphics[width=1\linewidth]{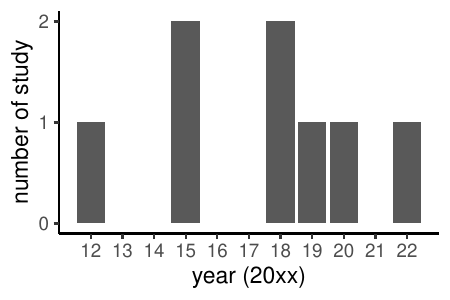}
    \caption{The selected studies that investigated the \reasons{}}
    \label{fig:publishedyearRQ1}
\end{subfigure}
\hspace{1cm} 
\begin{subfigure}{.5\textwidth}
    \includegraphics[width=.95\linewidth]{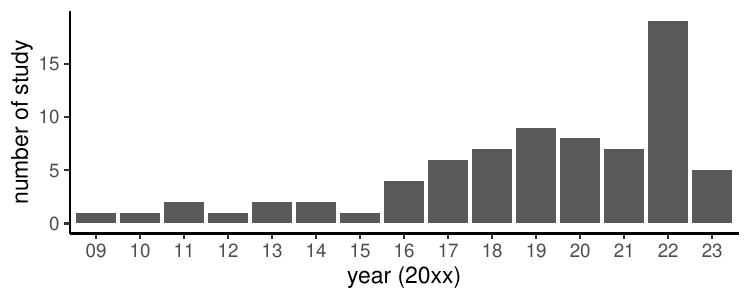}  
    \caption{The selected studies that proposed an approach to help the Agile practitioners estimate effort more accurate}
    \label{fig:publishedyearRQ2}
\end{subfigure}
\caption{The number of selected studies published in each year.}
\label{fig:publishedyear}
\end{figure}

We retrieved \litrevsearchresultunique{} unique studies from the search in five digital libraries and \litrevtotalpaper{} of them passed our three study selection phases.
\revised{R2-3}{
We extracted information from each study, including the purpose of the approach, the desired planning level for operation, the estimating artifact, the technique used, the dataset or participants involved, the study results, the evaluation method, and the outcome of the GRADE uncertainty assessment.
We listed all selected studies along with the extracted information in the Appendix (Section~\ref{sec:litrevappendix}).}
Of the \litrevtotalpaper{} selected studies, \litrevreasonspaper{} of them investigated the \reasons{} (IC-3A) and \litrevapproachpaper{} of them proposed an approach for effort estimation (IC-3B).
Noted that we found one study satisfied both IC-3A and IC-3B.
Figure~\ref{fig:litrevapproach} shows the number of studies that passed each study selection phase.
The majority of these studies were published in 2016 and thereafter (see Figure~\ref{fig:publishedyear}).
Below, we describe the details of the three screening phases.
A list of excluded papers with the reasons for exclusions is provided in our supplementary material~\cite{supplementary}.

\textbf{In phase A,} \litrevphaseascreening{} out of \litrevsearchresultunique{} studies satisfied the inclusion criteria based on the title, abstract, and metadata (see Figure~\ref{fig:litrevapproach}).
We found that no study was excluded due to IC-1 as all of the studies were peer-reviewed, while 58 studies did not satisfy IC-2 as the studies were not conducted in the Agile context.
We found that \litrevexcludeicfour{} studies did not satisfy IC-3A and IC-3B as they did not investigate the \reasons{} or propose an \approach{}.
In particular, the objectives of these \litrevexcludeicfour{} studies are related to software quality and testing (41 studies), development practices (32 studies), human aspect (18 studies), Agile adoption (17 studies), requirements engineering (13 studies), Agile planning (5 studies), factors or predictors considered during the estimation - not the \reasons{} (4 studies), and others (20 studies).
\revised{SEGRESS 16b}{
For example, Altaleb et al.~\cite{altaleb2020industrial} investigated the effort estimation predictors and Logue and McDaid~\cite{logue2008agile} proposed an approach to handling uncertainty in release planning.
These studies nearly met all criteria but were excluded because the predictors may not cause effort estimation inaccuracies and the approach is not proposed for effort estimation.}
\revised{SEGERSS 20d}{On the other hand, a study by Vetro et al.~\cite{vetro2018combining} satisfied both IC3A and IC3B (heterogeneity case) because it studied the root causes for wrong estimations and proposed a new estimation process to tackle these issues.}


\textbf{In phase B,} \litrevphasebscreening{} out of \litrevphaseascreening{} studies passed the exclusion criteria and inclusion criteria based on full text.
Particularly, we could not access the full paper of eight studies (EC-1).
We sent a direct request to the authors of these studies via email and retrieved the full-text copies of two studies.
After that, we excluded six duplicate or continuation studies (EC-2), 13 literature review studies (EC-3), seven experience report or replication studies (EC-4), and 23 studies that evaluated the existing effort estimation techniques.
Noted that all the remaining studies still satisfied the inclusion criteria based on the full text.

\textbf{In phase C,} \litrevtotalpaper{} out of \litrevphasebscreening{} studies passed all the quality criteria.
A total of 27 studies were excluded as not satisfy one of the quality criteria.
In particular, these studies did not describe techniques (QC-2; 2 studies), datasets, participants, or case studies (QC-3; 12 studies), evaluation or validation methods (QC-4; 11 studies), or results (QC-5; 2 studies).
\revised{SEGRESS 16b}{Few approaches nearly met all criteria but were excluded due to insufficient explanation of validation methods (e.g., \cite{owais2016effort, popli2014cost}).
We observed that the essential aspects of validation were notably absent, e.g., comparing predicted effort with actual effort, employing cross-validation techniques, or benchmarking against expert estimations.}

\revised{SEGRESS 20a}{\textbf{In phase D,} 
Tables~\ref{tab:table_litrev_studylist_rq1reasons}, ~\ref{tab:table_litrev_studylist_rq2estimate}, and ~\ref{tab:table_litrev_studylist_rq2support} show the results of our uncertainty assessment.
Specifically, we identified 15 out of 82 (18\%) studies subjected to high risk~\cite{kitchenham2022segress}, i.e., five studies relying only on students or student projects, five studies have unclear characteristics of the datasets or participants, three studies using artificial datasets, one study conducted based on findings from the pre-Agile manifesto, and one study used inconsistent approaches from the literature.
Nevertheless, we found that a minority of our selected studies were conducted on a small scale.
Thus, the impact of the risks from the selected studies on our findings is minimal.
}

\subsection{RQ1 results: \litrevrqone{}}\label{sec:litrevreasonsection}

\begin{table}
\caption{Reasons for inaccurate estimations reported in the literature (RQ1). The ranking numbers indicate the rankings of a reason in comparison to the other reasons reported in \underline{the same study}. The ``\textbf{n}'', ``\textbf{a}'', ``\textbf{o}'', and ``\textbf{x}'' indicates that the ranking was based on ``\textbf{n}umber of answered participants'', ``\textbf{a}greement scores provided by the participants'', ``\textbf{o}ccurrences reported by the participants'', and ``no ranking'', respectively.}
\label{tab:table_litrev_reasons}
\centering
\resizebox{0.9\linewidth}{!}{
\begin{tabular}{lll}
\hline
\textbf{Reasons}                                                            & \textbf{Ranks} & \textbf{Studies} \\
\hline \hline
\multicolumn{3}{l}{\textbf{R1) \litrevreasonqualitylarge{}}}                                                           \\
R1.1) Unclear information                                & 1n, 1n, 2a, 7o, x      & \cite{conoscenti2019combining,sandeep2022effort,usman2018effort,Britto2015,usman2015effort} \\
R1.2) Unstable information                                                 & 1n, 1n, 1a, x        & \cite{sandeep2022effort,usman2018effort,Britto2015,usman2015effort} \\
R1.3) Error in the information                                                 & 1n, 5o              & \cite{Britto2015, conoscenti2019combining}                \\
\hline
\multicolumn{3}{l}{\textbf{R2) Team-related}}                                                             \\
R2.1) Lack of experience of team members                                     & 1o, 2n, 3n, 6a, x              & \cite{conoscenti2019combining,sandeep2022effort,Britto2015,usman2015effort,usman2018effort}                \\
R2.2) Insufficient stakeholder participation                                & 2n, 2n, 12a
      & \cite{sandeep2022effort,Britto2015,usman2015effort} \\
R2.3) Knowledge sharing problem                                              & 3n, 5a, x            & \cite{usman2015effort,sandeep2022effort, vetro2018combining} \\
R2.4) Dominant personality                                                   & 14a             & \cite{sandeep2022effort} \\
\hline
\multicolumn{3}{l}{\textbf{R3) Estimation Practice}}                                                             \\
R3.1) Factors overlooking                                                     & 1n, 2o, 3a, x        & \cite{conoscenti2019combining, sandeep2022effort, usman2018effort, usman2015effort}                \\
R3.2) Considering unnecessary work                                            & 2o, x              & \cite{conoscenti2019combining, usman2018developing}                \\
R3.3) Lack of an estimation process                      & 5n, 9a           & \cite{usman2015effort,sandeep2022effort} \\
R3.4) Inappropirate estimation scale                                                 & x, x              & \cite{tamrakar2012does, vetro2018combining} \\
\hline
\multicolumn{3}{l}{\textbf{R4) Project Management}}                                                      \\
R4.1) Poor change management                                                 & 2n, 7a            & \cite{usman2015effort,sandeep2022effort}                \\
R4.2) Poor human-resource management                                         & 2n, 6o, x, x        & \cite{conoscenti2019combining, usman2018effort, usman2015effort, karna2020effects}                \\
R4.3) Communication overhead in  distributed team settings      & 3n, 16a, x         & \cite{sandeep2022effort,usman2018effort,Britto2015} \\
\hline
\multicolumn{3}{l}{\textbf{R5) Business Influence}}                                                      \\
R5.1) Overoptimism due to project bidding pressure                           & 4n, 5a            & \cite{usman2015effort,sandeep2022effort} \\
R5.2) Pressure of timeline                                                   & 10a              & \cite{sandeep2022effort} \\
\hline
\end{tabular}
}
\end{table}

In this RQ, we aim to better understand the common \reasons{} in Agile.
We identified \litrevreasonspaper{} studies that investigated the \reasons{} (see Section~\ref{sec:litrevtaxonomy}).
Based on our card sorting process, we identified five themes of the \reasons{}, i.e., \litrevreasonqualitylarge{} (R1), Team-related (R2), Estimation Practice (R3), Project Management (R4), and Business Influence (R5), which are described below.
Table~\ref{tab:table_litrev_reasons} lists the \reasons{} reported by these \litrevreasonspaper{} studies.
We also obtained the ranking of each reason from the selected studies (see Section~\ref{sec:litrevrankingmethod}).
For each study, we directly obtained the ranking of each reason (or its category) from the result section.
We found that the reasons were ranked based on the number of answered participants~\cite{usman2015effort, Britto2015}, the mean value of agreement scores provided by the participants~\cite{conoscenti2019combining}, and the occurrences reported by the participants~\cite{usman2018developing}.
Table~\ref{tab:table_litrev_reasons} also reported the rankings retrieved from these studies.
A higher ranking number indicates a higher ranking of a reason compared to the others in the same study, where the following alphabet is used to indicate the ranking methods.
In particular, the ``\textbf{n}'', ``\textbf{a}'', ``\textbf{o}'', and ``\textbf{x}'' indicate that the ranking was based on the ``\textbf{n}umber of answered participants'', ``\textbf{a}greement scores provided by the participants'', ``\textbf{o}ccurrences reported by the participants'', and ``no ranking'', respectively.
For example, a rank ``1\textbf{n}'' indicates that the reason was the most commonly reported based on the ``\textbf{n}umber of answered participants'' in the study.
Below, we describe each category of the \reasons{}.

\subsubsection{\textbf{Reasons for inaccurate estimations}}
\textbf{(R1) \litrevreasonqualitylarge{}: }
We found that five out of \litrevreasonspaper{} studies reported the (R1) \litrevreasonquality{} as the \reasons{}~\cite{conoscenti2019combining,sandeep2022effort,usman2018effort,Britto2015,usman2015effort}.
The reasons in this category are often reported in high ranking, e.g., ranked first and second based on \litrevreasonnumber{} and  \litrevreasonagreement{}.
Britto et al.~\cite{Britto2015} and Conoscenti et al.~\cite{conoscenti2019combining} reported that, with the poor quality of the available information, the team may inaccurately anticipate the effort to implement a desired functionality.
They also discussed that the quality issues were late discovered during the implementation.
Based on our card sorting process, three sub-themes of the quality issues emerged, i.e., unclear information (R1.1), unstable information (R1.2), or error in the information (R1.3).
Unclear information (R1.1) refers to the lack of detail or unclear information during effort estimation (e.g., user stories, acceptance criteria, requirements)~\cite{conoscenti2019combining, sandeep2022effort, usman2018effort, usman2015effort, Britto2015}.
Conoscenti et al.~\cite{conoscenti2019combining} described that the \textit{``understandability problems of the story''} or \textit{``unclear definition of the user acceptance criteria''} (i.e., the available information) would lead to inaccurate estimation.
Unstable information (R1.2) refers to the available information that keeps changing or changed after the estimation is done, thus may affect the estimation accuracy~\cite{sandeep2022effort, usman2018effort, usman2015effort, Britto2015}.
Usman et al.~\cite{usman2015effort} reported that \textit{``both changing [the existing requirements] and [introducing] new requirements were found to be one of the top reasons for effort overruns''}.
Error in the information (R1.3) refers to the requirements~\cite{Britto2015} and estimation logging~\cite{conoscenti2019combining} that were mis-documented.
Britto et al.~\cite{Britto2015} noted that \textit{``mis-documented requirements affect the accuracy of the effort estimates''}, which could lead to unpredicted activities during the estimation process.

These quality issues could apply to any level of information.
In particular, three studies reported quality issues for requirements.
Such a piece of information is considered a high-level abstraction of information~\cite{sandeep2022effort, usman2018effort, Britto2015}.
The other three studies reported quality issues for user stories~\cite{usman2015effort, sandeep2022effort} and acceptance criteria~\cite{conoscenti2019combining}, which are considered detailed information.


\textbf{(R2) \litrevreasonteamlarge{}:}
We found that six out of eight studies reported the (R2) team-related issues as the \reasons{}.
However, we found that these issues were reported with overall lower ranks than R1.
These issues were ranked first, second, third, and fourteenth (or lower) based on \litrevreasonobserve{}, \litrevreasonnumber{}, and \litrevreasonagreement{}.
During our card sorting process, four sub-themes of \litrevreasonteam{} issues emerged, i.e., lack of experience of team members (R2.1), insufficient stakeholder participation (R2.2), knowledge sharing problem (R2.3), and dominant personality (R2.4).
The lack of experience of team members (R2.1) refers to the lack of experience in the technology, domain knowledge, and effort estimation practices of team members~\cite{conoscenti2019combining, sandeep2022effort, usman2015effort, Britto2015}.
This also includes onboarding novice team members~\cite{conoscenti2019combining, sandeep2022effort, usman2018effort, usman2015effort} and the team with low cohesion (i.e., the team worked little time together)~\cite{Britto2015}.
Britto et al.~\cite{Britto2015} and Conoscenti et al.~\cite{conoscenti2019combining} reported that these issues might lead to a wrong assumption of the size of functionality or the team's ability to deliver, which affect the effort estimation accuracy.
Conoscenti et al.~\cite{conoscenti2019combining} also reported that the lack of experience occurred the most across all their studied projects, accounting for 26 of 83 over- and under-estimations (31\% of the time).
Insufficient stakeholder participation (R2.2) refers to the absence of stakeholders (i.e., development team, clients, and scrum master) during the estimation process.
Three scenarios of stakeholders' absence were reported in the studies.
First, the effort was not estimated by the development team~\cite{Britto2015}.
Second, the clients did not attend the estimation session to provide the details~\cite{usman2015effort, sandeep2022effort}.
Third, the effort was estimated considering the active participation of the clients, who in turn they did not participate~\cite{Britto2015}.
Knowledge sharing problem (R2.3) refers to the limited knowledge shared among the team members~\cite{usman2015effort,sandeep2022effort, vetro2018combining}.
For example, Vetro et al.~\cite{vetro2018combining} reported that \textit{``estimating new user stories without an explicit and shared reflection on previous estimations can lead to extreme wrong estimation."}
Usman et al.~\cite{usman2015effort} suggested that \textit{``knowledge sharing problems in the team and the presence of unskilled members in the team''} should be considered in the estimation and managed during the implementation.
Lastly, dominant personality (R2.4) refers to a team member with a dominant personality that could influence the team estimations~\cite{sandeep2022effort}.

\textbf{(R3) \litrevreasonestimationlarge{}:}
We found that five out of eight studies reported the (R3) \litrevreasonestimation{} issues as the \reasons{}.
These issues were ranked first, second, and fifth (or lower) based on \litrevreasonnumber{} and \litrevreasonobserve{}.
During our card sorting process, four sub-themes of \litrevreasonestimation{} issues emerged, i.e., factors overlooking (R3.1), considering unnecessary work (R3.2), lack of an estimation process (R3.3), and inappropriate estimation scale (R3.4).
Factors overlooking (R3.1) refers to the practice where the team did not consider the factors related to the \wi{} during the estimation.
For example, in the estimation session, the team overlooked side tasks~\cite{conoscenti2019combining}, overlooked non-functional requirements~\cite{sandeep2022effort, usman2015effort}, under-estimated the complexity of a developing function~\cite{conoscenti2019combining, usman2018effort}, ignored test effort~\cite{sandeep2022effort, usman2015effort}, ignored code review effort~\cite{conoscenti2019combining}, was not aware of technical problems~\cite{conoscenti2019combining}, did not considered the usability of a standard function~\cite{conoscenti2019combining}, or overlooked hardware problem~\cite{sandeep2022effort}.
Overlooking these factors could lead to an underestimation.
On the other hand, some practitioners may consider unnecessary work in the estimation.
Considering unnecessary work (R3.2) refers to the practice when the team considers unnecessary work in a \wi{} during the estimation.
The reported that unnecessary work is the work that was not required for the \wi{} (i.e., \textit{``Gold plating''})~\cite{conoscenti2019combining}, the duplicate or redundant functions included in the estimation but were implemented as part of other \wis{}~\cite{conoscenti2019combining}, or the functions that were over-estimated in terms of complexity~\cite{conoscenti2019combining, usman2018developing}.
Considering this unnecessary work could lead to an overestimation.
Lack of an estimation process (R3.3) refers to the practice where the team did not use any estimation process~\cite{sandeep2022effort, usman2015effort}, e.g., Planning Poker.
Although the Agile Manifesto prefers interactions among people over processes and tools~\cite{fowler2001agile}, Usman et al.~\cite{usman2015effort} argued that \textit{``this does not mean that agile practices advocate for the estimation process dimension to be completely ignored.''}
Inappropriate estimation scale (R3.4) refers to the practice where the practitioners use an estimation scale that may not reflect the actual effort.
Tamrakar et al.~\cite{tamrakar2012does} found that the use of the Fibonacci scale in the estimation could lead to poor estimation accuracy.
On the other hand, Vetro et al.~\cite{vetro2018combining} reported that having too many items in the numerical scale or using a misleading numerical scale could lead to wrong estimations.

\textbf{(R4) \litrevreasonprojectlarge{}:}
We found that six out of eight studies reported the (R4) \litrevreasonproject{} issues as the \reasons{}.
These issues were ranked second and third (or lower) based on \litrevreasonnumber{}.
During our card sorting process, three sub-themes of \litrevreasonproject{} issues emerged, i.e., poor change management (R4.1), poor human-resource management (R4.2), and communication overhead in distributed team settings (R4.3).
Poor change management (R4.1) refers to when the scope of work keeps changing due to poor change control (i.e., \textit{``scope creep''}).
Sandeep et al.~\cite{sandeep2022effort} and Usman et al.~\cite{usman2015effort} reported that these issues may negatively impact the development time and project cost.
Poor human resource management (R4.2) refers to delays due to the dependency on external resources~\cite{conoscenti2019combining, usman2018effort} and turnover issues~\cite{usman2015effort, karna2020effects}. 
Conoscenti et al.~\cite{conoscenti2019combining} and Usman et al.~\cite{usman2018effort} reported that the dependency on external human resources (e.g., product architects) could delay the implementation process and might also introduce communication overhead.
Usman et al.~\cite{usman2015effort} reported that a high employee turnover would affect the estimation accuracy.
In addition, Karna et al.~\cite{karna2020effects} reported that not all turnover can have a negative impact on the estimation accuracy, while unplanned turnover can have a significant negative impact on the reliability of the expert estimation.
Communication overhead in distributed team settings (R4.3) refers to the management problems when the team is working across multiple sites.
Britto et al.~\cite{Britto2015} and Sandeep et al.~\cite{sandeep2022effort} reported that distributed teams with different time zones, languages, and cultures may require additional effort for communication between the team members and the clients.
Such communication effort should be considered in the estimation.
Complementing this finding, Usman et al.~\cite{usman2018effort} reported their analysis based on six product customization that the work being done in multiple sites tends to be largely underestimated.

\textbf{(R5) \litrevreasonbusinesslarge{}:}
We found that two out of eight studies reported the (R5) \litrevreasonbusiness{} issues as the \reasons{}.
These issues were ranked fourth and tenth (or lower) based on \litrevreasonnumber{} and \litrevreasonagreement{}.
During our card sorting process, two sub-themes of \litrevreasonbusiness{} issues emerged, i.e., overoptimism due to project bidding pressure (R5.1) and pressure of timeline (R5.2).
Overoptimism due to project bidding pressure (R5.1) refers to when the team intentionally underestimates the effort by considering only the best-case scenario to obtain a contract~\cite{usman2015effort, sandeep2022effort}.
Usman et al.~\cite{usman2015effort} argued that ``\textit{purposeful underestimation is an unfair practice and is a clear breach of the code of ethics for software engineers as described in~\cite{gotterbarn1999computer}.}''
Pressure of timeline (R5.2) was reported by Sandeep et al.~\cite{sandeep2022effort} as one of the reasons for the inaccurate estimations.
However, no further explanation was provided.

\resultBox{
    \textbf{Findings:} 
    We identified five main \reasons{}, i.e., \litrevreasonqualitysmall{}, \litrevreasonteam{}, \litrevreasonestimation{}, \litrevreasonproject{}, and \litrevreasonbusiness{}.
    The \litrevreasonqualitysmall{} are commonly reported and often ranked as the top \reasons{}.
    
}

\subsection{RQ2 results: \litrevrqtwo{}}

\begin{figure}
    \centering
    \includegraphics[width=.9\linewidth]{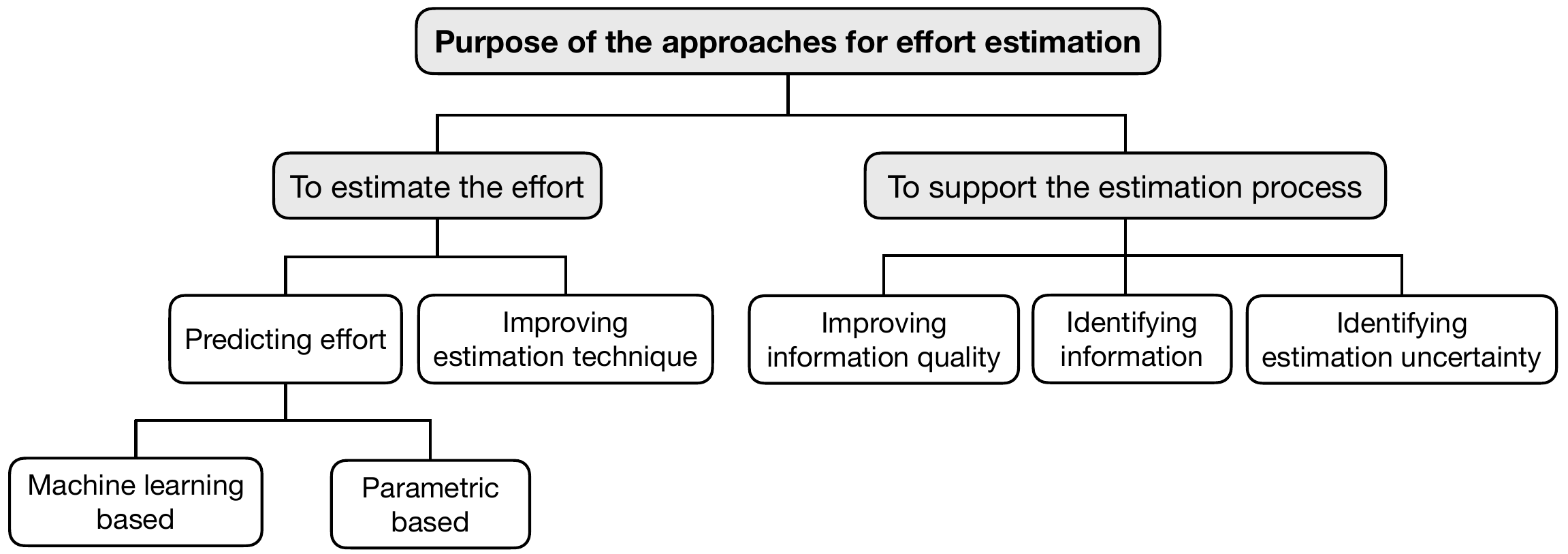}
    \caption{Categorization of the \approaches{} based on the purposes of the approaches.}
    \label{fig:litrevrq2_purpose}
\end{figure}



In this RQ, we identified \litrevapproachpaper{} studies that proposed the \approaches{} in Agile.
Based on our card-sorting process, we categorized these approaches into themes based on their purposes (see Figure~\ref{fig:litrevrq2_purpose}).
Section~\ref{sec:litrevpurposeestimate} presents the approaches that aim to \purposeestimate{}, while Section~\ref{sec:litrevpurposesupport} presents the approaches that aim to \purposesupport{}.
In addition, as described in Section~\ref{sec:litrevplanninglevel}, we identified the planning levels for which the approaches were designed to be used.
Lastly, Section~\ref{sec:litrevplanninglevelresult} discusses the planning levels of these approaches.

\subsubsection{\textbf{\purposeestimatelarge{}}}\label{sec:litrevpurposeestimate}

\begin{table}
\caption{The studies that proposed an approach to \purposeestimate{} at different planning levels (rows), different purposes and techniques (columns), and different estimating artifacts (cells). We summarized these studies in Appendix (Table~\ref{tab:table_litrev_studylist_rq2estimate}).}
\label{tab:table_litrev_result_approachestimation}
\resizebox{1\linewidth}{!}{
\setlength\tabcolsep{2pt}
\begin{tabular}{r|l|p{5.8cm}|p{3.5cm}|l}
\cline{3-5}
\multicolumn{2}{r}{\textbf{ }} & \multicolumn{3}{c}{\textbf{Purposes (and techniques)}} \\
\hline
& & \textbf{\purposeestimatepredictinglarge{}} & \textbf{\purposeestimatepredictinglarge{}} & \textbf{Improving estimation} \\
\multicolumn{1}{r|}{\textbf{Planning level}} & \# & \textbf{(Machine learning-based)} & \textbf{(Parametric-based)} & \textbf{technique (manual)}\\
\hline
\hline

\textbf{Sprint planning}  & \litrevestimatepredictpapersprintplanning{}   
& \wis{} \cite{alostad2017fuzzy, scott2018, malgonde2019ensemble, dantas2019effort, Choetkiertikul2019, fu2022gpt2sp, tawosi2022investigating}, sprints~\cite{ramessur2021predictive} 
& 
& \wis{} \cite{vetro2018combining, altaleb2020pair, madya2022prep, alsaadi2021scrum, rola2019application} \\

\textbf{Release planning}      & 4    
& \wis{}~\cite{winska2021reducing, Choetkiertikul2019} 
& sprints~\cite{kang2010model} 
& \wis{}~\cite{rosa2023data} \\

\textbf{Project planning}      & 9    
& \wis{} \cite{premalatha2019effort, dragicevic2017bayesian}, projects~\cite{sanchez2022software, rodriguez2023effort, sharma2022combined, bilgaiyan2018chaos} 
& projects~\cite{raslan2018enhanced, singal2022integrating} 
& projects~\cite{butt2022software}\\

\textbf{Project bidding}       & 3    
& 
& projects~\cite{parvez2013efficiency, rosa2017early, rosa2021empirical} 
& \\

\textbf{Maintenance}     & 2    
& requirement changes~\cite{sakhrawi2021support} 
& maintenance phases~\cite{choudhari2012phase} 
& \\

\hline
\textbf{Not specified} & \litrevestimatenotspecified{}   
& \wis{}~\cite{Porru2016, marapelli2020rnn, dan2020nlp, gultekin2020story, moharreri2016cost, miranda2021analysis, hemrajanipredicting, phan2022heterogeneous, phan2022story, kassem2023story, kassem2022software, prykhodko2019multiple}, sprints~\cite{Hearty2009, karna2020data}, requirement changes~\cite{sakhrawi2022software}, projects~\cite{adnan2019ontology, bilgaiyan2019effort, khuat2018novel, farahneh2011linear, satapathy2014story, sharma2020linear, ardiansyah2022mucpso, kaushik2020fuzzified, kaushik2022role, gupta2022automated, najm2022enhanced, panda2015empirical} 
& requirement changes~\cite{basri2016algorithmic}, releases~\cite{paz2014approach}, projects~\cite{aslam2017risk, govil2022estimation, arora2022efficient, butt2023prediction} 
& \wis{} \cite{alhamed2021playing, el2023ifejm}, projects~\cite{nunes2011iucp}\\
\hline

\end{tabular}
}
\end{table}

We identified \litrevestimatepaper{} studies that proposed an approach to \purposeestimatepredict{} (\litrevestimatepaperpredict{} studies) and to improve the estimation techniques (\litrevestimatepaperpractice{} studies).
In addition, we also categorized the approaches to \purposeestimatepredict{} based on the techniques used, i.e., machine learning-based and parametric-based.
Table~\ref{tab:table_litrev_result_approachestimation} lists these \litrevestimatepaper{} studies.

\smallheading{\textbf{\purposeestimatepredictinglarge{} (Machine Learning-based)}}
We identified 44 approaches aimed to \purposeestimatepredict{} using machine learning models.
The prediction models were built based on historical data to \purposeestimatepredict{} of \wis{}, sprints, releases, projects, requirement changes, and maintenance phases.
The ``effort'' predicted could be in the form of size, time, or cost.
These approaches used several techniques, i.e., traditional machine learning models (e.g., Support Vector Machine, Bayesian Network, k-Nearest Neighbors)~\cite{Porru2016, moharreri2016cost, scott2018, dantas2019effort, ramessur2021predictive, sakhrawi2021support, hemrajanipredicting, gultekin2020story, karna2020data, adnan2019ontology, satapathy2014story, sharma2020linear, farahneh2011linear, sanchez2022software, rodriguez2023effort, tawosi2022investigating, sharma2022combined, najm2022enhanced, winska2021reducing, sakhrawi2022software, prykhodko2019multiple}, artificial neural network~\cite{Choetkiertikul2019, premalatha2019effort, dragicevic2017bayesian, panda2015empirical, marapelli2020rnn, dan2020nlp, Hearty2009, bilgaiyan2019effort, kaushik2022role, phan2022heterogeneous, phan2022story, gupta2022automated, kassem2023story, fu2022gpt2sp, kassem2022software, bilgaiyan2018chaos}, Particle Swarm Optimization~\cite{ardiansyah2022mucpso, khuat2018novel}, ensemble-based model~\cite{malgonde2019ensemble, sakhrawi2022software}, fuzzy logic~\cite{alostad2017fuzzy, kaushik2020fuzzified}, and Monte Carlo simulation~\cite{miranda2021analysis}.
Most of these approaches used a regression technique (e.g., predicting Story Points value).
On the other hand, two approaches used a classification technique to predict the effort, e.g., classifying Story Points in the Fibonacci scale (Porru et al.~\cite{Porru2016}) or classifying the range of man-hours (Dan et al.~\cite{dan2020nlp}).
\makeitred{Lastly, only one approach used a clustering technique to provide estimations based on the closest \wi{} in the cluster (Tawosi et al.~\cite{tawosi2021multi}).}

\smallheading{\textbf{\purposeestimatepredictinglarge{} (Parametric-based)}}
We identified 13 approaches that used a parametric-based method to \purposeestimatepredict{} of sprints, projects, and maintenance phases~\cite{kang2010model, raslan2018enhanced, govil2022estimation, arora2022efficient, butt2023prediction, singal2022integrating, parvez2013efficiency, rosa2017early, rosa2021empirical, choudhari2012phase, basri2016algorithmic, paz2014approach, aslam2017risk}.
These approaches defined an equation or calculation method to calculate the effort based on different parameters, e.g., the parameters related to the stories, projects, or developers.
For example, effort in person-months $=1.3 \times REQ^{0.512} \times STAFF ^{0.478} \times SD^{1.001}$~\cite{rosa2017early}, where REQ is the number of requirements, STAFF is the number of team members available, and SD is the category of application domain (e.g., support=1, engineering=3).
These approaches could be used manually or implemented as an automated program.

\smallheading{\textbf{\purposeestimatepracticelarge{}}}
We identified 10 studies that proposed new (or improved) effort estimation techniques.
Altaleb et al.~\cite{altaleb2020pair} proposed a pair-estimation technique that requires team members to estimate in pairs to create a deeper communication about the work to be done.
Alhamed and Storer~\cite{alhamed2021playing} proposed a technique that simulates the Planning Poker estimation by recruiting the estimators from the Amazon Mechanical Turk platform.
This technique could achieve a similar accuracy as expert estimation with a lower cost.
Vetro et al.~\cite{vetro2018combining} proposed an expert-based estimation process that uses a shorter non-numerical estimation scale, compares the estimation with past user stories, and reviews past inaccurate estimations.
\makeitred{Madya et al.~\cite{madya2022prep} proposed a framework to improve the quality of user stories and to estimate the effort during sprint planning.
Butt et al.~\cite{butt2022software} proposed a web system to collect important information to facilitate the estimation process and to reduce experts' bias in project planning.
Nunes and Constantine~\cite{nunes2011iucp} proposed a method to estimate the size of software projects in the Interactive Use Case Points unit.
Rosa and Jardine~\cite{rosa2023data} proposed a method to use two new size measures (i.e., ``Functional Story'' and ``issues'') in predicting the effort during release planning.

While these approaches proposed new techniques for individual estimations, other approaches focused on calculating the final estimation (i.e., estimation consensus) for the team.
Alsaadi et al.~\cite{alsaadi2021scrum} proposed a tool to provide appropriate estimation points and calculate average estimations.
Beggar~\cite{el2023ifejm} proposed a fuzzy expert judgment method to help practitioners arrive consensus when performing expert judgment estimation.
Rola and Kuchta~\cite{rola2019application} proposed an estimation method using fuzzy numbers and rules to help practitioners form estimation consensus in sprint planning.
}


\subsubsection{\textbf{\purposesupportlarge{}}} \label{sec:litrevpurposesupport}

\begin{table}
\caption{The studies that proposed approaches to support the effort estimation process. We summarized the context and scope of these studies in Appendix (Table~\ref{tab:table_litrev_studylist_rq2support}).
}
\label{tab:table_litrev_result_approachsupport}
\centering
\resizebox{0.9\linewidth}{!}{
\setlength\tabcolsep{2pt}
\begin{tabular}{l|l|l}
\hline 
\textbf{Purposes} & \textbf{Planning level} & \textbf{Technique}         \\
\hline
\textbf{\litrevsupportcatone{}} & \\
- applying INVEST criteria to ensure the quality of user stories~\cite{buglione2013improving} & Sprint planning & manual\\
- using developer story to document technical information~\cite{algarni2019applying} & Not specified & manual \\

\textbf{\litrevsupportcattwo{}}   & \\
- proposing a checklist of relevant information for effort estimation~\cite{usman2018developing} & Sprint planning & manual \\
- helping the team decide to collect data based on the return on investment~\cite{taibi2017operationalizing} & Sprint planning & automated \\
- displaying the possible change impact on the current system~\cite{tanveer2017utilizing} & Not specified & automated \\
\textbf{\litrevsupportcatthree{}}                              & \\
- collecting and displaying risks of the estimation~\cite{grapenthin2016supporting} & Not specified & manual \\
- proposing a three-points estimation to illustrate the uncertainty~\cite{hannay2019agile} & Not specified & manual \\
- proposing a model to predict the changes of Story Points of a \wi{}~\cite{pasuksmit2022story} & Sprint planning & automated \\
- proposing a model to predict the changes of \wis{} description~\cite{pasuksmit2022towards} & Sprint planning & automated \\

\hline
\end{tabular}
}
\end{table}

We identified \litrevsupportpapertext{} studies that proposed the approaches to \purposesupport{}.
More specifically, these approaches were proposed to support the existing effort estimation processes or techniques by \litrevsupportcatonesmall{}, \litrevsupportcattwosmall{}, and \litrevsupportcatthreesmall{}.
majority of these approaches are manual and require the team's effort to operate. 
Table~\ref{tab:table_litrev_result_approachsupport} lists the approaches in this category.
Below, we present the approaches based on their purposes.

\smallheading{\textbf{\litrevsupportcatone{}}}
Two approaches were proposed to improve the quality of the available information that will be used during effort estimation.
Buglione and Abran~\cite{buglione2013improving} proposed the use of INVEST criteria (i.e., Independent, Negotiable, Valuable, Estimable, Small, Testable) to improve the quality of user stories for effort estimation and sprint planning.
The process requires the team and the customers to determine whether a user story meets the INVEST criteria before effort estimation.
The authors evaluated the approach in one company case study and reported the lessons learned.
Algarni and Magel~\cite{algarni2019applying} proposed a documentation form called ``developer story'' to capture the technical information for effort estimation.
Unlike a user story, a developer story typically includes the class design story with a list of methods and the method's contact criteria.
The authors evaluated the usefulness of developer stories by applying them as an input parameter to predict the source code size in 30 open-source Java systems.

\smallheading{\textbf{\litrevsupportcattwo{}}}
Three approaches were proposed to identify additional information to be used during effort estimation.
Usman et al.~\cite{usman2018developing} proposed a checklist to help the team recall relevant factors that should be considered during effort estimation, e.g., the team's skills, team domain knowledge, clarity of the requirements, and team recent productivity.
The authors evaluated the checklist with three companies.
However, the developers of one company decided to opt out of the checklist as it requires a lot of manual effort to operate.
Taibi et al.~\cite{taibi2017operationalizing} adapted the concept of Return on Invested Time (ROIT) to help the teams decide whether a metric (e.g., actual time spent, estimation error) should be collected for effort estimation.
The ROIT can be calculated as $\frac{TS - TC}{TC}$, where TS is the time saved (i.e., the smaller estimation error) with the help of the collected metric and TC is the time spent to collect the metric.
In this concept, a metric should be collected when the ROIT is positive.
The authors conducted a multiple-cases study in seven Agile projects and found that the approach could improve the estimation accuracy.
Tanveer et al.~\cite{tanveer2017utilizing} proposed a framework with a mock-up system that integrates the change impact analysis to provide additional information for effort estimation.
Given a user story, the system shows the methods (i.e., functions) that may be impacted by the implementation.
For each method, the system also shows its historical changes, dependency graph, and relevant code metrics such as size and complexity.
The authors evaluated this framework with six participants from three Agile teams at a company.
The authors reported that the framework helps visualize the impact and complexity of a change during effort estimation using Planning Poker.

\smallheading{\textbf{\litrevsupportcatthree{}}}
Two approaches were proposed to identify the uncertainty in effort estimation.
Grapenthin et al.~\cite{grapenthin2016supporting} suggested a practice to annotate the risks and effort drivers of user stories for effort estimation.
This practice requires the team to manually annotate the risks or effort drivers of user stories before the estimation session.
The authors evaluated the approach with a software features dataset from student teams and reported that using this approach could increase the effort estimation accuracy.
Hannay et al.~\cite{hannay2019agile} proposed a three-points estimation method to illustrate the uncertainty when estimating the effort of Epics (i.e., large and vague \wis{}).
In this method, the teams have to provide three estimates, i.e., bad, neutral, and good cases.
With the three-points estimation, the authors used Monte-Carlo simulation to illustrate the possible outcome of eight epics.
\makeitred{Pasuksmit et al.~\cite{pasuksmit2022story, pasuksmit2022towards} proposed two machine learning-based approaches to predict the future changes of Story Points~\cite{pasuksmit2022story} and the future changes of descriptions~\cite{pasuksmit2022towards} of a \wi{}.
The authors stated that, to avoid data leakage, these approaches  \textit{``use the information available when each \wi{} was assigned to the sprint to align with the realistic usage scenario where the information may be incomplete.''}
They found that the past tendency and the \wis{} description are the most influential prediction factors, and the correctly predicted description changes are related to scope modifications~\cite{pasuksmit2022towards}.
}

\subsubsection{\textbf{Planning levels}} \label{sec:litrevplanninglevelresult}
Table~\ref{tab:table_litrev_result_approachestimation} and Table~\ref{tab:table_litrev_result_approachsupport} show the planning levels for which the approaches were designed to be used.
We identified five planning levels: sprint planning, release planning, project planning, project bidding, and maintenance.
\revised{SEGRESS 20d}{Most of the studies proposed the approaches for a single planning level.
However, Choetkiertikul et al.~\cite{Choetkiertikul2019} proposed a Story Points prediction approach and noted that the predicted Story Points could be utilized for both sprint planning and release planning. 
Hence, we considered this approach for both planning levels (heterogeneity case).}
Below, we discuss the approaches based on the associated planning levels.
Since effort estimation in Agile is mainly conducted at the sprint planning level~\cite{usman2017effort}, our discussion focuses on the approaches proposed for sprint planning.
Then, we discuss the approaches designed for other planning levels (i.e., release planning, project planning, project bidding, and maintenance) and the approaches that did not explicitly specify the planning level.


\smallheading{\textbf{Sprint planning level}}
Table~\ref{tab:table_litrev_result_approachestimation} lists \litrevestimatepredictpapersprintplanning{} approaches that were proposed to \purposeestimate{} for (or during) sprint planning or iteration planning.
To help the teams in sprint planning, seven prediction models were proposed to \purposeestimate{} of a \wi{}~\cite{alostad2017fuzzy, scott2018, malgonde2019ensemble, dantas2019effort, Choetkiertikul2019, fu2022gpt2sp, tawosi2022investigating} or the total effort of a sprint~\cite{ramessur2021predictive}.
Five manual estimation techniques were proposed to \purposeestimate{} of \wis{} during sprint planning~\cite{vetro2018combining, altaleb2020pair, madya2022prep, alsaadi2021scrum, rola2019application}.
These approaches were evaluated using the information extracted from the \wis{} or sprints.
The extracted information includes the properties of \wis{} or sprints~\cite{alostad2017fuzzy, scott2018, malgonde2019ensemble}, the textual description of \wis{}~\cite{scott2018, Choetkiertikul2019, ramessur2021predictive, fu2022gpt2sp}, the experience and workload of developers~\cite{alostad2017fuzzy, scott2018, malgonde2019ensemble, dantas2019effort, ramessur2021predictive, tawosi2022investigating}, or the developing function~\cite{dantas2019effort}.
For example, Choetkiertikul et al.~\cite{Choetkiertikul2019} trained a neural network model to predict the \spfull{} of \wis{} using the textual description.
This model is proposed to be used either as an automated \spfull{} prediction model or as a decision support system.
Malgonde and Chari~\cite{malgonde2019ensemble} trained an ensemble-based model to predict the \spfull{} of \wis{} using priority, size, sprint, subtasks, and developer's experience.
The authors noted that these kinds of information were chosen as they assumed that they would be \textit{``readily available when a story is created.''}
Nevertheless, as information keeps changing in Agile~\cite{hoda2016multilevel, masood2020real}, it is unclear whether the authors of the prior studies used the information available during the sprint planning or used the latest information version (which is considered future data).


Table~\ref{tab:table_litrev_result_approachsupport} lists \litrevsupportforsprintplanning{} approaches that were proposed to \purposesupport{} during sprint planning.
\makeitred{These approaches were evaluated in Agile settings at the sprint planning level.}
Buglione and Abran~\cite{buglione2013improving} proposed the use of INVEST criteria to ensure the quality of user stories before effort estimation and sprint planning processes.
Usman et al.~\cite{usman2018developing} developed a checklist to help the team recall relevant factors during effort estimation at the sprint planning level.
Taibi et al.~\cite{taibi2017operationalizing} adapted the concept of Return On Invested Time (ROIT) to help the teams decide to collect the metrics for effort estimation at the sprint planning level.
\makeitred{Pasuksmit et al.~\cite{pasuksmit2022story, pasuksmit2022towards} proposed two machine learning-based approaches to predict the changes of Story Points~\cite{pasuksmit2022story} and the changes of \wi{} descriptions~\cite{pasuksmit2022towards} that occurred after the sprint had started.
The authors trained and evaluated these two prediction approaches using only \textit{``the information available when each \wi{} was assigned to the sprint''} to avoid data leakage.}


\smallheading{\textbf{Other levels}}
We identified \litrevestimateforotherlevels{} studies that proposed the approaches to \purposeestimate{} for release planning, project planning, project bidding, or maintenance phase (see Table~\ref{tab:table_litrev_result_approachestimation}).
\makeitred{Four studies proposed the approaches that could be used for release planning.
Kang et al.~\cite{kang2010model} and Winska et al.~\cite{winska2021reducing} proposed the approaches to estimate and track the effort.
Rosa and Jardine~\cite{rosa2023data} proposed an estimation method based on two new measures, i.e., ``Functional Story'' and ``Issues.''
Choetkiertikul et al.~\cite{Choetkiertikul2019} proposed a Story Points prediction approach and noted that the predicted Story Points can also be used for release planning.}
\litrevestimatepaperprojectplanningtextbig{} studies proposed the approaches to estimate the effort for project planning~\cite{premalatha2019effort, dragicevic2017bayesian, sanchez2022software, rodriguez2023effort, sharma2022combined, bilgaiyan2018chaos, raslan2018enhanced, singal2022integrating, butt2022software}.
Three studies proposed parametric-based approaches for the project bidding phase~\cite{parvez2013efficiency, rosa2017early, rosa2021empirical}.
Two studies proposed the approaches to predict the effort required for software maintenance phases~\cite{choudhari2012phase} and functional size of changes~\cite{sakhrawi2021support}.
At these levels, the effort is typically estimated in man-days or monetary cost in order to support business decisions or project planning.
Since detailed information usually be absent in these early planning levels~\cite{usman2015effort}, these studies used different techniques to overcome the lack of information problem.
For example, using the early design factors (e.g., risk resolution, team cohesion) as the model inputs~\cite{raslan2018enhanced} or using an approach that is robust to small inputs~\cite{premalatha2019effort}.

\smallheading{\textbf{Planning level not specified or insufficient context}}
There are \litrevestimatenotspecified{} studies that did not clearly specify the planning level for the approaches to be used (see Table~\ref{tab:table_litrev_result_approachestimation} and Table~\ref{tab:table_litrev_result_approachsupport}).
From these studies, \litrevestimatenotspecified{} of them proposed the approaches to \purposeestimate{} of \wis{}~\cite{Porru2016, marapelli2020rnn, dan2020nlp, gultekin2020story, moharreri2016cost, miranda2021analysis, hemrajanipredicting, phan2022heterogeneous, phan2022story, kassem2023story, kassem2022software, prykhodko2019multiple, alhamed2021playing, el2023ifejm}, sprints~\cite{Hearty2009, karna2020data}, releases~\cite{paz2014approach}, requirement changes~\cite{sakhrawi2022software, basri2016algorithmic}, or projects~\cite{adnan2019ontology, bilgaiyan2019effort, khuat2018novel, farahneh2011linear, satapathy2014story, sharma2020linear, ardiansyah2022mucpso, kaushik2020fuzzified, kaushik2022role, gupta2022automated, najm2022enhanced, aslam2017risk, nunes2011iucp, panda2015empirical, govil2022estimation, arora2022efficient, butt2023prediction}.
The other \litrevsupportnotspecified{} studies proposed the approaches to \purposesupport{} by proposing a developer story form~\cite{algarni2019applying}, displaying the change impact~\cite{tanveer2017utilizing}, displaying the estimation risks~\cite{grapenthin2016supporting}, and proposing a three-point estimation technique~\cite{hannay2019agile}.

\makeitred{From these approaches, we also identified a few of them that were proposed to ``assist'' or ``replace'' the Planning Poker estimation technique but did not clearly specify the planning level.
For example, Grapenthin et al.~\cite{grapenthin2016supporting} suggested a practice to annotate the risks and effort drivers of user stories for Planning Poker estimation, Tanveer et al.~\cite{tanveer2017utilizing} proposed an integration of change impact analysis to provide additional information for Planning Poker estimation, and Alhamed and Storer~\cite{alhamed2021playing} proposed a crowd-based technique to mimic the Planning Poker estimation performed by the team of experts.
Furthermore, four effort prediction approaches were proposed to complement the Planning Poker estimation without clearly specifying the planning level, i.e., Moharreri et al.~\cite{moharreri2016cost}, Phan and Jannesari~\cite{phan2022heterogeneous}, and Kassem et al.~\cite{kassem2023story, kassem2022software}.
Nevertheless, Mahnic et al.~\cite{Mahnic2012} suggested that Planning Poker can be used for sprint planning or release planning~\cite{Mahnic2012, rubin2012essential}.
Thus, we can imply that these proposed approaches might be suitable for sprint planning or release planning levels.
}


\begin{tcolorbox}[boxsep=1pt,left=2pt,right=2pt,top=2pt,bottom=2pt]
    \textbf{Findings:} We identified \litrevapproachpaper{} studies that proposed the approaches for effort estimation in Agile, which can be categorized into two main purposes.
    We found that \litrevestimatepaper{} approaches aim to \purposeestimate{}, while only \litrevsupportpaper{} approaches aim to \purposesupport{} (i.e., \litrevsupportcatonesmall{}, \litrevsupportcattwosmall{}, \litrevsupportcatthreesmall{}).
    Majority of these approaches were proposed to be used for sprint planning.
    However, \makeitred{for many of them}, we observed that it is unclear whether they have been evaluated based on the information available during the sprint planning or not.
    
    
\end{tcolorbox}
\section{Discussions}\label{sec:litrevdiscussion}

\revised{R2-7}{
We conducted a systematic literature review on \litrevtotalpaper{} studies where \litrevreasonspaper{} studies investigated the \reasons{} in Agile (RQ1) and \litrevapproachpaper{} studies proposed the \approaches{} in Agile (RQ2).
Note that we found one study included in both RQs.
These studies were published from year 2008 to 2023.
We now discuss broader implications and provide recommendations for practitioners and researchers based on our findings.

\subsection{Implications}\label{sec:litrevimplications}

\hspace{\parindent}\textbf{Poor information quality is the common \reason{}.}
In RQ1, we observed that the \litrevreasonquality{} (i.e., whether the information is unclear, unstable, or error) are the common \reason{} that are often reported in high ranking.
Some may argue that such quality issues are generally expected since the information in Agile is typically sketched out and changed (refined) over time~\cite{ernst2012case, hoda2010much, heck2017framework}.
However, prior studies reported that information quality issues can be discovered late during the sprint implementation, which could cause the estimation to become inaccurate (RQ1)~\cite{Britto2015, conoscenti2019combining}.
Even though the team can re-estimate to maintain the estimation accuracy~\cite{hoda2016multilevel, Bick2018, jpsurvey}, doing so after the sprint planning may cause the sprint plan to become unreliable~\cite{Cohn2006}.
In response to these issues, the practitioners may become \textit{``more conservative in their estimates mainly due to a high level of uncertainty or lack of detail''}~\cite{usman2018effort} (i.e., intentionally overestimate the effort).
These findings highlight that the available information quality is critical for effort estimation in Agile iterative development.

\textbf{Little has investigated the approaches to improve the information quality.}
While our RQ1 found that the quality of the available information is important for the estimation, our RQ2 only identified \litrevsupportpaper{} studies that proposed an approach to \purposesupport{} by \litrevsupportcatonesmall{}, \litrevsupportcattwosmall{}, and \litrevsupportcatthreesmall{}.
Furthermore, the majority of these approaches require team's manual effort of the developers to operate.
As reported by Usman et al.~\cite{usman2018developing}, the team may be reluctant to use such approaches due to the additional overhead.
These observations suggest that there is a need for an automated approach to \purposesupport{}, especially to improve the quality of the available information.

\hspace{\parindent}\textbf{Challenges related to team and estimation practices could impact effort estimation accuracy.}
Although in lower ranks, our study identified frequently recurrent themes (see RQ1), particularly the lack of experience of team members (R2.1; reported by five studies) and factors overlooking (R3.1; reported by four studies).
More specifically, an inexperienced team member (R2.1) tends to \textit{``exhibit the tendency to consider the best case scenario only''} (i.e., overoptimism)~\cite{usman2015effort}.
Intuitively, considering only best-case scenarios could lead the team to overlook critical information (R3.1), e.g., non-functional requirements~\cite{sandeep2022effort, usman2015effort}, test effort~\cite{sandeep2022effort, usman2015effort}, or the complexity of a developing function~\cite{conoscenti2019combining, usman2018effort}.
These findings point out the importance of equipping teams with the necessary skills and knowledge for effort estimation.

\hspace{\parindent}\textbf{The majority of the proposed approaches aimed to estimate the effort.}
In our RQ2, we observed a significant trend where the majority of the proposed approaches focused on effort estimation (\litrevestimatepaper{} studies) rather than support mechanisms for estimation processes (\litrevsupportpaper{} studies).
A significant observation is the predominance of machine learning-based techniques, including both traditional machine learning and deep learning models (e.g., regression using LSTM + RHWN~\cite{Choetkiertikul2019}, classification using SVMs~\cite{scott2018}, Deep Attention Neural network~\cite{kassem2023story}).
Some recent studies leveraged recent approaches (in 2022) like FastTexts~\cite{phan2022heterogeneous} or GPT-2~\cite{fu2022gpt2sp} that overcome the previous baselines.
This highlights an active research trend on the automated approaches for effort estimation. 
With the emerging advanced AI trends (e.g., GPT-4), we may expect better prediction accuracy while requiring less training data in the near future.

\textbf{Many effort estimation approaches for sprint planning may not be validated using realistic information.}
In RQ2, we found that \litrevestimatepaperwithplanninglevel{} studies specified the planning level for which the approaches to be used, where the majority of them (\litrevestimatepapersprintplanning{} studies) proposed to be used for sprint planning.
Intuitively, an effort prediction approach (especially machine learning-based) should be validated using the available information at the specified planning stage.
For example, Pasuksmit et al.~\cite{pasuksmit2022story, pasuksmit2022towards} validated their prediction approaches using the information available during sprint planning.
However, it is unclear whether other approaches were validated using only the available information or the future data (i.e., subjected to data leakage).
For example, Malgonde and Chari~\cite{malgonde2019ensemble} built an effort prediction model using the \wi{} variables (e.g., priority, size, subtasks) while assuming that the variables will be \textit{``readily available when a story is created.''}
While their assumption is correct, it is still unclear which version of information they used.
On the other hand, using the latest and complete information may not reflect a realistic usage scenario as the information in Agile is typically sketched out and refined over time~\cite{hoda2016multilevel, Bick2018, masood2020real}.
Therefore, the performance of these approaches may be sub-optimal in industrial settings.


\textbf{Validating the approach using artificial datasets or student project data may pose a risk to generalizability.}
In RQ2, we identified seven studies at high risk of uncertainty due to their reliance on generated datasets~\cite{kaushik2020fuzzified, ramessur2021predictive, aslam2017risk} or student project data~\cite{alsaadi2021scrum, nunes2011iucp, paz2014approach, grapenthin2016supporting}.
These data sources may lack the complexity, scale, and real-world Agile expertise necessary to accurately represent the generalizability of the approach in industrial contexts~\cite{kitchenham2022segress}.
Nevertheless, they may be acceptable for primitively validating the performance of an approach, especially when data is scarce.





\subsection{Recommendations for Agile Practitioners}

\hspace{\parindent}\textbf{Practitioners should prioritize enhancing the quality of the available information used in effort estimations.}
Particularly, the team should be aware of- and mitigate the common information quality issues (i.e., unclear, unstable, or error information).
The reported quality issues are related to user stories~\cite{usman2015effort, conoscenti2019combining}, user acceptance criteria~\cite{conoscenti2019combining}, and requirements~\cite{usman2015effort, usman2018effort, Britto2015}.
As recommended in the literature~\cite{jpsurvey, madampe2020multi}, the practitioners should perform detail analyses or confirm the information with stakeholders prior to effort estimation, especially to ensure the quality of user stories, user acceptance criteria, requirements, and test plans.

\textbf{Practitioners should apply the proposed approaches to improve the quality of the available information for effort estimation.}
The practitioners can consider the approaches for \litrevsupportcatonesmall{} or \litrevsupportcattwosmall{} (see Section~\ref{sec:litrevpurposesupport}), i.e., applying INVEST criteria on user stories~\cite{buglione2013improving}, adopting developer story for technical information~\cite{algarni2019applying}, using an estimation checklist~\cite{usman2018effort}, surfacing the change impact~\cite{tanveer2017utilizing}, or consider the return on investment in data collection~\cite{taibi2017operationalizing}.
In addition, the practitioners may also consider the approaches to identify the uncertainty (changes) in the information of \wis{}~\cite{pasuksmit2022towards}.
Although not directly proposed in the Agile estimation context, some approaches might be worth exploring for improving the information quality, e.g., identifying the missing information~\cite{Zimmermann2010, Chaparro2017}, extracting quality attributes~\cite{gilson2019extracting}, or generating use cases or test cases~\cite{gilson2020generating, fischbach2020specmate}.

\textbf{Practitioners should equip themselves with relevant domain and technical knowledge for effort estimation.}
The lack of experience of team members on the technology, domain knowledge, and effort estimation practices might lead to inaccurate estimations~\cite{conoscenti2019combining, sandeep2022effort, usman2015effort, Britto2015} (see Section~\ref{sec:litrevreasonsection}).
This also includes novice team members~\cite{conoscenti2019combining, sandeep2022effort, usman2018effort, usman2015effort} and the team with low cohesion~\cite{Britto2015}.
In particular, this problem could lead the estimating team to overlook critical information during effort estimations (e.g., non-functional requirements~\cite{sandeep2022effort, usman2015effort}, testing effort~\cite{sandeep2022effort, usman2015effort}, or function complexity~\cite{conoscenti2019combining, usman2018effort}), which eventually causes inaccurate estimations.
To address this challenge, Usman et al.~\cite{usman2018effort} recommended that \textit{``mature teams should be involved in the effort estimation process as they have architectural knowledge and expertise''}~\cite{usman2018effort}.
They noted that expert mentoring, especially from product architects, is \textit{``critical to achieve technical consistency''} in projects with teams of varying maturity levels.

\textbf{Practitioners should assess the generalizability of the automated approaches prior to implementation.}
Our SLR points out the potential data leakage issue in the validation process of some automated approaches (see Sections \ref{sec:litrevplanninglevelresult} and \ref{sec:litrevimplications}).
In particular, these approaches may not have been validated only with the available information.
Since information can be changed in Agile~\cite{Bick2018, hoda2016multilevel, masood2020real}, relying on final or updated information for validation could result in data leakage, potentially leading to sub-optimal performance in industrial adoption. 
Hence, practitioners should adopt approaches that effectively mitigate data leakage and are validated in realistic settings.

\subsection{Implications for Future Research}

\hspace{\parindent}\textbf{Future research should prioritize the development of automated approaches for improving the information quality and identifying the additional information for effort estimation.}
Our SLR identified only two manual approaches focused on improving information quality, and three approaches (one manual) aimed to assist in identifying additional information for effort estimation.
This scarcity highlights the need for automated systems to identify or improve the quality of the information for effort estimation (e.g., the information elements suggested by Pasuksmit et al.~\cite{jpsurvey}).
The future approaches may adopt the existing solutions in other contexts.
For example, identifying missing information~\cite{Zimmermann2010, Chaparro2017} or generating information~\cite{gilson2020generating}.
In addition, future work may utilize a large language model to suggest related content (e.g., user stories~\cite{marczak2023using}).
These approaches will potentially enrich the quality of information used for effort estimation in Agile.

\textbf{Future research should focus on validating approaches using realistic scenario information.}
Our RQ2 highlighted the risk of information leakage in the validation of several approaches.
For instance, approaches intended for sprint planning, where information is subject to change~\cite{hoda2016multilevel, Bick2018}, may be validated using the latest information (see Section~\ref{sec:litrevimplications}).
This raises concerns about the generalizability and practical relevance of these methods. 
Hence, future research should validate the approaches using only information available at the intended time of practical use. 
For example, when proposing a prediction approach for sprint planning at the \wis{} level, such as Jira issues~\cite{pasuksmit2022story, pasuksmit2022towards}, researchers should utilize historical logs to revert issue fields to their state during sprint planning. 
This approach would provide a more accurate dataset for evaluating the effectiveness of proposed methods.

\textbf{Future research should validate research findings in industrial contexts.}
Our RQ2 noted that some studies utilized a generated dataset or student projects to validate the proposed approaches.
While useful for initial validations, such datasets may not fully reflect the complexities of industrial scenarios~\cite{kitchenham2022segress}.
When it is challenging to find a representative dataset, an effective strategy would be to utilize the crowd-sourcing approach used by Alhamed and Storer~\cite{alhamed2021playing} or use the open-source project datasets shared by Tawosi et al.~\cite{tawosi2022versatile} or Choetkiertikul et al.~\cite{Choetkiertikul2019}).
These methods would substantially improve the reliability and relevance of research, bridging the gap between academic findings and practical industry challenges.

}
\section{Threats to Validity}\label{sec:litrevthreats}

This section discusses the potential threats to the validity of this systematic literature review (SLR).

\textbf{Construct validity} is related to the process of identifying the studies of this SLR.
There might be a chance that some relevant studies were not retrieved when using our search terms.
Changing the search terms (e.g., removing the ``Agile'' keyword) may impact our search results by including other studies in the literature review.
Yet, we designed our search terms to focus our literature review on the studies in a specific context.
To mitigate the risk of excluding relevant studies, we strive to extend our search by adding alternatives and synonyms of the main search terms.
Table~\ref{tab:searchterm} lists our search terms and their synonyms, which we believe that they are sufficient for covering the two research areas.

\revised{R1-3}{
The five search engines we used may not include every study related to the scope of our RQs.
In this study, we opted to use the five common search engines to align with the past SLRs in the related areas (i.e., \cite{alsaadi2022data, Usman2014, dantas2018effort, heck2018systematic}), which appeared to cover the major software engineering journals and conferences.
Nevertheless, some Agile studies may not be included in this paper.
For example, a study by Majchrzak and Madeyski~\cite{majchrzak2016factors} on the \reasons{} was not listed on any of the five search engines as they were published in a non-software engineering publication venue.
Future work that aims to review Agile studies in other contexts (e.g., management and economics) may need to consider additional search engines.
}


\textbf{Internal validity} is related to the confounding factors that might impact our study.
\revised{SEGRESS 23b}{We assessed the risk of uncertainty and bias assessment using GRADE as suggested by Kitchenham et al.~\cite{kitchenham2022segress} (see Section~\ref{sec:litrevphased}).
A few of our selected studies have a high risk of uncertainty (see Table~\ref{tab:table_litrev_studylist_rq1reasons}, Table~\ref{tab:table_litrev_studylist_rq2estimate}, and Table~\ref{tab:table_litrev_studylist_rq2support}).
However, even when we excluded these studies with the high risk out, our findings (themes and taxonomies) remained unchanged.
}

\revised{SEGRESS 23c}{
The first author conducted the first iteration of card sorting and manual analyses to discover the thematic taxonomies, other related information, and the risks of uncertainty and bias of individual studies (see Section~\ref{sec:litrevdataanalysis}).
The subjective opinion and personal experience of the first author might influence the analyses.
To mitigate this risk, the second author (with a different background and expertise) reviewed the results and discussed the disagreements with the first author.
The first author then conducted another round of analyses and reviewed all results until both reached a consensus.
Hence, our card sorting process is not subjective only to the first author.
}

We retrieved the ranks of \reasons{} from the selected studies.
These ranks originated from different experimental designs and ranking methods (i.e., the number of participants, occurrences observed, and agreement scores).
Comparing these rankings directly may misled our findings.
Therefore, we limited the interpretation of the ranking only to support the consensus of the selected studies on the common \reasons{}.
We believe that the risk of misinterpretation of the rankings on our findings is minimal.

\section{Conclusions}\label{sec:litrevconclusions}

Effort estimation is an integral part of Agile iterative development.
Accurate effort estimations help teams achieve reliable sprint planning, ensuring reliable delivery of software increments.
While many studies have investigated the \reasons{} and proposed \approaches{} in the Agile context, there was a gap in aggregating and synthesizing evidence regarding these research areas.
Our systematic literature review addressed this gap by focusing on two key research questions:

\begin{list}{}{\leftmargin=0pt}
\item \textbf{(RQ1) \litrevrqone{}}
\item \textbf{(RQ2) \litrevrqtwo{}}
\end{list}

Our RQ1 revealed that the quality of available information is a commonly reported \reason{} impacting estimation accuracy. 
In addition, team-related, estimation practice, project management, and business influence were also notable reasons. 
In RQ2, we observed that approaches for estimating the effort were predominantly explored, with fewer studies focusing on supporting the effort estimation process.
However, the validation process of some automated approaches was questioned, particularly regarding potential data leakage in validation and the use of indirect validation datasets or participants.

These findings highlight the need for future work to focus on improving the quality of available information for effort estimation with minimal overhead.
Practitioners should consider adopting an automated approach to help them improve the information quality that has been carefully evaluated in realistic scenarios.
It is crucial for future research to validate the proposed approaches using only the available information, while ensuring the datasets or participants reflect the complexities of industrial Agile environments.
Future research should also revisit the previously proposed effort prediction models using only the available information, as it remains uncertain whether they considered this aspect adequately.
This will bridge the gap between academic research and industry practice, enhancing the practical application of effort estimation in Agile iterative development.

\bibliographystyle{ACM-Reference-Format}
\bibliography{main}


\begin{thebibliography}{134}


\ifx \showCODEN    \undefined \def \showCODEN     #1{\unskip}     \fi
\ifx \showDOI      \undefined \def \showDOI       #1{#1}\fi
\ifx \showISBNx    \undefined \def \showISBNx     #1{\unskip}     \fi
\ifx \showISBNxiii \undefined \def \showISBNxiii  #1{\unskip}     \fi
\ifx \showISSN     \undefined \def \showISSN      #1{\unskip}     \fi
\ifx \showLCCN     \undefined \def \showLCCN      #1{\unskip}     \fi
\ifx \shownote     \undefined \def \shownote      #1{#1}          \fi
\ifx \showarticletitle \undefined \def \showarticletitle #1{#1}   \fi
\ifx \showURL      \undefined \def \showURL       {\relax}        \fi
\providecommand\bibfield[2]{#2}
\providecommand\bibinfo[2]{#2}
\providecommand\natexlab[1]{#1}
\providecommand\showeprint[2][]{arXiv:#2}

\bibitem[Adnan et~al\mbox{.}(2019)]%
        {adnan2019ontology}
\bibfield{author}{\bibinfo{person}{Muhammad Adnan}, \bibinfo{person}{Muhammad
  Afzal}, {and} \bibinfo{person}{Khadim~Hussain Asif}.}
  \bibinfo{year}{2019}\natexlab{}.
\newblock \showarticletitle{{Ontology-Oriented Software Effort Estimation
  System for E-commerce Applications Based on Extreme Programming and Scrum
  Methodologies}}.
\newblock \bibinfo{journal}{\emph{{The Computer Journal}}}
  \bibinfo{volume}{62}, \bibinfo{number}{11} (\bibinfo{year}{2019}),
  \bibinfo{pages}{1605--1624}.
\newblock


\bibitem[Algarni and Magel(2019)]%
        {algarni2019applying}
\bibfield{author}{\bibinfo{person}{Asaad Algarni} {and}
  \bibinfo{person}{Kenneth Magel}.} \bibinfo{year}{2019}\natexlab{}.
\newblock \showarticletitle{{Applying Software Design Metrics to Developer
  Story: A Supervised Machine Learning Analysis}}. In
  \bibinfo{booktitle}{\emph{Proceedings of the IEEE International Conference on
  Cognitive Machine Intelligence (CogMI)}}. \bibinfo{pages}{156--159}.
\newblock


\bibitem[Alhamed and Storer(2021)]%
        {alhamed2021playing}
\bibfield{author}{\bibinfo{person}{Mohammed Alhamed} {and} \bibinfo{person}{Tim
  Storer}.} \bibinfo{year}{2021}\natexlab{}.
\newblock \showarticletitle{{Playing Planning Poker in Crowds: Human
  Computation of Software Effort Estimates}}. In
  \bibinfo{booktitle}{\emph{Proceedings of the International Conference on
  Software Engineering (ICSE)}}. \bibinfo{pages}{1--12}.
\newblock


\bibitem[Alostad et~al\mbox{.}(2017)]%
        {alostad2017fuzzy}
\bibfield{author}{\bibinfo{person}{Jasem~M Alostad}, \bibinfo{person}{Laila~RA
  Abdullah}, {and} \bibinfo{person}{Lamya~Sulaiman Aali}.}
  \bibinfo{year}{2017}\natexlab{}.
\newblock \showarticletitle{{A Fuzzy based Model for Effort Estimation in Scrum
  Projects}}.
\newblock \bibinfo{journal}{\emph{International Journal of Advanced Computer
  Science and Applications}} \bibinfo{volume}{8}, \bibinfo{number}{9}
  (\bibinfo{year}{2017}), \bibinfo{pages}{270--277}.
\newblock


\bibitem[Alsaadi et~al\mbox{.}(2021)]%
        {alsaadi2021scrum}
\bibfield{author}{\bibinfo{person}{Bushra Alsaadi}, \bibinfo{person}{Bashaer
  Alsaadi}, \bibinfo{person}{Mashaal Alfhaid}, \bibinfo{person}{Athir
  Alghamdi}, \bibinfo{person}{Nedaa Almuallim}, {and} \bibinfo{person}{Bahjat
  Fakieh}.} \bibinfo{year}{2021}\natexlab{}.
\newblock \showarticletitle{Scrum Poker Estimator: A Planning Poker Tool for
  Accurate Story Point Estimation.}
\newblock \bibinfo{journal}{\emph{International Journal of Computer Information
  Systems \& Industrial Management Applications}}  \bibinfo{volume}{13}
  (\bibinfo{year}{2021}).
\newblock


\bibitem[Alsaadi and Saeedi(2022)]%
        {alsaadi2022data}
\bibfield{author}{\bibinfo{person}{Bashaer Alsaadi} {and}
  \bibinfo{person}{Kawther Saeedi}.} \bibinfo{year}{2022}\natexlab{}.
\newblock \showarticletitle{Data-driven effort estimation techniques of agile
  user stories: a systematic literature review}.
\newblock \bibinfo{journal}{\emph{Artificial Intelligence Review}}
  (\bibinfo{year}{2022}), \bibinfo{pages}{1--32}.
\newblock


\bibitem[Altaleb et~al\mbox{.}(2020a)]%
        {altaleb2020pair}
\bibfield{author}{\bibinfo{person}{Abdullah Altaleb}, \bibinfo{person}{Muna
  Altherwi}, {and} \bibinfo{person}{Andy Gravell}.}
  \bibinfo{year}{2020}\natexlab{a}.
\newblock \showarticletitle{{A Pair Estimation Technique of Effort Estimation
  in Mobile App Development for Agile Process: Case Study}}. In
  \bibinfo{booktitle}{\emph{Proceedings of the International Conference on
  Information Science and Systems (ICISS)}}. \bibinfo{pages}{29--37}.
\newblock


\bibitem[Altaleb et~al\mbox{.}(2020b)]%
        {altaleb2020industrial}
\bibfield{author}{\bibinfo{person}{Abdullah Altaleb}, \bibinfo{person}{Muna
  Altherwi}, {and} \bibinfo{person}{Andy Gravell}.}
  \bibinfo{year}{2020}\natexlab{b}.
\newblock \showarticletitle{{An Industrial Investigation into Effort Estimation
  Predictors for Mobile App Development in Agile Processes}}. In
  \bibinfo{booktitle}{\emph{Proceedings of the International Conference on
  Industrial Technology and Management (ICITM)}}. \bibinfo{pages}{291--296}.
\newblock


\bibitem[Andrew and Selamat(2012)]%
        {andrew2012systematic}
\bibfield{author}{\bibinfo{person}{Beatrice Andrew} {and} \bibinfo{person}{Ali
  Selamat}.} \bibinfo{year}{2012}\natexlab{}.
\newblock \showarticletitle{{Systematic Literature Review of Missing Data
  Imputation Techniques for Effort Prediction}}. In
  \bibinfo{booktitle}{\emph{Proceedings of the International Conference on
  Information and Knowledge Management}}. \bibinfo{pages}{222--226}.
\newblock


\bibitem[Ardiansyah et~al\mbox{.}(2022)]%
        {ardiansyah2022mucpso}
\bibfield{author}{\bibinfo{person}{Ardiansyah Ardiansyah},
  \bibinfo{person}{Ridi Ferdiana}, {and} \bibinfo{person}{Adhistya~Erna
  Permanasari}.} \bibinfo{year}{2022}\natexlab{}.
\newblock \showarticletitle{{MUCPSO: A Modified Chaotic Particle Swarm
  Optimization with Uniform Initialization for Optimizing Software Effort
  Estimation}}.
\newblock \bibinfo{journal}{\emph{Applied Sciences}} \bibinfo{volume}{12},
  \bibinfo{number}{3} (\bibinfo{year}{2022}), \bibinfo{pages}{1081}.
\newblock


\bibitem[Arora et~al\mbox{.}(2022)]%
        {arora2022efficient}
\bibfield{author}{\bibinfo{person}{Mohit Arora}, \bibinfo{person}{Sahil Verma},
  \bibinfo{person}{Kavita}, \bibinfo{person}{Marcin Wozniak},
  \bibinfo{person}{Jana Shafi}, {and} \bibinfo{person}{Muhammad~Fazal Ijaz}.}
  \bibinfo{year}{2022}\natexlab{}.
\newblock \showarticletitle{An efficient ANFIS-EEBAT approach to estimate
  effort of Scrum projects}.
\newblock \bibinfo{journal}{\emph{Scientific Reports}} \bibinfo{volume}{12},
  \bibinfo{number}{1} (\bibinfo{year}{2022}), \bibinfo{pages}{7974}.
\newblock


\bibitem[Aslam et~al\mbox{.}(2017)]%
        {aslam2017risk}
\bibfield{author}{\bibinfo{person}{Waqar Aslam}, \bibinfo{person}{Farah Ijaz},
  \bibinfo{person}{Muhammad Ikram~Ullah Lali}, {and} \bibinfo{person}{Waqar
  Mehmood}.} \bibinfo{year}{2017}\natexlab{}.
\newblock \showarticletitle{Risk Aware and Quality Enriched Effort Estimation
  for Mobile Applications in Distributed Agile Software Development.}
\newblock \bibinfo{journal}{\emph{Journal of Information Science and
  Engineering}} \bibinfo{volume}{33}, \bibinfo{number}{6}
  (\bibinfo{year}{2017}), \bibinfo{pages}{1481--1500}.
\newblock


\bibitem[Basri et~al\mbox{.}(2016)]%
        {basri2016algorithmic}
\bibfield{author}{\bibinfo{person}{Sufyan Basri}, \bibinfo{person}{Nazri Kama},
  \bibinfo{person}{Haslina~Md Sarkan}, \bibinfo{person}{Saiful Adli}, {and}
  \bibinfo{person}{Faizura Haneem}.} \bibinfo{year}{2016}\natexlab{}.
\newblock \showarticletitle{An Algorithmic-Based Change Effort Estimation Model
  for Software Development}. In \bibinfo{booktitle}{\emph{Proceedings of the
  Asia-Pacific Software Engineering Conference (APSEC)}}.
  \bibinfo{pages}{177--184}.
\newblock


\bibitem[{Bick} et~al\mbox{.}(2018)]%
        {Bick2018}
\bibfield{author}{\bibinfo{person}{S. {Bick}}, \bibinfo{person}{K. {Spohrer}},
  \bibinfo{person}{R. {Hoda}}, \bibinfo{person}{A. {Scheerer}}, {and}
  \bibinfo{person}{A. {Heinzl}}.} \bibinfo{year}{2018}\natexlab{}.
\newblock \showarticletitle{{Coordination Challenges in Large-Scale Software
  Development: A Case Study of Planning Misalignment in Hybrid Settings}}.
\newblock \bibinfo{journal}{\emph{Transactions of Software Engineering (TSE)}}
  \bibinfo{volume}{44}, \bibinfo{number}{10} (\bibinfo{year}{2018}),
  \bibinfo{pages}{932--950}.
\newblock


\bibitem[Bilgaiyan et~al\mbox{.}(2018)]%
        {bilgaiyan2018chaos}
\bibfield{author}{\bibinfo{person}{Saurabh Bilgaiyan}, \bibinfo{person}{Kunwar
  Aditya}, \bibinfo{person}{Samaresh Mishra}, {and}
  \bibinfo{person}{Madhabananda Das}.} \bibinfo{year}{2018}\natexlab{}.
\newblock \showarticletitle{Chaos-based modified morphological genetic
  algorithm for software development cost estimation}. In
  \bibinfo{booktitle}{\emph{Proceedings of the Progress in Computing, Analytics
  and Networking}}. Springer, \bibinfo{pages}{31--40}.
\newblock


\bibitem[Bilgaiyan et~al\mbox{.}(2019)]%
        {bilgaiyan2019effort}
\bibfield{author}{\bibinfo{person}{Saurabh Bilgaiyan},
  \bibinfo{person}{Samaresh Mishra}, {and} \bibinfo{person}{Madhabananda Das}.}
  \bibinfo{year}{2019}\natexlab{}.
\newblock \showarticletitle{{Effort estimation in agile software development
  using experimental validation of neural network models}}.
\newblock \bibinfo{journal}{\emph{International Journal of Information
  Technology}} \bibinfo{volume}{11}, \bibinfo{number}{3}
  (\bibinfo{year}{2019}), \bibinfo{pages}{569--573}.
\newblock


\bibitem[Britto et~al\mbox{.}(2015)]%
        {Britto2015}
\bibfield{author}{\bibinfo{person}{Ricardo Britto}, \bibinfo{person}{Emilia
  Mendes}, {and} \bibinfo{person}{J{\"u}rgen B{\"o}rstler}.}
  \bibinfo{year}{2015}\natexlab{}.
\newblock \showarticletitle{{An Empirical Investigation on Effort Estimation in
  Agile Global Software Development}}. In \bibinfo{booktitle}{\emph{Proceedings
  of the International Conference on Global Software Engineering}}.
  \bibinfo{pages}{38--45}.
\newblock


\bibitem[Buglione and Abran(2013)]%
        {buglione2013improving}
\bibfield{author}{\bibinfo{person}{Luigi Buglione} {and} \bibinfo{person}{Alain
  Abran}.} \bibinfo{year}{2013}\natexlab{}.
\newblock \showarticletitle{Improving the User Story Agile Technique Using the
  INVEST Criteria}. In \bibinfo{booktitle}{\emph{Proceedings of the
  International Workshop on Software Measurement}}. \bibinfo{pages}{49--53}.
\newblock


\bibitem[Butt et~al\mbox{.}(2023)]%
        {butt2023prediction}
\bibfield{author}{\bibinfo{person}{Shariq~Aziz Butt}, \bibinfo{person}{Tuncay
  Ercan}, \bibinfo{person}{Muhammad Binsawad}, \bibinfo{person}{Paola-Patricia
  Ariza-Colpas}, \bibinfo{person}{Jorge Diaz-Martinez},
  \bibinfo{person}{Gabriel Pineres-Espitia}, \bibinfo{person}{Emiro
  De-La-Hoz-Franco}, \bibinfo{person}{Marlon Alberto~Pineres Melo},
  \bibinfo{person}{Roberto~Morales Ortega}, {and} \bibinfo{person}{Juan-David
  De-La-Hoz-Hernandez}.} \bibinfo{year}{2023}\natexlab{}.
\newblock \showarticletitle{Prediction based cost estimation technique in agile
  development}.
\newblock \bibinfo{journal}{\emph{Advances in Engineering Software}}
  \bibinfo{volume}{175} (\bibinfo{year}{2023}), \bibinfo{pages}{103329}.
\newblock


\bibitem[Butt et~al\mbox{.}(2022)]%
        {butt2022software}
\bibfield{author}{\bibinfo{person}{Shariq~Aziz Butt}, \bibinfo{person}{Ayesha
  Khalid}, \bibinfo{person}{Tuncay Ercan}, \bibinfo{person}{Paola~Patricia
  Ariza-Colpas}, \bibinfo{person}{Acosta-Coll Melisa}, \bibinfo{person}{Gabriel
  Pineres-Espitia}, \bibinfo{person}{Emiro De-La-Hoz-Franco},
  \bibinfo{person}{Marlon Alberto~Pineres Melo}, {and}
  \bibinfo{person}{Roberto~Morales Ortega}.} \bibinfo{year}{2022}\natexlab{}.
\newblock \showarticletitle{A software-based cost estimation technique in scrum
  using a developer's expertise}.
\newblock \bibinfo{journal}{\emph{Advances in Engineering Software}}
  \bibinfo{volume}{171} (\bibinfo{year}{2022}), \bibinfo{pages}{103159}.
\newblock


\bibitem[Chaparro et~al\mbox{.}(2017)]%
        {Chaparro2017}
\bibfield{author}{\bibinfo{person}{Oscar Chaparro}, \bibinfo{person}{Jing Lu},
  \bibinfo{person}{Fiorella Zampetti}, \bibinfo{person}{Laura Moreno},
  \bibinfo{person}{Massimiliano Di~Penta}, \bibinfo{person}{Andrian Marcus},
  \bibinfo{person}{Gabriele Bavota}, {and} \bibinfo{person}{Vincent Ng}.}
  \bibinfo{year}{2017}\natexlab{}.
\newblock \showarticletitle{{Detecting Missing Information in Bug
  Descriptions}}. In \bibinfo{booktitle}{\emph{Proceedings of the joint meeting
  on Foundations of Software Engineering (ESEC/FSE)}}.
  \bibinfo{pages}{396--407}.
\newblock


\bibitem[Choetkiertikul et~al\mbox{.}(2019)]%
        {Choetkiertikul2019}
\bibfield{author}{\bibinfo{person}{Morakot Choetkiertikul},
  \bibinfo{person}{Hoa~Khanh Dam}, \bibinfo{person}{Truyen Tran},
  \bibinfo{person}{Trang Thi~Minh Pham}, \bibinfo{person}{Aditya Ghose}, {and}
  \bibinfo{person}{Tim Menzies}.} \bibinfo{year}{2019}\natexlab{}.
\newblock \showarticletitle{{A deep learning model for estimating story
  points}}.
\newblock \bibinfo{journal}{\emph{Transactions of Software Engineering (TSE)}}
  \bibinfo{volume}{45}, \bibinfo{number}{07} (\bibinfo{year}{2019}),
  \bibinfo{pages}{637--656}.
\newblock


\bibitem[Choudhari and Suman(2012)]%
        {choudhari2012phase}
\bibfield{author}{\bibinfo{person}{Jitender Choudhari} {and}
  \bibinfo{person}{Ugrasen Suman}.} \bibinfo{year}{2012}\natexlab{}.
\newblock \showarticletitle{{Phase wise Effort Estimation for Software
  Maintenance: An Extended SMEEM Model}}. In
  \bibinfo{booktitle}{\emph{Proceedings of the CUBE International Information
  Technology Conference}}. \bibinfo{pages}{397--402}.
\newblock


\bibitem[Coelho and Basu(2012)]%
        {coelho2012effort}
\bibfield{author}{\bibinfo{person}{Evita Coelho} {and} \bibinfo{person}{Anirban
  Basu}.} \bibinfo{year}{2012}\natexlab{}.
\newblock \showarticletitle{{Effort Estimation in Agile Software Development
  using Story Points}}.
\newblock \bibinfo{journal}{\emph{International Journal of Applied Information
  Systems (IJAIS)}} \bibinfo{volume}{3}, \bibinfo{number}{7}
  (\bibinfo{year}{2012}), \bibinfo{pages}{7--10}.
\newblock


\bibitem[Cohn(2006)]%
        {Cohn2006}
\bibfield{author}{\bibinfo{person}{Mike Cohn}.}
  \bibinfo{year}{2006}\natexlab{}.
\newblock \bibinfo{booktitle}{\emph{{Agile estimating and planning}}}.
\newblock \bibinfo{publisher}{{Pearson Education}}.
\newblock


\bibitem[Conoscenti et~al\mbox{.}(2019)]%
        {conoscenti2019combining}
\bibfield{author}{\bibinfo{person}{Marco Conoscenti}, \bibinfo{person}{Veronika
  Besner}, \bibinfo{person}{Antonio Vetr{\`o}}, {and}
  \bibinfo{person}{Daniel~M{\'e}ndez Fern{\'a}ndez}.}
  \bibinfo{year}{2019}\natexlab{}.
\newblock \showarticletitle{{Combining data analytics and developers feedback
  for identifying reasons of inaccurate estimations in agile software
  development}}.
\newblock \bibinfo{journal}{\emph{Journal of Systems and Software (JSS)}}
  \bibinfo{volume}{156} (\bibinfo{year}{2019}), \bibinfo{pages}{126--135}.
\newblock


\bibitem[Dan et~al\mbox{.}(2020)]%
        {dan2020nlp}
\bibfield{author}{\bibinfo{person}{Iftinc{\u{a}} Dan}, \bibinfo{person}{Rusu
  C{\u{a}}t{\u{a}}lin}, {and} \bibinfo{person}{Oswald Oliver}.}
  \bibinfo{year}{2020}\natexlab{}.
\newblock \showarticletitle{{An NLP Approach to Estimating Effort in a Work
  Environment}}. In \bibinfo{booktitle}{\emph{International Conference on
  Software, Telecommunications and Computer Networks}}. \bibinfo{pages}{1--6}.
\newblock


\bibitem[Dantas et~al\mbox{.}(2019)]%
        {dantas2019effort}
\bibfield{author}{\bibinfo{person}{Emanuel Dantas}, \bibinfo{person}{Antonio
  Alexandre~Moura Costa}, \bibinfo{person}{Marcus Vinicius},
  \bibinfo{person}{Mirko Perkusich}, \bibinfo{person}{Hyggo~Oliveira de
  Almeida}, {and} \bibinfo{person}{Angelo Perkusich}.}
  \bibinfo{year}{2019}\natexlab{}.
\newblock \showarticletitle{{An Effort Estimation Support Tool for Agile
  Software Development: An Empirical Evaluation}}. In
  \bibinfo{booktitle}{\emph{International Journal of Software Engineering and
  Knowledge Engineering (IJSEKE)}}. \bibinfo{pages}{82--116}.
\newblock


\bibitem[Dantas et~al\mbox{.}(2018)]%
        {dantas2018effort}
\bibfield{author}{\bibinfo{person}{Emanuel Dantas}, \bibinfo{person}{Mirko
  Perkusich}, \bibinfo{person}{Ednaldo Dilorenzo}, \bibinfo{person}{Danilo~FS
  Santos}, \bibinfo{person}{Hyggo Almeida}, {and} \bibinfo{person}{Angelo
  Perkusich}.} \bibinfo{year}{2018}\natexlab{}.
\newblock \showarticletitle{{Effort Estimation in Agile Software Development:
  an Updated Review}}.
\newblock \bibinfo{journal}{\emph{International Journal of Software Engineering
  and Knowledge Engineering (IJSEKE)}} \bibinfo{volume}{28},
  \bibinfo{number}{11n12} (\bibinfo{year}{2018}), \bibinfo{pages}{1811--1831}.
\newblock


\bibitem[Dave and Dutta(2014)]%
        {dave2014neural}
\bibfield{author}{\bibinfo{person}{Vachik~S Dave} {and}
  \bibinfo{person}{Kamlesh Dutta}.} \bibinfo{year}{2014}\natexlab{}.
\newblock \showarticletitle{{Neural network based models for software effort
  estimation: a review}}.
\newblock \bibinfo{journal}{\emph{Artificial Intelligence Review}}
  \bibinfo{volume}{42}, \bibinfo{number}{2} (\bibinfo{year}{2014}),
  \bibinfo{pages}{295--307}.
\newblock


\bibitem[Dragicevic et~al\mbox{.}(2017)]%
        {dragicevic2017bayesian}
\bibfield{author}{\bibinfo{person}{Srdjana Dragicevic}, \bibinfo{person}{Stipe
  Celar}, {and} \bibinfo{person}{Mili Turic}.} \bibinfo{year}{2017}\natexlab{}.
\newblock \showarticletitle{{Bayesian network model for task effort estimation
  in agile software development}}.
\newblock \bibinfo{journal}{\emph{Journal of Systems and Software (JSS)}}
  \bibinfo{volume}{127} (\bibinfo{year}{2017}), \bibinfo{pages}{109--119}.
\newblock


\bibitem[El~Beggar(2023)]%
        {el2023ifejm}
\bibfield{author}{\bibinfo{person}{Omar El~Beggar}.}
  \bibinfo{year}{2023}\natexlab{}.
\newblock \showarticletitle{IFEJM: New Intuitionistic Fuzzy Expert Judgment
  Method for Effort Estimation in Agile Software Development}.
\newblock \bibinfo{journal}{\emph{Arabian Journal for Science and Engineering}}
  (\bibinfo{year}{2023}), \bibinfo{pages}{1--22}.
\newblock


\bibitem[Ernst and Murphy(2012)]%
        {ernst2012case}
\bibfield{author}{\bibinfo{person}{Neil~A Ernst} {and} \bibinfo{person}{Gail~C
  Murphy}.} \bibinfo{year}{2012}\natexlab{}.
\newblock \showarticletitle{{Case Studies in Just-In-Time Requirements
  Analysis}}. In \bibinfo{booktitle}{\emph{Proceedings of the International
  Workshop on Empirical Requirements Engineering (EmpiRE)}}.
  \bibinfo{pages}{25--32}.
\newblock


\bibitem[Farahneh and Issa(2011)]%
        {farahneh2011linear}
\bibfield{author}{\bibinfo{person}{Hasan~O Farahneh} {and}
  \bibinfo{person}{Ayman~A Issa}.} \bibinfo{year}{2011}\natexlab{}.
\newblock \showarticletitle{{A Linear Use Case Based Software Cost Estimation
  Model}}.
\newblock \bibinfo{journal}{\emph{International Journal of Computer and
  Information Engineering}} \bibinfo{volume}{5}, \bibinfo{number}{1}
  (\bibinfo{year}{2011}), \bibinfo{pages}{31--35}.
\newblock


\bibitem[Fern{\'a}ndez-Diego et~al\mbox{.}(2020)]%
        {fernandez2020update}
\bibfield{author}{\bibinfo{person}{Marta Fern{\'a}ndez-Diego},
  \bibinfo{person}{Erwin~R M{\'e}ndez}, \bibinfo{person}{Fernando
  Gonz{\'a}lez-Ladr{\'o}n-De-Guevara}, \bibinfo{person}{Silvia Abrah{\~a}o},
  {and} \bibinfo{person}{Emilio Insfran}.} \bibinfo{year}{2020}\natexlab{}.
\newblock \showarticletitle{{An Update on Effort Estimation in Agile Software
  Development: a Systematic Literature Review}}.
\newblock \bibinfo{journal}{\emph{IEEE Access}}  \bibinfo{volume}{8}
  (\bibinfo{year}{2020}), \bibinfo{pages}{166768--166800}.
\newblock


\bibitem[Fischbach et~al\mbox{.}(2020)]%
        {fischbach2020specmate}
\bibfield{author}{\bibinfo{person}{Jannik Fischbach}, \bibinfo{person}{Andreas
  Vogelsang}, \bibinfo{person}{Dominik Spies}, \bibinfo{person}{Andreas
  Wehrle}, \bibinfo{person}{Maximilian Junker}, {and} \bibinfo{person}{Dietmar
  Freudenstein}.} \bibinfo{year}{2020}\natexlab{}.
\newblock \showarticletitle{{SPECMATE: Automated Creation of Test Cases from
  Acceptance Criteria}}. In \bibinfo{booktitle}{\emph{2020 IEEE 13th
  International Conference on Software Testing, Validation and Verification
  (ICST)}}. \bibinfo{pages}{321--331}.
\newblock


\bibitem[Fowler and Highsmith(2001)]%
        {fowler2001agile}
\bibfield{author}{\bibinfo{person}{Martin Fowler} {and} \bibinfo{person}{Jim
  Highsmith}.} \bibinfo{year}{2001}\natexlab{}.
\newblock \showarticletitle{The agile manifesto}.
\newblock \bibinfo{journal}{\emph{Software Development}} \bibinfo{volume}{9},
  \bibinfo{number}{8} (\bibinfo{year}{2001}), \bibinfo{pages}{28--32}.
\newblock


\bibitem[Fu and Tantithamthavorn(2022)]%
        {fu2022gpt2sp}
\bibfield{author}{\bibinfo{person}{Michael Fu} {and} \bibinfo{person}{Chakkrit
  Tantithamthavorn}.} \bibinfo{year}{2022}\natexlab{}.
\newblock \showarticletitle{GPT2SP: A transformer-based agile story point
  estimation approach}.
\newblock \bibinfo{journal}{\emph{IEEE Transactions on Software Engineering}}
  \bibinfo{volume}{49}, \bibinfo{number}{2} (\bibinfo{year}{2022}),
  \bibinfo{pages}{611--625}.
\newblock


\bibitem[Gautam and Singh(2018)]%
        {gautam2018state}
\bibfield{author}{\bibinfo{person}{Swarnima~Singh Gautam} {and}
  \bibinfo{person}{Vrijendra Singh}.} \bibinfo{year}{2018}\natexlab{}.
\newblock \showarticletitle{{The state-of-the-art in software development
  effort estimation}}.
\newblock \bibinfo{journal}{\emph{Journal of Software: Evolution and Process}}
  \bibinfo{volume}{30}, \bibinfo{number}{12} (\bibinfo{year}{2018}),
  \bibinfo{pages}{e1983}.
\newblock


\bibitem[Gilson et~al\mbox{.}(2019)]%
        {gilson2019extracting}
\bibfield{author}{\bibinfo{person}{Fabian Gilson}, \bibinfo{person}{Matthias
  Galster}, {and} \bibinfo{person}{Fran{\c{c}}ois Georis}.}
  \bibinfo{year}{2019}\natexlab{}.
\newblock \showarticletitle{{Extracting Quality Attributes from User Stories
  for Early Architecture Decision Making}}. In
  \bibinfo{booktitle}{\emph{Proceedings of the International Conference on
  Software Architecture Companion (ICSA-C)}}. \bibinfo{pages}{129--136}.
\newblock


\bibitem[Gilson et~al\mbox{.}(2020)]%
        {gilson2020generating}
\bibfield{author}{\bibinfo{person}{Fabian Gilson}, \bibinfo{person}{Matthias
  Galster}, {and} \bibinfo{person}{Fran{\c{c}}ois Georis}.}
  \bibinfo{year}{2020}\natexlab{}.
\newblock \showarticletitle{{Generating Use Case Scenarios from User Stories}}.
  In \bibinfo{booktitle}{\emph{Proceedings of the International Conference on
  Software and System Processes}}. \bibinfo{pages}{31--40}.
\newblock


\bibitem[Gotterbarn et~al\mbox{.}(1999)]%
        {gotterbarn1999computer}
\bibfield{author}{\bibinfo{person}{Don Gotterbarn}, \bibinfo{person}{Keith
  Miller}, {and} \bibinfo{person}{Simon Rogerson}.}
  \bibinfo{year}{1999}\natexlab{}.
\newblock \showarticletitle{{Computer Society and ACM Approve Software
  Engineering Code of Ethics}}.
\newblock \bibinfo{journal}{\emph{Computer}} \bibinfo{volume}{32},
  \bibinfo{number}{10} (\bibinfo{year}{1999}), \bibinfo{pages}{84--88}.
\newblock


\bibitem[Govil and Sharma(2022)]%
        {govil2022estimation}
\bibfield{author}{\bibinfo{person}{Nikhil Govil} {and} \bibinfo{person}{Ashish
  Sharma}.} \bibinfo{year}{2022}\natexlab{}.
\newblock \showarticletitle{Estimation of cost and development effort in
  Scrum-based software projects considering dimensional success factors}.
\newblock \bibinfo{journal}{\emph{Advances in Engineering Software}}
  \bibinfo{volume}{172} (\bibinfo{year}{2022}), \bibinfo{pages}{103209}.
\newblock


\bibitem[Grapenthin et~al\mbox{.}(2016)]%
        {grapenthin2016supporting}
\bibfield{author}{\bibinfo{person}{Simon Grapenthin}, \bibinfo{person}{Matthias
  Book}, \bibinfo{person}{Thomas Richter}, {and} \bibinfo{person}{Volker
  Gruhn}.} \bibinfo{year}{2016}\natexlab{}.
\newblock \showarticletitle{{Supporting Feature Estimation with Risk and Effort
  Annotations}}. In \bibinfo{booktitle}{\emph{Proceedings of the Euromicro
  Conference on Software Engineering and Advanced Application}}.
  \bibinfo{pages}{17--24}.
\newblock


\bibitem[Grimstad et~al\mbox{.}(2006)]%
        {grimstad2006software}
\bibfield{author}{\bibinfo{person}{Stein Grimstad}, \bibinfo{person}{Magne
  J{\o}rgensen}, {and} \bibinfo{person}{Kjetil Mol{\o}kken-{\O}stvold}.}
  \bibinfo{year}{2006}\natexlab{}.
\newblock \showarticletitle{{Software effort estimation terminology: The tower
  of Babel}}.
\newblock \bibinfo{journal}{\emph{Information and Software Technology (IST)}}
  \bibinfo{volume}{48}, \bibinfo{number}{4} (\bibinfo{year}{2006}),
  \bibinfo{pages}{302--310}.
\newblock


\bibitem[Gultekin and Kalipsiz(2020)]%
        {gultekin2020story}
\bibfield{author}{\bibinfo{person}{Muaz Gultekin} {and} \bibinfo{person}{Oya
  Kalipsiz}.} \bibinfo{year}{2020}\natexlab{}.
\newblock \showarticletitle{{Story Point-Based E®ort Estimation Model with
  Machine Learning Techniques}}.
\newblock \bibinfo{journal}{\emph{International Journal of Software Engineering
  and Knowledge Engineering (IJSEKE)}} \bibinfo{volume}{30},
  \bibinfo{number}{01} (\bibinfo{year}{2020}), \bibinfo{pages}{43--66}.
\newblock


\bibitem[Gupta and Mahapatra(2022)]%
        {gupta2022automated}
\bibfield{author}{\bibinfo{person}{Neha Gupta} {and}
  \bibinfo{person}{Rajendra~Prasad Mahapatra}.}
  \bibinfo{year}{2022}\natexlab{}.
\newblock \showarticletitle{Automated software effort estimation for agile
  development system by heuristically improved hybrid learning}.
\newblock \bibinfo{journal}{\emph{Concurrency and Computation: Practice and
  Experience}} \bibinfo{volume}{34}, \bibinfo{number}{25}
  (\bibinfo{year}{2022}), \bibinfo{pages}{e7267}.
\newblock


\bibitem[Guyatt et~al\mbox{.}(2008)]%
        {guyatt2008grade}
\bibfield{author}{\bibinfo{person}{Gordon~H Guyatt}, \bibinfo{person}{Andrew~D
  Oxman}, \bibinfo{person}{Gunn~E Vist}, \bibinfo{person}{Regina Kunz},
  \bibinfo{person}{Yngve Falck-Ytter}, \bibinfo{person}{Pablo Alonso-Coello},
  {and} \bibinfo{person}{Holger~J Sch{\"u}nemann}.}
  \bibinfo{year}{2008}\natexlab{}.
\newblock \showarticletitle{{GRADE: an emerging consensus on rating quality of
  evidence and strength of recommendations}}.
\newblock \bibinfo{journal}{\emph{British Medical Journal (BMJ)}}
  \bibinfo{volume}{336}, \bibinfo{number}{7650} (\bibinfo{year}{2008}),
  \bibinfo{pages}{924--926}.
\newblock


\bibitem[Hacalo{\u{g}}lu and Demir{\"o}rs(2018)]%
        {hacalouglu2018challenges}
\bibfield{author}{\bibinfo{person}{Tuna Hacalo{\u{g}}lu} {and}
  \bibinfo{person}{Onur Demir{\"o}rs}.} \bibinfo{year}{2018}\natexlab{}.
\newblock \showarticletitle{{Challenges of Using Software Size in Agile
  Software Development: A Systematic Literature Review}}. In
  \bibinfo{booktitle}{\emph{Proceedings of the International Workshop on
  Software Measurement and 12th International Conference on Software Process
  and Product Measurement (IWSM-MENSURA)}}.
\newblock


\bibitem[Hannay et~al\mbox{.}(2019)]%
        {hannay2019agile}
\bibfield{author}{\bibinfo{person}{Jo~Erskine Hannay},
  \bibinfo{person}{Hans~Christian Benestad}, {and} \bibinfo{person}{Kjetil
  Strand}.} \bibinfo{year}{2019}\natexlab{}.
\newblock \showarticletitle{{Agile Uncertainty Assessment for Benefit Points
  and Story Points}}.
\newblock \bibinfo{journal}{\emph{IEEE Software}} \bibinfo{volume}{36},
  \bibinfo{number}{4} (\bibinfo{year}{2019}), \bibinfo{pages}{50--62}.
\newblock


\bibitem[Hearty et~al\mbox{.}(2009)]%
        {Hearty2009}
\bibfield{author}{\bibinfo{person}{Peter Hearty}, \bibinfo{person}{Norman
  Fenton}, \bibinfo{person}{David Marquez}, {and} \bibinfo{person}{Martin
  Neil}.} \bibinfo{year}{2009}\natexlab{}.
\newblock \showarticletitle{{Predicting project velocity in XP using a learning
  dynamic Bayesian network model}}.
\newblock \bibinfo{journal}{\emph{Transactions of Software Engineering (TSE)}}
  \bibinfo{volume}{35}, \bibinfo{number}{1} (\bibinfo{year}{2009}),
  \bibinfo{pages}{124--137}.
\newblock


\bibitem[Heck and Zaidman(2017)]%
        {heck2017framework}
\bibfield{author}{\bibinfo{person}{Petra Heck} {and} \bibinfo{person}{Andy
  Zaidman}.} \bibinfo{year}{2017}\natexlab{}.
\newblock \showarticletitle{{A framework for quality assessment of just-in-time
  requirements: the case of open source feature requests}}.
\newblock \bibinfo{journal}{\emph{Requirements Engineering (RE)}}
  \bibinfo{volume}{22}, \bibinfo{number}{4} (\bibinfo{year}{2017}),
  \bibinfo{pages}{453--473}.
\newblock


\bibitem[Heck and Zaidman(2018)]%
        {heck2018systematic}
\bibfield{author}{\bibinfo{person}{Petra Heck} {and} \bibinfo{person}{Andy
  Zaidman}.} \bibinfo{year}{2018}\natexlab{}.
\newblock \showarticletitle{{A Systematic Literature Review on Quality Criteria
  for Agile Requirements Specifications}}.
\newblock \bibinfo{journal}{\emph{Software Quality Journal}}
  \bibinfo{volume}{26}, \bibinfo{number}{1} (\bibinfo{year}{2018}),
  \bibinfo{pages}{127--160}.
\newblock


\bibitem[Hemrajani and N(2021)]%
        {hemrajanipredicting}
\bibfield{author}{\bibinfo{person}{M Hemrajani} {and} \bibinfo{person}{Vyas
  N}.} \bibinfo{year}{2021}\natexlab{}.
\newblock \showarticletitle{Predicting Effort of Agile Software Projects Using
  Linear Regression, Ridge Regression and Logistic Regression}.
\newblock  (\bibinfo{year}{2021}), \bibinfo{pages}{14--19}.
\newblock


\bibitem[Hoda and Murugesan(2016)]%
        {hoda2016multilevel}
\bibfield{author}{\bibinfo{person}{Rashina Hoda} {and}
  \bibinfo{person}{Latha~K. Murugesan}.} \bibinfo{year}{2016}\natexlab{}.
\newblock \showarticletitle{{Multi-Level Agile Project Management Challenges: A
  Self-Organizing Team Perspective}}.
\newblock \bibinfo{journal}{\emph{Journal of Systems and Software (JSS)}}
  \bibinfo{volume}{117} (\bibinfo{year}{2016}), \bibinfo{pages}{245--257}.
\newblock


\bibitem[Hoda et~al\mbox{.}(2010)]%
        {hoda2010much}
\bibfield{author}{\bibinfo{person}{Rashina Hoda}, \bibinfo{person}{James
  Noble}, {and} \bibinfo{person}{Stuart Marshall}.}
  \bibinfo{year}{2010}\natexlab{}.
\newblock \showarticletitle{{How Much is Just Enough? Some Documentation
  Patterns on Agile Projects}}. In \bibinfo{booktitle}{\emph{Proceedings of the
  European Conference on Pattern Languages of Programs}}.
  \bibinfo{pages}{1--13}.
\newblock


\bibitem[Idri et~al\mbox{.}(2015)]%
        {idri2015analogy}
\bibfield{author}{\bibinfo{person}{Ali Idri}, \bibinfo{person}{Fatima azzahra
  Amazal}, {and} \bibinfo{person}{Alain Abran}.}
  \bibinfo{year}{2015}\natexlab{}.
\newblock \showarticletitle{{Analogy-based software development effort
  estimation: A systematic mapping and review}}.
\newblock \bibinfo{journal}{\emph{Information and Software Technology (IST)}}
  \bibinfo{volume}{58} (\bibinfo{year}{2015}), \bibinfo{pages}{206--230}.
\newblock


\bibitem[Idri et~al\mbox{.}(2016)]%
        {idri2016systematic}
\bibfield{author}{\bibinfo{person}{Ali Idri}, \bibinfo{person}{Mohamed Hosni},
  {and} \bibinfo{person}{Alain Abran}.} \bibinfo{year}{2016}\natexlab{}.
\newblock \showarticletitle{{Systematic literature review of ensemble effort
  estimation}}.
\newblock \bibinfo{journal}{\emph{Journal of Systems and Software (JSS)}}
  \bibinfo{volume}{118} (\bibinfo{year}{2016}), \bibinfo{pages}{151--175}.
\newblock


\bibitem[J{\o}rgensen(2004)]%
        {jorgensen2004review}
\bibfield{author}{\bibinfo{person}{Magne J{\o}rgensen}.}
  \bibinfo{year}{2004}\natexlab{}.
\newblock \showarticletitle{{A review of studies on expert estimation of
  software development effort}}.
\newblock \bibinfo{journal}{\emph{Journal of Systems and Software (JSS)}}
  \bibinfo{volume}{70}, \bibinfo{number}{1-2} (\bibinfo{year}{2004}),
  \bibinfo{pages}{37--60}.
\newblock


\bibitem[Kang et~al\mbox{.}(2010)]%
        {kang2010model}
\bibfield{author}{\bibinfo{person}{Sungjoo Kang}, \bibinfo{person}{Okjoo Choi},
  {and} \bibinfo{person}{Jongmoon Baik}.} \bibinfo{year}{2010}\natexlab{}.
\newblock \showarticletitle{{Model-Based Dynamic Cost Estimation and Tracking
  Method for Agile Software Development}}. In
  \bibinfo{booktitle}{\emph{Proceedings of the International Conference on
  Computer and Information Science}}. \bibinfo{pages}{743--748}.
\newblock


\bibitem[Karna et~al\mbox{.}(2020a)]%
        {karna2020data}
\bibfield{author}{\bibinfo{person}{Hrvoje Karna}, \bibinfo{person}{Sven
  Gotovac}, {and} \bibinfo{person}{Linda Vickovi{\'c}}.}
  \bibinfo{year}{2020}\natexlab{a}.
\newblock \showarticletitle{{Data Mining Approach to Effort Modeling On Agile
  Software Projects}}.
\newblock \bibinfo{journal}{\emph{Informatica}} \bibinfo{volume}{44},
  \bibinfo{number}{2} (\bibinfo{year}{2020}).
\newblock


\bibitem[Karna et~al\mbox{.}(2020b)]%
        {karna2020effects}
\bibfield{author}{\bibinfo{person}{Hrvoje Karna}, \bibinfo{person}{Sven
  Gotovac}, \bibinfo{person}{Linda Vickovi{\'c}}, {and} \bibinfo{person}{Luka
  Mihanovi{\'c}}.} \bibinfo{year}{2020}\natexlab{b}.
\newblock \showarticletitle{{The Effects of Turnover on Expert Effort
  Estimation}}.
\newblock \bibinfo{journal}{\emph{Journal of Information and Organizational
  Sciences}} \bibinfo{volume}{44}, \bibinfo{number}{1} (\bibinfo{year}{2020}),
  \bibinfo{pages}{51--81}.
\newblock


\bibitem[KASSEM et~al\mbox{.}(2022)]%
        {kassem2022software}
\bibfield{author}{\bibinfo{person}{HAITHEM KASSEM}, \bibinfo{person}{KHALED
  MAHAR}, {and} \bibinfo{person}{AMANI SAAD}.} \bibinfo{year}{2022}\natexlab{}.
\newblock \showarticletitle{Software Effort Estimation Using Hierarchical
  Attention Neural Network}.
\newblock \bibinfo{journal}{\emph{Journal of Theoretical and Applied
  Information Technology}} \bibinfo{volume}{100}, \bibinfo{number}{18}
  (\bibinfo{year}{2022}).
\newblock


\bibitem[Kassem et~al\mbox{.}(2023)]%
        {kassem2023story}
\bibfield{author}{\bibinfo{person}{Haithem Kassem}, \bibinfo{person}{Khaled
  Mahar}, {and} \bibinfo{person}{Amani~A Saad}.}
  \bibinfo{year}{2023}\natexlab{}.
\newblock \showarticletitle{Story Point Estimation Using Issue Reports With
  Deep Attention Neural Network}.
\newblock \bibinfo{journal}{\emph{e-Informatica Software Engineering Journal}}
  \bibinfo{volume}{17}, \bibinfo{number}{1} (\bibinfo{year}{2023}).
\newblock


\bibitem[Kaushik et~al\mbox{.}(2020)]%
        {kaushik2020fuzzified}
\bibfield{author}{\bibinfo{person}{Anupama Kaushik},
  \bibinfo{person}{Devendra~Kr Tayal}, {and} \bibinfo{person}{Kalpana Yadav}.}
  \bibinfo{year}{2020}\natexlab{}.
\newblock \showarticletitle{A fuzzified story point approach for agile
  projects}.
\newblock \bibinfo{journal}{\emph{International Journal of Agile Systems and
  Management}} \bibinfo{volume}{13}, \bibinfo{number}{2}
  (\bibinfo{year}{2020}), \bibinfo{pages}{103--129}.
\newblock


\bibitem[Kaushik et~al\mbox{.}(2022)]%
        {kaushik2022role}
\bibfield{author}{\bibinfo{person}{Anupama Kaushik},
  \bibinfo{person}{Devendra~Kumar Tayal}, {and} \bibinfo{person}{Kalpana
  Yadav}.} \bibinfo{year}{2022}\natexlab{}.
\newblock \showarticletitle{The role of neural networks and metaheuristics in
  agile software development effort estimation}.
\newblock In \bibinfo{booktitle}{\emph{Research Anthology on Artificial Neural
  Network Applications}}. \bibinfo{pages}{306--328}.
\newblock


\bibitem[Khuat and Le(2018)]%
        {khuat2018novel}
\bibfield{author}{\bibinfo{person}{Thanh~Tung Khuat} {and}
  \bibinfo{person}{My~Hanh Le}.} \bibinfo{year}{2018}\natexlab{}.
\newblock \showarticletitle{{A Novel Hybrid ABC-PSO Algorithm for Effort
  Estimation of Software Projects Using Agile Methodologies}}.
\newblock \bibinfo{journal}{\emph{Journal of Intelligent Systems}}
  \bibinfo{volume}{27}, \bibinfo{number}{3} (\bibinfo{year}{2018}),
  \bibinfo{pages}{489--506}.
\newblock


\bibitem[Kitchenham and Charters(2007)]%
        {kitchenham2007guidelines}
\bibfield{author}{\bibinfo{person}{Barbara Kitchenham} {and}
  \bibinfo{person}{Stuart Charters}.} \bibinfo{year}{2007}\natexlab{}.
\newblock \showarticletitle{{Guidelines for performing Systematic Literature
  Reviews in Software Engineering}}.
\newblock  (\bibinfo{year}{2007}).
\newblock


\bibitem[Kitchenham et~al\mbox{.}(2022)]%
        {kitchenham2022segress}
\bibfield{author}{\bibinfo{person}{Barbara Kitchenham}, \bibinfo{person}{Lech
  Madeyski}, {and} \bibinfo{person}{David Budgen}.}
  \bibinfo{year}{2022}\natexlab{}.
\newblock \showarticletitle{SEGRESS: Software engineering guidelines for
  reporting secondary studies}.
\newblock \bibinfo{journal}{\emph{IEEE Transactions on Software Engineering}}
  \bibinfo{volume}{49}, \bibinfo{number}{3} (\bibinfo{year}{2022}),
  \bibinfo{pages}{1273--1298}.
\newblock


\bibitem[Logue and McDaid(2008)]%
        {logue2008agile}
\bibfield{author}{\bibinfo{person}{Kevin Logue} {and} \bibinfo{person}{Kevin
  McDaid}.} \bibinfo{year}{2008}\natexlab{}.
\newblock \showarticletitle{{Agile release planning: Dealing with uncertainty
  in development time and business value}}. In \bibinfo{booktitle}{\emph{Proc
  of the IEEE International Conference and Workshop on the Engineering of
  Computer Based Systems}}. \bibinfo{pages}{437--442}.
\newblock


\bibitem[Madampe et~al\mbox{.}(2020)]%
        {madampe2020multi}
\bibfield{author}{\bibinfo{person}{Kashumi Madampe}, \bibinfo{person}{Rashina
  Hoda}, {and} \bibinfo{person}{John Grundy}.} \bibinfo{year}{2020}\natexlab{}.
\newblock \showarticletitle{{A Multi-dimensional Study of Requirements Changes
  in Agile Software Development Projects}}.
\newblock \bibinfo{journal}{\emph{arXiv preprint arXiv:2012.03423}}
  (\bibinfo{year}{2020}).
\newblock


\bibitem[Madya et~al\mbox{.}(2022)]%
        {madya2022prep}
\bibfield{author}{\bibinfo{person}{Gusti~Raditia Madya}, \bibinfo{person}{Eko~K
  Budiardjo}, {and} \bibinfo{person}{Kodrat Mahatma}.}
  \bibinfo{year}{2022}\natexlab{}.
\newblock \showarticletitle{PREP: A Post-Requirements Effort Estimation Method
  in Scrum's Sprint Grooming}. In \bibinfo{booktitle}{\emph{2022 International
  Conference on Data and Software Engineering (ICoDSE)}}. IEEE,
  \bibinfo{pages}{132--137}.
\newblock


\bibitem[Mahni{\v{c}} and Hovelja(2012)]%
        {Mahnic2012}
\bibfield{author}{\bibinfo{person}{Viljan Mahni{\v{c}}} {and}
  \bibinfo{person}{Toma{\v{z}} Hovelja}.} \bibinfo{year}{2012}\natexlab{}.
\newblock \showarticletitle{{On using planning poker for estimating user
  stories}}.
\newblock \bibinfo{journal}{\emph{Journal of Systems and Software (JSS)}}
  \bibinfo{volume}{85}, \bibinfo{number}{9} (\bibinfo{year}{2012}),
  \bibinfo{pages}{2086--2095}.
\newblock


\bibitem[MAJCHRZAK and Madeyski(2016)]%
        {majchrzak2016factors}
\bibfield{author}{\bibinfo{person}{Marek MAJCHRZAK} {and} \bibinfo{person}{Lech
  Madeyski}.} \bibinfo{year}{2016}\natexlab{}.
\newblock \showarticletitle{{Factors influencing user story estimations: an
  industrial interview and a conceptual model}}.
\newblock \bibinfo{journal}{\emph{Central and Eastern European Journal of
  Management and Economics (CEEJME)}} \bibinfo{number}{4}
  (\bibinfo{year}{2016}), \bibinfo{pages}{261--280}.
\newblock


\bibitem[Malgonde and Chari(2019)]%
        {malgonde2019ensemble}
\bibfield{author}{\bibinfo{person}{Onkar Malgonde} {and}
  \bibinfo{person}{Kaushal Chari}.} \bibinfo{year}{2019}\natexlab{}.
\newblock \showarticletitle{{An ensemble-based model for predicting agile
  software development effort}}.
\newblock \bibinfo{journal}{\emph{Empirical Software Engineering (EMSE)}}
  \bibinfo{volume}{24}, \bibinfo{number}{2} (\bibinfo{year}{2019}),
  \bibinfo{pages}{1017--1055}.
\newblock


\bibitem[Marapelli et~al\mbox{.}(2020)]%
        {marapelli2020rnn}
\bibfield{author}{\bibinfo{person}{Bhaskar Marapelli}, \bibinfo{person}{Anil
  Carie}, {and} \bibinfo{person}{Sardar~MN Islam}.}
  \bibinfo{year}{2020}\natexlab{}.
\newblock \showarticletitle{{RNN-CNN MODEL: A Bi-directional Long Short-Term
  Memory Deep Learning Network For Story Point Estimation}}. In
  \bibinfo{booktitle}{\emph{Proceedings of the International Conference on
  Innovative Technologies in Intelligent Systems and Industrial Applications
  (CITISIA)}}. \bibinfo{pages}{1--7}.
\newblock


\bibitem[Marczak-Czajka and Cleland-Huang(2023)]%
        {marczak2023using}
\bibfield{author}{\bibinfo{person}{Agnieszka Marczak-Czajka} {and}
  \bibinfo{person}{Jane Cleland-Huang}.} \bibinfo{year}{2023}\natexlab{}.
\newblock \showarticletitle{{Using ChatGPT to Generate Human-Value User Stories
  as Inspirational Triggers}}. In \bibinfo{booktitle}{\emph{2023 IEEE 31st
  International Requirements Engineering Conference Workshops (REW)}}.
  \bibinfo{pages}{52--61}.
\newblock


\bibitem[Masood et~al\mbox{.}(2020)]%
        {masood2020real}
\bibfield{author}{\bibinfo{person}{Zainab Masood}, \bibinfo{person}{Rashina
  Hoda}, {and} \bibinfo{person}{Kelly Blincoe}.}
  \bibinfo{year}{2020}\natexlab{}.
\newblock \showarticletitle{Real World Scrum A Grounded Theory of Variations in
  Practice}.
\newblock \bibinfo{journal}{\emph{Transactions on Software Engineering (TSE)}}
  \bibinfo{volume}{48}, \bibinfo{number}{5} (\bibinfo{year}{2020}),
  \bibinfo{pages}{1579--1591}.
\newblock


\bibitem[Miranda et~al\mbox{.}(2021)]%
        {miranda2021analysis}
\bibfield{author}{\bibinfo{person}{Pedro Miranda}, \bibinfo{person}{J~Pascoal
  Faria}, \bibinfo{person}{Filipe~F Correia}, \bibinfo{person}{Ahmed Fares},
  \bibinfo{person}{Ricardo Gra{\c{c}}a}, {and} \bibinfo{person}{Jo{\~a}o~Mendes
  Moreira}.} \bibinfo{year}{2021}\natexlab{}.
\newblock \showarticletitle{{An Analysis of Monte Carlo Simulations for
  Forecasting Software Projects}}. In \bibinfo{booktitle}{\emph{Proceedings of
  the Annual ACM Symposium on Applied Computing}}. \bibinfo{pages}{1550--1558}.
\newblock


\bibitem[Moharreri et~al\mbox{.}(2016)]%
        {moharreri2016cost}
\bibfield{author}{\bibinfo{person}{Kayhan Moharreri},
  \bibinfo{person}{Alhad~Vinayak Sapre}, \bibinfo{person}{Jayashree
  Ramanathan}, {and} \bibinfo{person}{Rajiv Ramnath}.}
  \bibinfo{year}{2016}\natexlab{}.
\newblock \showarticletitle{{Cost-Effective Supervised Learning Models for
  Software Effort Estimation in Agile Environments}}. In
  \bibinfo{booktitle}{\emph{Proceedings of the IEEE Annual computer software
  and applications conference (COMPSAC)}}, Vol.~\bibinfo{volume}{2}.
  \bibinfo{pages}{135--140}.
\newblock


\bibitem[Najm et~al\mbox{.}(2022)]%
        {najm2022enhanced}
\bibfield{author}{\bibinfo{person}{Assia Najm}, \bibinfo{person}{Abdelali
  Zakrani}, {and} \bibinfo{person}{Abdelaziz Marzak}.}
  \bibinfo{year}{2022}\natexlab{}.
\newblock \showarticletitle{An enhanced support vector regression model for
  agile projects cost estimation}.
\newblock \bibinfo{journal}{\emph{IAES International Journal of Artificial
  Intelligence (IJ-AI)}} \bibinfo{volume}{11}, \bibinfo{number}{1}
  (\bibinfo{year}{2022}), \bibinfo{pages}{265}.
\newblock


\bibitem[Nunes et~al\mbox{.}(2011)]%
        {nunes2011iucp}
\bibfield{author}{\bibinfo{person}{Nuno Nunes}, \bibinfo{person}{Larry
  Constantine}, {and} \bibinfo{person}{Rick Kazman}.}
  \bibinfo{year}{2011}\natexlab{}.
\newblock \showarticletitle{{iUCP: Estimating Interactive-Software Project Size
  with Enhanced Use-Case Points}}.
\newblock \bibinfo{journal}{\emph{IEEE software}} \bibinfo{volume}{28},
  \bibinfo{number}{4} (\bibinfo{year}{2011}).
\newblock


\bibitem[Owais and Ramakishore(2016)]%
        {owais2016effort}
\bibfield{author}{\bibinfo{person}{Mohd Owais} {and} \bibinfo{person}{R
  Ramakishore}.} \bibinfo{year}{2016}\natexlab{}.
\newblock \showarticletitle{{Effort, duration and cost estimation in agile
  software development}}. In \bibinfo{booktitle}{\emph{2016 Ninth International
  Conference on Contemporary Computing (IC3)}}. \bibinfo{pages}{1--5}.
\newblock


\bibitem[Page et~al\mbox{.}(2021)]%
        {page2021prisma}
\bibfield{author}{\bibinfo{person}{Matthew~J Page}, \bibinfo{person}{Joanne~E
  McKenzie}, \bibinfo{person}{Patrick~M Bossuyt}, \bibinfo{person}{Isabelle
  Boutron}, \bibinfo{person}{Tammy~C Hoffmann}, \bibinfo{person}{Cynthia~D
  Mulrow}, \bibinfo{person}{Larissa Shamseer}, \bibinfo{person}{Jennifer~M
  Tetzlaff}, \bibinfo{person}{Elie~A Akl}, \bibinfo{person}{Sue~E Brennan},
  {et~al\mbox{.}}} \bibinfo{year}{2021}\natexlab{}.
\newblock \showarticletitle{{The PRISMA 2020 statement: an updated guideline
  for reporting systematic reviews}}.
\newblock \bibinfo{journal}{\emph{International Journal of Surgery}}
  \bibinfo{volume}{88} (\bibinfo{year}{2021}), \bibinfo{pages}{105906}.
\newblock


\bibitem[Panda et~al\mbox{.}(2015)]%
        {panda2015empirical}
\bibfield{author}{\bibinfo{person}{Aditi Panda},
  \bibinfo{person}{Shashank~Mouli Satapathy}, {and}
  \bibinfo{person}{Santanu~Kumar Rath}.} \bibinfo{year}{2015}\natexlab{}.
\newblock \showarticletitle{{Empirical Validation of Neural Network Models for
  Agile Software Effort Estimation based on Story Points}}.
\newblock \bibinfo{journal}{\emph{Procedia Computer Science}}
  \bibinfo{volume}{57} (\bibinfo{year}{2015}), \bibinfo{pages}{772--781}.
\newblock


\bibitem[Parvez(2013)]%
        {parvez2013efficiency}
\bibfield{author}{\bibinfo{person}{Abu Wahid Md~Masud Parvez}.}
  \bibinfo{year}{2013}\natexlab{}.
\newblock \showarticletitle{{Efficiency Factor and Risk Factor Based User Case
  Point Test Effort Estimation Model Compatible with Agile Software
  Development}}. In \bibinfo{booktitle}{\emph{Proceedings of the International
  Conference on Information Technology and Electrical Engineering (ICITEE)}}.
  IEEE, \bibinfo{pages}{113--118}.
\newblock


\bibitem[Pasuksmit(2023)]%
        {supplementary}
\bibfield{author}{\bibinfo{person}{Jirat Pasuksmit}.}
  \bibinfo{year}{2023}\natexlab{}.
\newblock \showarticletitle{{{Supplementary material: A Systematic Literature
  Review on Reasons and Approaches for Accurate EffortEstimations in Agile}}}.
\newblock  (\bibinfo{date}{12} \bibinfo{year}{2023}).
\newblock
\urldef\tempurl%
\url{https://doi.org/10.6084/m9.figshare.22122245.v6}
\showDOI{\tempurl}


\bibitem[Pasuksmit et~al\mbox{.}(2021)]%
        {jpsurvey}
\bibfield{author}{\bibinfo{person}{Jirat Pasuksmit}, \bibinfo{person}{Patanamon
  Thongtanunam}, {and} \bibinfo{person}{Shanika Karunasekera}.}
  \bibinfo{year}{2021}\natexlab{}.
\newblock \showarticletitle{{Towards Just-Enough Documentation for Agile Effort
  Estimation: What Information Should Be Documented?}}. In
  \bibinfo{booktitle}{\emph{{Proceedings of the IEEE International Conference
  on Software Maintenance and Evolution (ICSME)}}}. \bibinfo{pages}{114--125}.
\newblock


\bibitem[Pasuksmit et~al\mbox{.}(2022a)]%
        {pasuksmit2022story}
\bibfield{author}{\bibinfo{person}{Jirat Pasuksmit}, \bibinfo{person}{Patanamon
  Thongtanunam}, {and} \bibinfo{person}{Shanika Karunasekera}.}
  \bibinfo{year}{2022}\natexlab{a}.
\newblock \showarticletitle{Story points changes in agile iterative
  development: An empirical study and a prediction approach}.
\newblock \bibinfo{journal}{\emph{Empirical Software Engineering (EMSE)}}
  \bibinfo{volume}{27}, \bibinfo{number}{6} (\bibinfo{year}{2022}),
  \bibinfo{pages}{156}.
\newblock


\bibitem[Pasuksmit et~al\mbox{.}(2022b)]%
        {pasuksmit2022towards}
\bibfield{author}{\bibinfo{person}{Jirat Pasuksmit}, \bibinfo{person}{Patanamon
  Thongtanunam}, {and} \bibinfo{person}{Shanika Karunasekera}.}
  \bibinfo{year}{2022}\natexlab{b}.
\newblock \showarticletitle{Towards reliable agile iterative planning via
  predicting documentation changes of work items}. In
  \bibinfo{booktitle}{\emph{Proceedings of the International Conference on
  Mining Software Repositories (MSR)}}. \bibinfo{pages}{35--47}.
\newblock


\bibitem[Paz et~al\mbox{.}(2014)]%
        {paz2014approach}
\bibfield{author}{\bibinfo{person}{Freddy Paz}, \bibinfo{person}{Claudia
  Zapata}, {and} \bibinfo{person}{Jos{\'e}~Antonio Pow-Sang}.}
  \bibinfo{year}{2014}\natexlab{}.
\newblock \showarticletitle{An Approach for Effort Estimation in Incremental
  Software Development using Cosmic Function Points}. In
  \bibinfo{booktitle}{\emph{Proceedings of the International Symposium on
  Empirical Software Engineering and Measurement (ESEM)}}.
  \bibinfo{pages}{1--4}.
\newblock


\bibitem[Perkusich et~al\mbox{.}(2020)]%
        {perkusich2020intelligent}
\bibfield{author}{\bibinfo{person}{Mirko Perkusich},
  \bibinfo{person}{Lenardo~Chaves e Silva}, \bibinfo{person}{Alexandre Costa},
  \bibinfo{person}{Felipe Ramos}, \bibinfo{person}{Renata Saraiva},
  \bibinfo{person}{Arthur Freire}, \bibinfo{person}{Ednaldo Dilorenzo},
  \bibinfo{person}{Emanuel Dantas}, \bibinfo{person}{Danilo Santos},
  \bibinfo{person}{Kyller Gorg{\^o}nio}, {et~al\mbox{.}}}
  \bibinfo{year}{2020}\natexlab{}.
\newblock \showarticletitle{{Intelligent software engineering in the context of
  agile software development: A systematic literature review}}.
\newblock \bibinfo{journal}{\emph{Information and Software Technology (IST)}}
  \bibinfo{volume}{119} (\bibinfo{year}{2020}), \bibinfo{pages}{106241}.
\newblock


\bibitem[Petticrew and Roberts(2008)]%
        {petticrew2008systematic}
\bibfield{author}{\bibinfo{person}{Mark Petticrew} {and} \bibinfo{person}{Helen
  Roberts}.} \bibinfo{year}{2008}\natexlab{}.
\newblock \bibinfo{booktitle}{\emph{Systematic reviews in the social sciences:
  A practical guide}}.
\newblock \bibinfo{publisher}{John Wiley \& Sons}.
\newblock


\bibitem[Phan and Jannesari(2022a)]%
        {phan2022heterogeneous}
\bibfield{author}{\bibinfo{person}{Hung Phan} {and} \bibinfo{person}{Ali
  Jannesari}.} \bibinfo{year}{2022}\natexlab{a}.
\newblock \showarticletitle{Heterogeneous Graph Neural Networks for Software
  Effort Estimation}. In \bibinfo{booktitle}{\emph{Proceedings of the 16th
  ACM/IEEE International Symposium on Empirical Software Engineering and
  Measurement}}. \bibinfo{pages}{103--113}.
\newblock


\bibitem[Phan and Jannesari(2022b)]%
        {phan2022story}
\bibfield{author}{\bibinfo{person}{Hung Phan} {and} \bibinfo{person}{Ali
  Jannesari}.} \bibinfo{year}{2022}\natexlab{b}.
\newblock \showarticletitle{Story point level classification by text level
  graph neural network}. In \bibinfo{booktitle}{\emph{Proceedings of the 1st
  International Workshop on Natural Language-based Software Engineering}}.
  \bibinfo{pages}{75--78}.
\newblock


\bibitem[Popli and Chauhan(2014)]%
        {popli2014cost}
\bibfield{author}{\bibinfo{person}{Rashmi Popli} {and} \bibinfo{person}{Naresh
  Chauhan}.} \bibinfo{year}{2014}\natexlab{}.
\newblock \showarticletitle{{Cost and effort estimation in agile software
  development}}. In \bibinfo{booktitle}{\emph{2014 international conference on
  reliability optimization and information technology (ICROIT)}}.
  \bibinfo{pages}{57--61}.
\newblock


\bibitem[Porru et~al\mbox{.}(2016)]%
        {Porru2016}
\bibfield{author}{\bibinfo{person}{Simone Porru}, \bibinfo{person}{Alessandro
  Murgia}, \bibinfo{person}{Serge Demeyer}, \bibinfo{person}{Michele Marchesi},
  {and} \bibinfo{person}{Roberto Tonelli}.} \bibinfo{year}{2016}\natexlab{}.
\newblock \showarticletitle{{Estimating Story Points from Issue Reports}}. In
  \bibinfo{booktitle}{\emph{Proceedings of the International Conference on
  Predictive Models and Data Analytics in Software Engineering}}.
  \bibinfo{pages}{2:1--2:10}.
\newblock


\bibitem[Premalatha and Srikrishna(2019)]%
        {premalatha2019effort}
\bibfield{author}{\bibinfo{person}{Hosahalli~Mahalingappa Premalatha} {and}
  \bibinfo{person}{Chimanahalli~Venkateshavittalachar Srikrishna}.}
  \bibinfo{year}{2019}\natexlab{}.
\newblock \showarticletitle{Effort estimation in agile software development
  using evolutionary cost-sensitive deep belief network}.
\newblock \bibinfo{journal}{\emph{International Journal of Information
  Technology}} \bibinfo{volume}{12}, \bibinfo{number}{2}
  (\bibinfo{year}{2019}), \bibinfo{pages}{261--269}.
\newblock


\bibitem[Prykhodko and Prykhodko(2019)]%
        {prykhodko2019multiple}
\bibfield{author}{\bibinfo{person}{NV Prykhodko} {and} \bibinfo{person}{SB
  Prykhodko}.} \bibinfo{year}{2019}\natexlab{}.
\newblock \showarticletitle{A multiple non-linear regression model to estimate
  the agile testing efforts for small web projects}.
\newblock \bibinfo{journal}{\emph{Radio Electronics, Computer Science,
  Control}} \bibinfo{number}{2} (\bibinfo{year}{2019}),
  \bibinfo{pages}{158--166}.
\newblock


\bibitem[Ramessur and Nagowah(2021)]%
        {ramessur2021predictive}
\bibfield{author}{\bibinfo{person}{Melvina~Autar Ramessur} {and}
  \bibinfo{person}{Soulakshmee~Devi Nagowah}.} \bibinfo{year}{2021}\natexlab{}.
\newblock \showarticletitle{A predictive model to estimate effort in a sprint
  using machine learning techniques}.
\newblock \bibinfo{journal}{\emph{International Journal of Information
  Technology}} \bibinfo{volume}{13}, \bibinfo{number}{3}
  (\bibinfo{year}{2021}), \bibinfo{pages}{1101--1110}.
\newblock


\bibitem[Raslan and Darwish(2018)]%
        {raslan2018enhanced}
\bibfield{author}{\bibinfo{person}{Atef~Tayh Raslan} {and}
  \bibinfo{person}{Nagy~Ramadan Darwish}.} \bibinfo{year}{2018}\natexlab{}.
\newblock \showarticletitle{An Enhanced Framework for Effort Estimation of
  Agile Projects}.
\newblock \bibinfo{journal}{\emph{International Journal of Intelligent
  Engineering and Systems}} \bibinfo{volume}{11}, \bibinfo{number}{3}
  (\bibinfo{year}{2018}), \bibinfo{pages}{205--214}.
\newblock


\bibitem[Rodr{\'\i}guez~S{\'a}nchez et~al\mbox{.}(2023)]%
        {rodriguez2023effort}
\bibfield{author}{\bibinfo{person}{Eduardo Rodr{\'\i}guez~S{\'a}nchez},
  \bibinfo{person}{Eduardo~Filem{\'o}n V{\'a}zquez~Santacruz}, {and}
  \bibinfo{person}{Humberto Cervantes~Maceda}.}
  \bibinfo{year}{2023}\natexlab{}.
\newblock \showarticletitle{Effort and Cost Estimation Using Decision Tree
  Techniques and Story Points in Agile Software Development}.
\newblock \bibinfo{journal}{\emph{Mathematics}} \bibinfo{volume}{11},
  \bibinfo{number}{6} (\bibinfo{year}{2023}), \bibinfo{pages}{1477}.
\newblock


\bibitem[Rola and Kuchta(2019)]%
        {rola2019application}
\bibfield{author}{\bibinfo{person}{Pawe{\l} Rola} {and} \bibinfo{person}{Dorota
  Kuchta}.} \bibinfo{year}{2019}\natexlab{}.
\newblock \showarticletitle{Application of fuzzy sets to the expert estimation
  of Scrum-based projects}.
\newblock \bibinfo{journal}{\emph{Symmetry}} \bibinfo{volume}{11},
  \bibinfo{number}{8} (\bibinfo{year}{2019}), \bibinfo{pages}{1032}.
\newblock


\bibitem[Rosa et~al\mbox{.}(2021)]%
        {rosa2021empirical}
\bibfield{author}{\bibinfo{person}{Wilson Rosa}, \bibinfo{person}{Bradford~K
  Clark}, \bibinfo{person}{Raymond Madachy}, {and} \bibinfo{person}{Barry~W
  Boehm}.} \bibinfo{year}{2021}\natexlab{}.
\newblock \showarticletitle{Empirical effort and schedule estimation models for
  agile processes in the US DoD}.
\newblock \bibinfo{journal}{\emph{IEEE Transactions on Software Engineering}}
  \bibinfo{volume}{48}, \bibinfo{number}{8} (\bibinfo{year}{2021}),
  \bibinfo{pages}{3117--3130}.
\newblock


\bibitem[Rosa and Jardine(2023)]%
        {rosa2023data}
\bibfield{author}{\bibinfo{person}{Wilson Rosa} {and} \bibinfo{person}{Sara
  Jardine}.} \bibinfo{year}{2023}\natexlab{}.
\newblock \showarticletitle{Data-driven agile software cost estimation models
  for DHS and DoD}.
\newblock \bibinfo{journal}{\emph{Journal of Systems and Software (JSS)}}
  \bibinfo{volume}{203} (\bibinfo{year}{2023}), \bibinfo{pages}{111739}.
\newblock


\bibitem[Rosa et~al\mbox{.}(2017)]%
        {rosa2017early}
\bibfield{author}{\bibinfo{person}{Wilson Rosa}, \bibinfo{person}{Raymond
  Madachy}, \bibinfo{person}{Bradford Clark}, {and} \bibinfo{person}{Barry
  Boehm}.} \bibinfo{year}{2017}\natexlab{}.
\newblock \showarticletitle{{Early Phase Cost Models for Agile Software
  Processes in the US DoD}}. In \bibinfo{booktitle}{\emph{Proceedings of the
  International Symposium on Empirical Software Engineering and Measurement
  (ESEM)}}. \bibinfo{pages}{30--37}.
\newblock


\bibitem[Rubin(2012)]%
        {rubin2012essential}
\bibfield{author}{\bibinfo{person}{Kenneth~S Rubin}.}
  \bibinfo{year}{2012}\natexlab{}.
\newblock \bibinfo{booktitle}{\emph{{Essential Scrum: A Practical Guide to the
  Most Popular Agile Process}}}.
\newblock


\bibitem[Sakhrawi et~al\mbox{.}(2021)]%
        {sakhrawi2021support}
\bibfield{author}{\bibinfo{person}{Zaineb Sakhrawi}, \bibinfo{person}{Asma
  Sellami}, {and} \bibinfo{person}{Nadia Bouassida}.}
  \bibinfo{year}{2021}\natexlab{}.
\newblock \showarticletitle{{Support vector regression for enhancement effort
  prediction of Scrum projects from COSMIC functional size}}.
\newblock \bibinfo{journal}{\emph{Innovations in Systems and Software
  Engineering}} (\bibinfo{year}{2021}), \bibinfo{pages}{1--17}.
\newblock


\bibitem[Sakhrawi et~al\mbox{.}(2022)]%
        {sakhrawi2022software}
\bibfield{author}{\bibinfo{person}{Zaineb Sakhrawi}, \bibinfo{person}{Asma
  Sellami}, {and} \bibinfo{person}{Nadia Bouassida}.}
  \bibinfo{year}{2022}\natexlab{}.
\newblock \showarticletitle{Software enhancement effort estimation using
  stacking ensemble model within the scrum projects: a proposed web interface}.
  In \bibinfo{booktitle}{\emph{Proceedings of the International Conference on
  Software Technologies (ICSOFT)}}. \bibinfo{pages}{91--100}.
\newblock


\bibitem[S{\'a}nchez et~al\mbox{.}(2022)]%
        {sanchez2022software}
\bibfield{author}{\bibinfo{person}{Eduardo~Rodr{\'\i}guez S{\'a}nchez},
  \bibinfo{person}{Humberto~Cervantes Maceda}, {and}
  \bibinfo{person}{Eduardo~Vazquez Santacruz}.}
  \bibinfo{year}{2022}\natexlab{}.
\newblock \showarticletitle{Software effort estimation for Agile Software
  Development using a strategy based on K-nearest neighbors algorithm}. In
  \bibinfo{booktitle}{\emph{IEEE Mexican International Conference on Computer
  Science (ENC)}}. \bibinfo{pages}{1--6}.
\newblock


\bibitem[Sandeep et~al\mbox{.}(2022)]%
        {sandeep2022effort}
\bibfield{author}{\bibinfo{person}{RC Sandeep}, \bibinfo{person}{Mary
  S{\'a}nchez-Gord{\'o}n}, \bibinfo{person}{Ricardo Colomo-Palacios}, {and}
  \bibinfo{person}{Monica Kristiansen}.} \bibinfo{year}{2022}\natexlab{}.
\newblock \showarticletitle{{Effort Estimation in Agile Software Development: A
  Exploratory Study of Practitioners’ Perspective}}. In
  \bibinfo{booktitle}{\emph{Proceedings of the International Conference on Lean
  and Agile Software Development}}. \bibinfo{pages}{136--149}.
\newblock


\bibitem[Satapathy et~al\mbox{.}(2014)]%
        {satapathy2014story}
\bibfield{author}{\bibinfo{person}{Shashank~Mouli Satapathy},
  \bibinfo{person}{Aditi Panda}, {and} \bibinfo{person}{Santanu Rath}.}
  \bibinfo{year}{2014}\natexlab{}.
\newblock \showarticletitle{Story Point Approach based Agile Software Effort
  Estimation using Various SVR Kernel Methods}.
\newblock \bibinfo{journal}{\emph{Proceedings of the International Conference
  on Software Engineering and Knowledge Engineering, SEKE}}
  \bibinfo{volume}{2014}.
\newblock


\bibitem[Scott and Pfahl(2018)]%
        {scott2018}
\bibfield{author}{\bibinfo{person}{Ezequiel Scott} {and}
  \bibinfo{person}{Dietmar Pfahl}.} \bibinfo{year}{2018}\natexlab{}.
\newblock \showarticletitle{{Using Developers’ Features to Estimate Story
  Points}}. In \bibinfo{booktitle}{\emph{Proceedings of the International
  Conference on Software and System Process}}. \bibinfo{pages}{106--110}.
\newblock


\bibitem[Sharma and Chaudhary(2020)]%
        {sharma2020linear}
\bibfield{author}{\bibinfo{person}{Amrita Sharma} {and} \bibinfo{person}{Neha
  Chaudhary}.} \bibinfo{year}{2020}\natexlab{}.
\newblock \showarticletitle{Linear regression model for agile software
  development effort estimation}. In \bibinfo{booktitle}{\emph{Proceedings of
  the International Conference on Recent Advances and Innovations in
  Engineering}}. \bibinfo{pages}{1--4}.
\newblock


\bibitem[Sharma and Chaudhary(2022)]%
        {sharma2022combined}
\bibfield{author}{\bibinfo{person}{Amrita Sharma} {and} \bibinfo{person}{Neha
  Chaudhary}.} \bibinfo{year}{2022}\natexlab{}.
\newblock \showarticletitle{The combined model for software development effort
  estimation using polynomial regression for heterogeneous projects}.
\newblock \bibinfo{journal}{\emph{Radioelectronic and computer systems}}
  \bibinfo{number}{2} (\bibinfo{year}{2022}), \bibinfo{pages}{75--82}.
\newblock


\bibitem[Sharma and Singh(2017)]%
        {sharma2017systematic}
\bibfield{author}{\bibinfo{person}{Pinkashia Sharma} {and}
  \bibinfo{person}{Jaiteg Singh}.} \bibinfo{year}{2017}\natexlab{}.
\newblock \showarticletitle{{Systematic Literature Review on Software Effort
  Estimation Using Machine Learning Approaches}}. In
  \bibinfo{booktitle}{\emph{Proceedings of the International Conference on Next
  Generation Computing and Information Systems}}. \bibinfo{pages}{43--47}.
\newblock


\bibitem[Singal et~al\mbox{.}(2022)]%
        {singal2022integrating}
\bibfield{author}{\bibinfo{person}{Prerna Singal}, \bibinfo{person}{Prabha
  Sharma}, {and} \bibinfo{person}{A~Charan Kumari}.}
  \bibinfo{year}{2022}\natexlab{}.
\newblock \showarticletitle{Integrating software effort estimation with risk
  management}.
\newblock \bibinfo{journal}{\emph{International Journal of System Assurance
  Engineering and Management}} \bibinfo{volume}{13}, \bibinfo{number}{5}
  (\bibinfo{year}{2022}), \bibinfo{pages}{2413--2428}.
\newblock


\bibitem[Taibi et~al\mbox{.}(2017)]%
        {taibi2017operationalizing}
\bibfield{author}{\bibinfo{person}{Davide Taibi}, \bibinfo{person}{Valentina
  Lenarduzzi}, \bibinfo{person}{Philipp Diebold}, {and} \bibinfo{person}{Ilaria
  Lunesu}.} \bibinfo{year}{2017}\natexlab{}.
\newblock \showarticletitle{{Operationalizing the Experience Factory for Effort
  Estimation in Agile Processes}}. In \bibinfo{booktitle}{\emph{Proceedings of
  the International Conference on Evaluation and Assessment in Software
  Engineering (EASE)}}. \bibinfo{pages}{31--40}.
\newblock


\bibitem[Tamrakar and J{\o}rgensen(2012)]%
        {tamrakar2012does}
\bibfield{author}{\bibinfo{person}{Ritesh Tamrakar} {and}
  \bibinfo{person}{Magne J{\o}rgensen}.} \bibinfo{year}{2012}\natexlab{}.
\newblock \showarticletitle{{Does the use of Fibonacci numbers in Planning
  Poker affect effort estimates?}}. In \bibinfo{booktitle}{\emph{Proceedings of
  the International Conference on Evaluation and Assessment in Software
  Engineering}}. \bibinfo{pages}{228--232}.
\newblock


\bibitem[Tanveer et~al\mbox{.}(2017)]%
        {tanveer2017utilizing}
\bibfield{author}{\bibinfo{person}{Binish Tanveer}, \bibinfo{person}{Anna~Maria
  Vollmer}, {and} \bibinfo{person}{Ulf~Martin Engel}.}
  \bibinfo{year}{2017}\natexlab{}.
\newblock \showarticletitle{Utilizing change impact analysis for effort
  estimation in agile development}. In \bibinfo{booktitle}{\emph{Proceedings of
  the Euromicro Conference on Software Engineering and Advanced Applications
  (SEAA)}}. \bibinfo{pages}{430--434}.
\newblock


\bibitem[Tawosi et~al\mbox{.}(2022b)]%
        {tawosi2022versatile}
\bibfield{author}{\bibinfo{person}{Vali Tawosi}, \bibinfo{person}{Afnan
  Al-Subaihin}, \bibinfo{person}{Rebecca Moussa}, {and}
  \bibinfo{person}{Federica Sarro}.} \bibinfo{year}{2022}\natexlab{b}.
\newblock \showarticletitle{{A versatile dataset of agile open source software
  projects}}. In \bibinfo{booktitle}{\emph{Proceedings of the International
  Conference on Mining Software Repositories (MSR)}}.
  \bibinfo{pages}{707--711}.
\newblock


\bibitem[Tawosi et~al\mbox{.}(2022a)]%
        {tawosi2022investigating}
\bibfield{author}{\bibinfo{person}{Vali Tawosi}, \bibinfo{person}{Afnan
  Al-Subaihin}, {and} \bibinfo{person}{Federica Sarro}.}
  \bibinfo{year}{2022}\natexlab{a}.
\newblock \showarticletitle{{Investigating the Effectiveness of Clustering for
  Story Point Estimation}}. In \bibinfo{booktitle}{\emph{Proceedings of the
  International Conference on Software Analysis, Evolution and Reengineering
  (SANER)}}. \bibinfo{pages}{827--838}.
\newblock


\bibitem[Tawosi et~al\mbox{.}(2021)]%
        {tawosi2021multi}
\bibfield{author}{\bibinfo{person}{Vali Tawosi}, \bibinfo{person}{Federica
  Sarro}, \bibinfo{person}{Alessio Petrozziello}, {and} \bibinfo{person}{Mark
  Harman}.} \bibinfo{year}{2021}\natexlab{}.
\newblock \showarticletitle{Multi-objective software effort estimation: A
  replication study}.
\newblock \bibinfo{journal}{\emph{Transactions on Software Engineering (TSE)}}
  \bibinfo{volume}{48}, \bibinfo{number}{8} (\bibinfo{year}{2021}),
  \bibinfo{pages}{3185--3205}.
\newblock


\bibitem[Trendowicz et~al\mbox{.}(2011)]%
        {trendowicz2011state}
\bibfield{author}{\bibinfo{person}{Adam Trendowicz},
  \bibinfo{person}{J{\"u}rgen M{\"u}nch}, {and} \bibinfo{person}{Ross
  Jeffery}.} \bibinfo{year}{2011}\natexlab{}.
\newblock \showarticletitle{{State of the Practice in Software Effort
  Estimation: A Survey and Literature Review}}. In
  \bibinfo{booktitle}{\emph{Proceedings of the Software Engineering Techniques:
  Third IFIP TC 2 Central and East European Conference (CEE-SET)}}.
  \bibinfo{pages}{232--245}.
\newblock


\bibitem[Usman et~al\mbox{.}(2017)]%
        {usman2017effort}
\bibfield{author}{\bibinfo{person}{Muhammad Usman}, \bibinfo{person}{J{\"u}rgen
  B{\"o}rstler}, {and} \bibinfo{person}{Kai Petersen}.}
  \bibinfo{year}{2017}\natexlab{}.
\newblock \showarticletitle{{An Effort Estimation Taxonomy for Agile Software
  Development}}.
\newblock \bibinfo{journal}{\emph{International Journal of Software Engineering
  and Knowledge Engineering (IJSEKE)}} \bibinfo{volume}{27},
  \bibinfo{number}{4} (\bibinfo{year}{2017}), \bibinfo{pages}{641--674}.
\newblock


\bibitem[Usman et~al\mbox{.}(2018a)]%
        {usman2018effort}
\bibfield{author}{\bibinfo{person}{Muhammad Usman}, \bibinfo{person}{Ricardo
  Britto}, \bibinfo{person}{Lars-Ola Damm}, {and} \bibinfo{person}{J{\"u}rgen
  B{\"o}rstler}.} \bibinfo{year}{2018}\natexlab{a}.
\newblock \showarticletitle{{Effort estimation in large-scale software
  development: An industrial case study}}.
\newblock \bibinfo{journal}{\emph{Information and Software Technology (IST)}}
  \bibinfo{volume}{99} (\bibinfo{year}{2018}), \bibinfo{pages}{21--40}.
\newblock


\bibitem[Usman et~al\mbox{.}(2015)]%
        {usman2015effort}
\bibfield{author}{\bibinfo{person}{Muhammad Usman}, \bibinfo{person}{Emilia
  Mendes}, {and} \bibinfo{person}{J{\"u}rgen B{\"o}rstler}.}
  \bibinfo{year}{2015}\natexlab{}.
\newblock \showarticletitle{{Effort Estimation in Agile Software Development: A
  Survey on the State of the Practice}}. In
  \bibinfo{booktitle}{\emph{Proceedings of the International Conference on
  Evaluation and Assessment in Software Engineering (EASE)}}.
  \bibinfo{pages}{1--10}.
\newblock


\bibitem[Usman et~al\mbox{.}(2014)]%
        {Usman2014}
\bibfield{author}{\bibinfo{person}{Muhammad Usman}, \bibinfo{person}{Emilia
  Mendes}, \bibinfo{person}{Francila Weidt}, {and} \bibinfo{person}{Ricardo
  Britto}.} \bibinfo{year}{2014}\natexlab{}.
\newblock \showarticletitle{{Effort Estimation in Agile Software Development: A
  Systematic Literature Review}}. In \bibinfo{booktitle}{\emph{Proceedings of
  the International Conference on Predictive Models in Software Engineering}}.
  \bibinfo{pages}{82--91}.
\newblock


\bibitem[Usman et~al\mbox{.}(2018b)]%
        {usman2018developing}
\bibfield{author}{\bibinfo{person}{Muhammad Usman}, \bibinfo{person}{Kai
  Petersen}, \bibinfo{person}{J{\"u}rgen B{\"o}rstler}, {and}
  \bibinfo{person}{Pedro~Santos Neto}.} \bibinfo{year}{2018}\natexlab{b}.
\newblock \showarticletitle{{Developing and using checklists to improve
  software effort estimation: A multi-case study}}.
\newblock \bibinfo{journal}{\emph{Journal of Systems and Software (JSS)}}
  \bibinfo{volume}{146} (\bibinfo{year}{2018}), \bibinfo{pages}{286--309}.
\newblock


\bibitem[Vetro et~al\mbox{.}(2018)]%
        {vetro2018combining}
\bibfield{author}{\bibinfo{person}{Antonio Vetro}, \bibinfo{person}{Rupert
  D{\"u}rre}, \bibinfo{person}{Marco Conoscenti},
  \bibinfo{person}{Daniel~M{\'e}ndez Fern{\'a}ndez}, {and}
  \bibinfo{person}{Magne J{\o}rgensen}.} \bibinfo{year}{2018}\natexlab{}.
\newblock \showarticletitle{{Combining Data Analytics with Team Feedback to
  Improve the Estimation Process in Agile Software Development}}.
\newblock \bibinfo{journal}{\emph{Foundations of Computing and Decision
  Sciences}} \bibinfo{volume}{43}, \bibinfo{number}{4} (\bibinfo{year}{2018}),
  \bibinfo{pages}{305--334}.
\newblock


\bibitem[Wen et~al\mbox{.}(2012)]%
        {wen2012systematic}
\bibfield{author}{\bibinfo{person}{Jianfeng Wen}, \bibinfo{person}{Shixian Li},
  \bibinfo{person}{Zhiyong Lin}, \bibinfo{person}{Yong Hu}, {and}
  \bibinfo{person}{Changqin Huang}.} \bibinfo{year}{2012}\natexlab{}.
\newblock \showarticletitle{{Systematic literature review of machine learning
  based software development effort estimation models}}.
\newblock \bibinfo{journal}{\emph{Information and Software Technology (IST)}}
  \bibinfo{volume}{54}, \bibinfo{number}{1} (\bibinfo{year}{2012}),
  \bibinfo{pages}{41--59}.
\newblock


\bibitem[Wi{\'n}ska et~al\mbox{.}(2021)]%
        {winska2021reducing}
\bibfield{author}{\bibinfo{person}{Ewelina Wi{\'n}ska}, \bibinfo{person}{Estera
  Kot}, {and} \bibinfo{person}{W{\l}odzimierz D{\k{a}}browski}.}
  \bibinfo{year}{2021}\natexlab{}.
\newblock \showarticletitle{Reducing the uncertainty of agile software
  development using a random forest classification algorithm}. In
  \bibinfo{booktitle}{\emph{International Conference on Lean and Agile Software
  Development}}. Springer, \bibinfo{pages}{145--155}.
\newblock


\bibitem[Ziauddin and Zia(2012)]%
        {ziauddin2012effort}
\bibfield{author}{\bibinfo{person}{Shahid Kamal~Tipu Ziauddin} {and}
  \bibinfo{person}{Shahrukh Zia}.} \bibinfo{year}{2012}\natexlab{}.
\newblock \showarticletitle{{An Effort Estimation Model for Agile Software
  Development}}.
\newblock \bibinfo{journal}{\emph{Advances in computer science and its
  applications}} \bibinfo{volume}{2}, \bibinfo{number}{1}
  (\bibinfo{year}{2012}), \bibinfo{pages}{314--324}.
\newblock


\bibitem[Zimmermann et~al\mbox{.}(2010)]%
        {Zimmermann2010}
\bibfield{author}{\bibinfo{person}{Thomas Zimmermann}, \bibinfo{person}{Rahul
  Premraj}, \bibinfo{person}{Nicolas Bettenburg}, \bibinfo{person}{Sascha
  Just}, \bibinfo{person}{Adrian Schroter}, {and} \bibinfo{person}{Cathrin
  Weiss}.} \bibinfo{year}{2010}\natexlab{}.
\newblock \showarticletitle{{What Makes a Good Bug Report?}}
\newblock \bibinfo{journal}{\emph{Transactions of Software Engineering (TSE)}}
  \bibinfo{volume}{36}, \bibinfo{number}{5} (\bibinfo{year}{2010}),
  \bibinfo{pages}{618--643}.
\newblock


\end{thebibliography}

\section{Appendix}\label{sec:litrevappendix}
The first part of this section provide the SEGRESS Systematic Literature Review checklist~\cite{kitchenham2022segress}.
The rest of this section summarize the purpose, approach, scope, and the results of the selected studies.
Table~\ref{tab:table_litrev_segress} listed the SEGRESS (The PRISMA 2020-Inspired Structured Checklist for Reporting SE Secondary Studies), along with our explanation on how this work comply with each checklist item.
Table~\ref{tab:table_litrev_studylist_rq1reasons} summarize the selected studies that investigated the \reasons{} in Agile.
Table~\ref{tab:table_litrev_studylist_rq2estimate} summarize the studies that proposed an approach to estimate effort.
Table~\ref{tab:table_litrev_studylist_rq2support} summarize the studies that proposed an approach to support the effort estimation practices.

\begin{tcolorbox}[boxsep=3pt,left=10pt,right=10pt,top=10pt,bottom=10pt]
    \centering \textbf{All the tables are listed on the next page.}
\end{tcolorbox}





\begin{scriptsize}
\begin{longtable}{p{.07\textwidth}p{.1\textwidth}p{.05\textwidth}p{.4\textwidth}p{.4\textwidth}}
\caption{SEGRESS checklists}
\label{tab:table_litrev_segress}
\\
\textbf{Section} & \textbf{Sub-section} & \textbf{PRISMA Item} & \textbf{Summary} & \textbf{Checklist} \\

\hline

Full Report & n/a & n/a & Consider referencing material in the protocol, publishing material in supplementary material, and reporting any large-scale model building exercise separately from the SR report. & \checkmark{} Yes. We stated the use of guideline in the methodology while included this checklist in the supplementary material.\footnotemark[\getrefnumber{supplementary}]\\

Title & Title & 1 & Identify the type of review and specify the topic being reviewed. & \checkmark{} Yes. We identified the type of review in the title. \\

Abstract & Structured abstract & 2 & Provide a structured summary, including Background, Objective, Methods, Results, Limitations, and Conclusion. & \checkmark{} Yes. We provided the abstract in a structured style. \\

Introduction & Opening & n/a & Introduce the larger problem the paper is targeting, lay out a broad context for the work, and highlight the importance of the work to a large audience. & \checkmark{} Yes. We set the context of the study and highlighted the importance of this work in the Introduction section (Section~\ref{sec:introduction}). \\

Introduction & Rationale & 3 & Describe the rationale of the study and how it contributes to the larger problem. & \checkmark{} Yes. We explained the rationale of the study and contribution in the Introduction section (Section~\ref{sec:introduction}). \\

Introduction & Objectives & 4 & Specify the research questions and explain how they contribute to the larger problem. & \checkmark{} Yes. We specified and justify the research questions in the Research Questions section (Section~\ref{sec:litrevrq}), next to the Introduction section). \\

Methods & Eligibility criteria & 5 & Use the study characteristics to define eligibility criteria based on the intervention or topic of interest. Criteria used to restrict the search must be specified and justified. & \checkmark{} Yes. We defined the eligibility criteria based on the past literature reviews in effort estimation. We also constructed the search criteria following the PICOC suggested by Kitchenham and Charters~\cite{kitchenham2007guidelines} \\

Methods & Information sources & 6 & Describe all information sources, databases, primary study references, and others (e.g., researchers) with search end dates & \checkmark{} Yes. We described the search end dates of our study in the Search Strategy section (Section~\ref{sec:searchstrategy}). \\

Methods & Search Strategy & 7 & Present full search strategy, e.g., electronic search strings, manual search, method(s) used to assess achieved completeness. & \checkmark{} Yes. We described the search strategy in the Search Strategy section (Section~\ref{sec:searchstrategy}) \\

Methods & Selection Process & 8 & State the process for selecting studies, i.e., phases, assessors, handling disagreements, and tools used. & \checkmark{} Yes. We described the process for study selection, assessing, and handling disagreements in the Study selection section (Section~\ref{sec:studyselection}).  \\

Methods & Data Collection Process & 9 & Specify the method used to collect data from reports, i.e., processes, reviewers, whether they worked independently, tools used. & \checkmark{} Yes. We specified the methods used to obtain the data and construct the thematic taxonomies in the Data Analysis section (Section~\ref{sec:litrevdataanalysis}). \\

Methods & Data items & 10a & List, define, and justify all outcomes for which data was sought, explaining their relationship to the research questions. & \checkmark{} Yes. We presented and justify the extracted data for each research question in the Data Analysis section (Section~\ref{sec:litrevdataanalysis}). \\

Methods & Data items & 10b & List and define all non-outcome variables (e.g., participant and intervention characteristics) and assumptions made about missing or unclear information. & \checkmark{} Yes. We described the non-outcome variables collected in the Quality criteria and Uncertainty assessment section (Section \textbackslash{}ref\textbackslash{}{sec:litrevphasec\textbackslash{}} and \textbackslash{}ref\textbackslash{}{sec:litrevphased\textbackslash{}}). \\

Methods & Study Risk Of Bias Assessment & 11 & Specify the methods used to assess risk of bias, i.e., tools used, reviewers, and whether they worked independently. & \checkmark{} Yes. We specified the process to assess the risk of bias of individual study along with the GRADE assessment in Uncertainty and risk of bias assessment section (Section~\ref{sec:litrevphased}). \\

Methods & Effect Measures & 12 & Specify for each outcome the effect measures used in the synthesis or presentation of results. Not required for qualitative reviews. & $\times$ No. Our qualitative review did not include any calculation of effect measure. \\

Methods & Analysis and Synthesis & 13a & Report the methods used for synthesis of primary study outcomes. & \checkmark{} Yes. We reported our Open Card Sorting process to derive the themes and taxonomy of the primary study outcomes (RQ1) in the Discovering the themes and taxonomy section (Section~\ref{sec:litrevtaxonomy}). \\

Methods & Methods & 13b & Describe any methods required to prepare the data for presentation or synthesis. Describe the coding processes adopted. & \checkmark{} Yes. We described in details the methods to collect and prepare the data in the Data Analysis section (Section~\ref{sec:litrevdataanalysis}). \\

Methods & Methods & 13c & Describe any methods used to tabulate or display results. & \checkmark{} Yes. We described the tools we used in the Data analysis section (Section~\ref{sec:litrevdataanalysis}). \\

Methods & Methods & 13d & Describe any methods used to synthesize results with rationale. Qualitative studies should, where necessary, identify constructs analyzed, explain how findings from different studies were compared, and specify how synthesized findings were validated. & \checkmark{} Yes. We described the validation method of the synthesized findings in the Data Analysis section (Section~\ref{sec:litrevdataanalysis}). \\

Methods & Methods & 13e & Describe any sensitivity analysis conducted to assess robustness of the synthesized results and, including any impact of deviant cases and exceptions. & \checkmark{} Yes. We described that we mark the studies with high risk of bias and describe the impact if we exclude them from our study in the Discovering the theme and taxonomy section (Section~\ref{sec:litrevtaxonomy}). \\

Methods & Methods & 13f & Describe any methods used to explore possible causes of heterogeneity among study results. & \checkmark{} Yes. We described that we exploring the heterogeneity among the study results in the Data analysis section (Section~\ref{sec:litrevdataanalysis}). \\

Methods & Reporting Bias Assessment & 14 & Describe any methods used to assess risk of bias due to publication bias. & \checkmark{} Yes. We described our publication bias assessment using GRADE in Phase D: Uncertainty and risk of bias assessment section (Section~\ref{sec:litrevphased}). \\

Methods & Certainty Assessment & 15 & Describe methods used to assess certainty (or confidence) in the body of evidence for an outcome. & \checkmark{} Yes. We described the process we applied GRADE certainty assessment to individual study in Phase D: Uncertainty and risk of bias assessment section (Section~\ref{sec:litrevphased}). \\

Results & Study selection & 16a & Describe the results of the search and selection process, ideally using a flow diagram. Qualitative studies should describe any iteration between selection and synthesis. & \checkmark{} Yes. We described the results of the search and section processes along with flow diagram in the Search Results section (Section~\ref{sec:litrevsearchresult}). \\

Results & Study selection & 16b & Cite studies that met many but not all inclusion criteria (’near-misses‘) and explain why they were excluded. & \checkmark{} Yes. We described the example studies that almost being included in the study along with the supporting reasons in the Search Results section (Section~\ref{sec:litrevsearchresult}). \\

Results & Study characteristics & 17 & Describe the characteristics of each included study, and provide citations. & \checkmark{} Yes. We described the characteristics with citation of each included study in the Appendix, i.e., Table~\ref{tab:table_litrev_studylist_rq1reasons}, Table~\ref{tab:table_litrev_studylist_rq2estimate}, and Table~\ref{tab:table_litrev_studylist_rq2support} (Section~\ref{sec:litrevappendix}). \\

Results & Risk of Bias in Studies & 18 & Present data on the risk assessment for each study. & \checkmark{} Yes. We presented the risk of bias assessment of each study in the Appendix, i.e., Table~\ref{tab:table_litrev_studylist_rq1reasons}, Table~\ref{tab:table_litrev_studylist_rq2estimate}, and Table~\ref{tab:table_litrev_studylist_rq2support} (Section~\ref{sec:litrevappendix}). \\

Results & Results of individual studies & 19 & For qualitative reviews, present the major findings from each study included in the synthesis. & \checkmark{} Yes. We presented the key findings from individual studies in the Appendix, i.e., Table~\ref{tab:table_litrev_studylist_rq1reasons}, Table~\ref{tab:table_litrev_studylist_rq2estimate}, and Table~\ref{tab:table_litrev_studylist_rq2support} (Section~\ref{sec:litrevappendix}). \\

Results & Results of Analyses and Syntheses & 20 & Describe results of syntheses and analyses. & \checkmark{} Yes. We presented the derived themes and taxonomies for each research question in the Results section (Section~\ref{sec:litrevresult}). \\

Results & Results of Analyses and Syntheses & 20a & Report each synthesis by briefly summarising the characteristics and risk of bias among contributing studies. For qualitative studies, define any derived themes, and focus on theory building and testing. Provide appropriate quotations with sources. & \checkmark{} Yes. We shortly summarised the characteristics along with risk of bias when presenting the themes and taxonomies where appropriate. We also provide quotations from studies to help illustrate the characteristics of the themes in the Results section (Section~\ref{sec:litrevresult}). \\

Results & Results of Analyses and Syntheses & 20b & Present results of all statistical syntheses conducted. Only required for quantitative reviews. & $\times$ No. Not applicable to our study. \\

Results & Results of Analyses and Syntheses & 20c & Present results of robustness assessment of the synthesized results. Qualitative studies should report deviant cases,  exceptions, and any additional validation. & \checkmark{} Yes. We discussed the deviant cases in our results in the Results section (Section~\ref{sec:litrevresult}). \\

Results & Results of Analyses and Syntheses & 20d & Present results of all investigations of possible causes of heterogeneity among study results. & \checkmark{} Yes. We reported one study included in both RQs and another study included in two taxonomies in the Search Results section (Section~\ref{sec:litrevresult}). \\

Results & Reporting Biases & 21 & Report results of assessing publication bias for each synthesis. Not usually required for qualitative studies. & \checkmark{} Yes. We presented the risk of bias for the reported themes and taxonomies in the Results section (Section~\ref{sec:litrevresult}). \\

Results & Certainty of Evidence & 22 & Present assessment of certainty in the body of evidence for each reported finding. & \checkmark{} Yes. We presented the assessment of certainty for each finding in the Results section (Section~\ref{sec:litrevresult}). \\

Discussion & Discussion & 23a & Provide a general interpretation of the results in the context of other evidence and compare the findings with other reviews on the same topic. & \checkmark{} Yes. We discussed interpretation of the results and compare the findings with other work in the Discussion section (Section~\ref{sec:litrevdiscussion}). \\

Discussion & Discussion & 23b & Discuss any limitations of the evidence included in the review. & \checkmark{} Yes. We discussed the limitations of the evidence included in the review in the Threats to Validity section (Section~\ref{sec:litrevthreats}). \\

Discussion & Discussion & 23c & Discuss any limitations of the review process used. & \checkmark{} Yes. We discussed the limitations of our review process in the Threats to Validity section (Section~\ref{sec:litrevthreats}). \\

Discussion & Discussion & 23d & Discuss implications of the results for practice, policy and future research. & \checkmark{} Yes. We discussed the implications of our results in context of software engineering in practical and the recommendations for practitioners and future research in the Discussion section (Section~\ref{sec:litrevdiscussion}). \\

Discussion & Registration and Protocol & 24a & Provide registration information for the review. Optional for all review types. & $\times$ No. We did not register this study. \\

Discussion & Registration and Protocol & 24b & Indicate where the review protocol can be accessed or state why no protocol is available. & $\times$ No. We did not register this study. \\

Discussion & Registration and Protocol & 24c & Describe and explain any amendments to information provided at registration or in the protocol. & $\times$ No. We did not register this study. \\

Discussion & Support & 25 & Describe sources of financial and non-financial support for the review and the role of the funders or sponsors of the review. & \checkmark{} Yes. We declared that our work has no financial sponsor in the Introduction section (Section~\ref{sec:introduction}). \\

Discussion & Competing Interests & 26 & Declare competing interests of the review authors. & \checkmark{} Yes. We declared that the first author is currently working at a public company in the Introduction section (Section~\ref{sec:introduction}). \\

Discussion & Availability Of Data Code and Other Materials & 27 & Report which of the following are publicly available and where they can be found: template data collection forms; data extracted from included studies; data used for all analyses; analytic code; any other materials used in the review or to produce the review. & \checkmark{} Yes. We provided all data and results in the supplementary material package. \\

\hline
\end{longtable}
\end{scriptsize}

\begin{scriptsize}
\begin{landscape}

\begin{longtable}{p{.06\textwidth}p{.15\textwidth}p{.3\textwidth}p{.1\textwidth}p{.20\textwidth}p{.1\textwidth}p{.07\textwidth}p{.08\textwidth}p{.07\textwidth}}

\caption{Summary of the selected studies that investigate the \reasons{} in Agile (Section~\ref{sec:litrevreasonsection}).}
\label{tab:table_litrev_studylist_rq1reasons}

\\

\textbf{Authors} & \textbf{Title} & \textbf{Purpose and techniques} & \textbf{Dataset used / Participants} & \textbf{Results} & GRADE-Individual Study Bias & GRADE-imprecision & GRADE-Inconsistency & GRADE-indirectness  \\ 
\hline

        Usman et al.~\cite{usman2015effort} & Effort Estimation in Agile Software Development: A Survey on the State of the Practice & Conducted an online survey on the state of the practice of Agile effort estimation. The reported \reasons{} are grouped into four categories and ranked based on the number of reported respondents. & 60 respondents & 28 from 60 respondents reported the \reasons{}. & none & none & none & none \\ 
        Britto et al.~\cite{Britto2015} & An Empirical Investigation on Effort Estimation in Agile Global Software Development & Conducted an online survey on the practice of Agile effort estimation in distributed team settings. These challenges were later grouped into seven categories and ranked based on the number of reported respondents. & 51 respondents & 24 from 51 respondents reported the challenges that impact the effort estimation accuracy. & none & none & none & none \\ 
        Usman et al.~\cite{usman2018effort} & Effort estimation in large-scale software development: An industrial case study & Conducted an exploratory longitudinal case study based on archival research and semi-structured interviews on the effort estimation practices in large-scale Agile software development. & Four interview participants & Reported five challenges that led to the inaccuracies in effort estimations. & moderate - small sample size & none & none & none \\ 
        Conoscenti et al.~\cite{conoscenti2019combining} & Combining data analytics and developers feedback for identifying reasons of inaccurate estimations in agile software development & Employed a tool to ask the developers whether the estimated effort is inaccurate or not along with the supporting reason. The tool provided a checklist of reasons for the developers to check for an inaccurate estimation. The authors also provided an ``Other'' check-box to collect additional reasons if any. The results were then ranked them based on their occurrences. & 70 developers in a medium-sized company with 124 user stories. & Reported 14 \reasons{} that were ranked based on their occurrences. & high - The reasons investigated were derived from the paper published before Agile manifesto (before year 2000). & none & none & none \\ 
        Sandeep et al.~\cite{sandeep2022effort} & Effort Estimation in Agile Software Development: A Exploratory Study of Practitioners’ Perspective & Conducted an online survey that is designed based on a literature review to provide empirical evidence on Agile effort estimation techniques. The reasons were then ranked using the average scores of five-points agreement scales. & 53 agile practitioners & Reported 20 \reasons{} in Agile software development. & none & none & none & none \\ 
        Tamrakar and Jørgensen~\cite{tamrakar2012does} & Does the use of Fibonacci numbers in Planning Poker affect effort estimates? & Conducted two empirical studies to estimate effort with and without using Fibonacci scale to understand the effect of the Fibonacci scale adoption in effort estimation. & 104 students and 16 professionals & Reported that the use of Fibonacci scale is likely to affect the estimates towards the lower values. & none & none & none & none \\ 
        Karna et al.~\cite{karna2020effects} & The Effects of Turnover on Expert Effort Estimation & Examined the data from a large private software project to determined the effects of the employee turnover on expert effort estimation. & One large Agile project & Reported that only unplanned turnover can have a significant negative impact on the reliability of the estimates. & moderate - only one project dataset & none & none & none \\ 
        Vetro et al.~\cite{vetro2018combining} & Combining Data Analytics with Team Feedback to Improve the Estimation Process in Agile Software Development & Combined software data analytic and feedback from the teams in a company to identify root causes of the wrong estimates. & Four software development projects & Reported that using estimation scale with 1) too many items in the estimation scale 2) misleading numerical scale and 3) lack of shared reflection on the previous estimations can lead to wrong estimations. & none & none & none & none \\ 
\hline
\end{longtable}
\end{landscape}
\end{scriptsize}

\begin{landscape}
\begin{scriptsize}
\begin{longtable}{p{.07\textwidth}p{.12\textwidth}p{.07\textwidth}p{.05\textwidth}p{.05\textwidth}p{.15\textwidth}p{.10\textwidth}p{.15\textwidth}p{.07\textwidth}p{.06\textwidth}p{.07\textwidth}p{.08\textwidth}}

\caption{Summary of the selected studies that proposed an approach to estimate the effort (Section~\ref{sec:litrevpurposeestimate}). The ``ML'' denotes that the purpose of the approach is to predict the effort using machine learning-based technique.}
\label{tab:table_litrev_studylist_rq2estimate}

\\
\textbf{Authors} & \textbf{Title} & \textbf{Purpose} & \textbf{Planning level} & \textbf{Artifact} & \textbf{Technique} &\textbf{Dataset used} & \textbf{Evaluation method performed} & \textbf{GRADE-Individual Study Bias} & \textbf{GRADE-imprecision} & \textbf{GRADE-Inconsistency} & \textbf{GRADE-indirectness}  \\

\hline
        Alhamed and Storer \cite{alhamed2021playing} & Playing Planning Poker in Crowds: Human Computation of Software Effort Estimates & Improving estimation technique (manual) & Sprint planning & work items & Using the opinions of the crowd recruited from Amazon Turk to estimate the completion time. & 5,000 software engineers on Amazon Turks, 39 software tasks & Conducted the crowd estimation for 30 trials with 39 stories. Comparing the prediction accuracy with the experts. & none & none & high - first work using crowd-sourcing & high - crowd estimators may not have sufficient estimation experience \\ 
        Alsaadi et al. \cite{alsaadi2021scrum} & Scrum Poker Estimator: A Planning Poker Tool for Accurate Story Point Estimation. & Improving estimation technique (manual) & Not specified & work items & Proposing a tool to assist the estimation process at work items level in determining the estimation method, measure weight of estimation factors, extract questions for the factors, and determine the estimated fibonacci number. & One undergraduate student senior project, consists of five user stories in four sprints & Applied the proposed approach with four students in their undergraduate senior project. & high - student projects & none & none & none \\ 
        Altaleb et al. \cite{altaleb2020pair} & A Pair Estimation Technique of Effort Estimation in Mobile App Development for Agile Process: Case Study & Improving estimation technique (manual) & Sprint planning & work items & Proposing a pair estimation technique based on learnings from a focus group and an observation to enhance a deeper discussion between developers before estimating the effort. & A case study with a company of 150 employees & Examined the prediction accuracy after adopting the approach using a correlation test. & none & none & none & none \\ 
        Madya et al. \cite{madya2022prep} & PREP: A Post-Requirements Effort Estimation Method in Scrum's Sprint Grooming & Improving estimation technique (manual) & Sprint planning & work items & Proposing a framework that guides the practitioners to develop high quality user stories, estimate the effort in detailed steps, and evaluate the approach after each iteration. & Seven sprints dataset from one project in one company & Compared the prediction accuracy with the actual effort, and measured the reduction of bugs. & none & none & none & none \\ 
        Omar El Beggar \cite{el2023ifejm} & IFEJM: New Intuitionistic Fuzzy Expert Judgment Method for Effort Estimation in Agile Software Development & Improving estimation technique (manual) & Not specified & work items & Using a non-linear model to adjust the Planning Poker estimations of a work item when a consensus is not reached. & One agile project data with two sprints estimated by students and professionals & Compared the prediction accuracy with the actual effort. & none & none & none & none \\ 
        Rola and Kuchta \cite{rola2019application} & Application of fuzzy sets to the expert estimation of Scrum-based projects & Improving estimation technique (manual) & Sprint planning & work items & Proposing a framework for a team to provide an estimation range, and apply a mathematical formula to calculate the final estimation of a work item. & Two case studies of an international IT project and an R and D project & Following symmetry principles, evaluated and compared the accuracy of the fuzzy approach and compare to the crisp estimate. & none & none & none & none \\ 
        Vetro et al. \cite{vetro2018combining} & Combining Data Analytics with Team Feedback to Improve the Estimation Process in Agile Software Development & Improving estimation technique (manual) & Sprint planning & work items & Proposing an estimation technique that requires the team to revisit the estimations of previous sprints and use a shorter numerical scale. & 708 user stories from 4 projects in a company & Applied the estimation process on a new project & none & none & none & none \\ 
        Adnan et al. \cite{adnan2019ontology} & Ontology-Oriented Software Effort Estimation System for E-commerce Applications Based on Extreme Programming and Scrum Methodologies & Predict the effort (ML) & Not specified & projects & Using a regression model to predict the project effort based on lesson-learned (textual data) from previous estimations. & 20 XP and Scrum projects & Used the model to predicted the project effort in a case study. & moderate - report only relative accuracy units & none & none & none \\ 
        Alostad et al. \cite{alostad2017fuzzy} & A Fuzzy based Model for Effort Estimation in Scrum Projects & Predict the effort (ML) & Sprint planning & work items & Using fuzzy-based model to estimate the Story Points. & 30 tasks from 3 sprints in a project & Applied the model on three sprints and observed the accuracy in each sprint. & moderate - report only relative accuracy units & none & none & none \\ 
        Ardiansyah et al. \cite{ardiansyah2022mucpso} & MUCPSO: A Modified Chaotic Particle Swarm Optimization with Uniform Initialization for Optimizing Software Effort Estimation & Predict the effort (ML) & not specified & work items & Proposing a modified chaotic particle swarm optimization to optimize three effort estimation methods, i.e., Use case points, COCOMO, and Agile. & Three datasets retrieved from the prior work for the three estimation methods. & Compared the prediction performance with three particle swarm optimization variants and a nature-inspired algorithm. The evaluation is conducted 30 rounds to eliminate coincidence. & moderate - old dataset & none & none & none \\ 
        Arora et al. \cite{arora2022efficient} & An efficient ANFIS-EEBAT approach to estimate effort of Scrum projects & Predict the effort (ML) & Not specified & projects & Using a parametric-based approaches, i.e., adaptive neuro-fuzzy inference system (ANFIS) with EEBAT technique, to predict the effort of a project & 21 Agile projects from 6 software houses proposed by Ziauddin, et al. & Compared the prediction accuracy with the state-of-the-art meta-heuristic and machine learning approach using nonparametric tests. & none & none & none & none \\ 
        Bilgaiyan et al. \cite{bilgaiyan2018chaos} & Chaos-based modified morphological genetic algorithm for software development cost estimation & Predict the effort (ML) & Project planning & projects & Using a dilation-erosion perceptron to predict the effort of a project. & Five datasets from literature, i.e., Albrecht, Desharnais, Kemerer, COCOMO, and KotenGray & Compared the prediction accuracy with the actual effort using an evaluation function (EF), PRED(x), and MMRE units. & moderate - report only relative accuracy units & none & none & none \\ 
        Bilgaiyan et al. \cite{bilgaiyan2019effort} & Effort estimation in agile software development using experimental validation of neural network models & Predict the effort (ML) & Not specified & projects & Using Elman Recurrent Network and feedforward Back-Propagation Network models to predict the project effort based on the features of user story and team. & 21 projects retrieved from Ziauddin and Zia \cite{ziauddin2012effort} & Compared the predicted effort with actual effort. & none
        & none & none & none \\ 
        Choetkiertikul et al. \cite{Choetkiertikul2019} & A deep learning model for estimating story points & Predict the effort (ML) & Sprint planning or release planning & work items & Using a combination of Long Short-Term Memory and Recurrent Highway Network to estimate the Story Points of a work item. & 23,313 issues from 16 open-source projects & Compared the prediction performance with baselines (i.e., LSTM +RF, BoW+RF, Doc2vec+RF) based on 60(train)-20(validate)-20(test) split, chronologically. Perform the evaluation with original and adjusted Story Points. & none & none & none & low - open-source projects may adopt Agile differently \\ 
        Dan et al. \cite{dan2020nlp} & An NLP Approach to Estimating Effort in a Work Environment & Predict the effort (ML) & Not specified & work items & Using a Bi-LSTM model to classify the time taken to complete a user story based on their description. & One company dataset & Compared the predicted effort with actual effort and compare with an LSTM model and a concatenated model. & high - the size of the dataset is not described & none & moderate - training a language model, but the language data is in Germany & high - it is unclear whether the projects were developed in Agile \\ 
        Dantas et al. \cite{dantas2019effort} & An Effort Estimation Support Tool for Agile Software Development: An Empirical Evaluation & Predict the effort (ML) & Sprint planning & work items & Using decision trees to predict the time (in hours) to complete a user story based on the features extracted from user stories and the features provided by the team. & 530 user stories and 1,879 tasks from a company & Compared the predicted effort with the estimation of 24 professionals and the actual effort based on 10-folds cross validation / we compared the values of Magnitude of Relative Error from the teams’ estimations with the values provided by the tool. & none & none & none & none \\ 
        Dragicevic et al. \cite{dragicevic2017bayesian} & Bayesian network model for task effort estimation in agile software development & Predict the effort (ML) & Project planning & work items & Using a Bayesian Network model to estimate the effort of a task based on software entities, tasks complexity, and knowledge and skills of the developers. & 160 tasks from a company & Evaluated the prediction performance based on 10-folds cross-validation. & moderate - small dataset & none & none & none \\ 
        Farahneh and Issa \cite{farahneh2011linear} & A Linear Use Case Based Software Cost Estimation Model & Predict the effort (ML) & Not specified & projects & Using a linear model to predict the project effort based on project size and programming language. & 66 projects retrieved from Issa et al. & Compared the predicted effort with actual effort and compare with a baseline. & none & none & none & high - it is unclear whether the projects were developed in Agile \\ 
        Fu and Tantithamthavorn \cite{fu2022gpt2sp} & GPT2SP: A transformer-based agile story point estimation approach & Predict the effort (ML) & Sprint planning & work items & Using a GPT-2 pre-trained language model to predict the Story Points of a work item. & 16 open-source projects from Choetkiertikul et al. & Evaluated the prediction accuracy with the actual effort (in MAE) and compared the performance with nine baseline approaches in within-project and cross-projects scenarios. & none & none & none & low - open-source projects may adopt Agile differently \\ 
        Govil and Sharma \cite{govil2022estimation} & Estimation of cost and development effort in Scrum-based software projects considering dimensional success factors & Predict the effort (ML) & Not specified & projects & Using a parametric-based algorithm that takes user story, story points, and success factors as inputs to calculate the effort and cost of the project & 30 Scrum-based projects, validated by industry professionals with experience in Agile. & Compare the accuracy of the proposed algorithm with the existing algorithm, and validating the result with annotators. & high - unclear explanation where the 30 projects come from, and it is unclear why the dataset needs to be validated (and how). & none & none & none \\ 
        Gultekin and Kalipsiz \cite{gultekin2020story} & Story Point-Based Effort Estimation Model with Machine Learning Techniques & Predict the effort (ML) & not specified & work items & Using regression-based models to estimate the Story Points based on the features of work items. & 56,687 issues, 3,834 iterations, from 5 OSS projects & Compared the predicted effort with actual effort. & none & none & none & none \\ 
        Gupta and Mahapatra \cite{gupta2022automated} & Automated software effort estimation for agile development system by heuristically improved hybrid learning & Predict the effort (ML) & Not specified & projects & Using a Deep Belief Network $\backslash$ & Artificial Neural Network - based technique to predict the effort of a project. & Three publicly available datasets, i.e., COCOMO, NASA93, and CHINA & moderate - old datasets & none & none & none \\ 
        Hearty et al. \cite{Hearty2009} & Predicting project velocity in XP using a learning dynamic Bayesian network model & Predict the effort (ML) & Not specified & sprints & Using a Bayesian Network model to predict the project velocity based on the process factors. & One company project & Applied the model to an industrial case study to predict the project velocity for each iteration. & moderate - report only relative accuracy units & none & none & none \\ 
        Karna et al. \cite{karna2020data} & Data Mining Approach to Effort Modeling On Agile Software Projects & Predict the effort (ML) & Not specified & sprints & Using a K-Nearest Neighbor technique to predict the effort of an iteration based on the features of work items and the estimator. & One company project & Compared the predicted effort with actual effort. & moderate - small dataset & none & none & none \\ 
        Kassem et al. \cite{kassem2022software} & Software Effort Estimation Using Hierarchical Attention Neural Network & Predict the effort (ML) & Not specified & work items & Using hierarchical attention networks to classify user stories into a range of story points. & 16 open-source projects from Choetkiertikul et al. & Evaluated the prediction accuracy with the ground truth story points and compared with four baselines. & none & none & none & low - open-source projects may adopt Agile differently \\ 
        Kassem et al. \cite{kassem2023story} & Story Point Estimation Using Issue Reports With Deep Attention Neural Network & Predict the effort (ML) & Not specified (planning poker) & work items & Using Hierarchical Attention Networks (HANs) at word and sentence levels for predicting Story Points of a work item. & 16 open-source projects from Choetkiertikul et al. & Compared the prediction accuracy with two baselines, i.e., Deep-SP and BERT. & none & none & none & low - open-source projects may adopt Agile differently \\ 
        Kaushik et al. \cite{kaushik2020fuzzified} & A fuzzified story point approach for agile projects & Predict the effort (ML) & Not specified & projects & Using the fuzzified Story Points approach to calculate the project effort based on the cost drivers. & 145 generated projects & Compared the prediction performance between fuzzy interval type 1 and type 2 & high - using a generated dataset & none & high - using a generated dataset & high - using a generated dataset \\ 
        Kaushik et al. \cite{kaushik2022role} & The role of neural networks and metaheuristics in agile software development effort estimation & Predict the effort (ML) & Not specified & projects & Estimating the effort using artificial neural networks (Radial Basis Function Neural Network and Functional Link Artificial Neural Network) and a metaheuristic technique (Whale Optimization Algorithm). & 21 projects retrieved from Ziauddin and Zia \cite{ziauddin2012effort} & Compare the predicted effort with actual effort, and compare with other proposed approaches. & moderate - report only relative accuracy units & none & none & none \\ 
        Khuat et al. \cite{khuat2018novel} & A Novel Hybrid ABC-PSO Algorithm for Effort Estimation of Software Projects Using Agile Methodologies & Predict the effort (ML) & Not specified & projects & Using a combination of the artificial bee colony and particle swarm optimization algorithms to estimate software project effort. & 21 projects retrieved from Ziauddin and Zia \cite{ziauddin2012effort} & Compared the predicted effort with actual effort. & moderate - report only relative accuracy units & none & none & none \\ 
        Malgonde and Chari \cite{malgonde2019ensemble} & An ensemble-based model for predicting agile software development effort & Predict the effort (ML) & Sprint planning & work items & Using an ensemble-based method to predict the effort of a story. & 503 user stories of 24 projects from university IT department & Conducted experiments to compare the prediction performance of the ensemble-based approach with other approaches based on blocked cross-validation technique. & none & none & none & none \\ 
        Marapelli et al. \cite{marapelli2020rnn} & RNN-CNN MODEL: A Bi-directional Long Short-Term Memory Deep Learning Network For Story Point Estimation & Predict the effort (ML) & Not specified & work items & Using a Bi-LSTM (RNN with CNN) model to predict Story Points based on the description of a user story. & 16 open-source projects retrieved from Choetkiertikul et al. & Compared the predicted effort with actual effort and compared the prediction performance with the approach of Choetkiertikul et al. & high - unclear explanation on the cross-validation method & none & none & low - open-source projects may adopt Agile differently \\ 
        Miranda et al. \cite{miranda2021analysis} & An Analysis of Monte Carlo Simulations for Forecasting Software Projects & Predict the effort (ML) & Not specified & work items & Using Monte-Carlo simulations to predict the effort and deliver time (based on Story Points) of a user story. & 71 projects, 12,000 user stories, 135,000 changes, 150 sprints, from a company dataset & Compared the model prediction performance with expert estimations based on 50$\backslash$\% hold-out. The prediction was evaluated  in 3 aspects, i.e., delivery date, effort (hours) for each work item, and the impact of the size of historical data used. & none & none & none & none \\ 
        Moharreri et al. \cite{moharreri2016cost} & Cost-Effective Supervised Learning Models for Software Effort Estimation in Agile Environments & Predict the effort (ML) & Sprint planning & work items & Using machine learning models to improve the Planning Poker accuracy (predict the effort) based on the features extracted from Agile story cards. & Story cards from 10 teams in a company & Compared the predicted effort with actual effort. & high - the characteristic of the 10 studied teams is unclear & none & none & none\\ 
        Najm et al. \cite{najm2022enhanced} & An enhanced support vector regression model for agile projects cost estimation & Predict the effort (ML) & Not specified & projects & Using Support Vector Regression with Radial Bias Function kernel (SVR-RBF) to predict the effort of a project. & 21 Agile projects from 6 software houses proposed by Ziauddin et al. & Compared the prediction accuracy with 10 baselines using a leave-one-out cross-validation. & moderate - report only relative accuracy units & none & none & none \\ 
        Nunes et al. \cite{nunes2011iucp} & iUCP: Estimating Interactive-Software Project Size with Enhanced Use-Case Points & Predict the effort (ML) & Not specified & projects & Proposing a technique to estimate the project size based on use-case points estimation with the usage-centered design. & 30 master students & Conducted an empirical experiment with master’s students. The students developed a group project using the proposed approaches. & high - tested the approach with master students only & none & none & high - students may not have sufficient experience in Agile software development \\ 
        Panda et al. \cite{panda2015empirical} & Empirical Validation of Neural Network Models for Agile Software Effort Estimation based on Story Points & Predict the effort (ML) & Not specified & projects & Evaluating different neural networks (GRNN, PNN, GMDH, and CCNN) to enhance the Story Points Approach estimation. & 21 projects retrieved from Ziauddin and Zia \cite{ziauddin2012effort} & Compared the prediction performance among the four neural networks and the actual effort, based on 3-folds cross validation. & none & none & none & none \\ 
        Phan and Jannesari \cite{phan2022heterogeneous} & Heterogeneous Graph Neural Networks for Software Effort Estimation & Predict the effort (ML) & Not specified & work items & Using a heterogeneous graph neural networks model with FastText to predict the effort of a work item. & 16 open-source projects from Choetkiertikul et al. & Compared the proposed approach with baselines (i.e., GPT2SP and Deep-SE) across three scenarios of estimation: within project, cross-project within the repository, and cross-project cross repository. & none & none & none & low - open-source projects may adopt Agile differently \\ 
        Phan and Jannesari \cite{phan2022story} & Story point level classification by text level graph neural network & Predict the effort (ML) & Not specified & work items & Using a Graph Neural Network to perform effort classification of a work item based on textual data. & 16 open-source projects from Choetkiertikul et al. & Evaluated the effectiveness of the approach through experiments and compared the results with TF-IDF approach. & none & none & none & low - open-source projects may adopt Agile differently \\ 
        Porru et al. \cite{Porru2016} & Estimating Story Points from Issue Reports & Predict the effort (ML) & Sprint planning & work items & Using naive-bayes classifiers to classify the Story Points of a work item. & 1 industrial project and 8 open source projects & Evaluated the prediction performance based on 10-folds cross-validation. & none & none & none & none \\ 
        Premalatha and Srikrishna \cite{premalatha2019effort} & Effort estimation in agile software development using evolutionary cost-sensitive deep belief network & Predict the effort (ML) & Project planning & projects & Using an Evolutionary Cost-Sensitive Deep Belief Network (ECS-DBN) to predict the effort of a project. & 160 work items from one small software company & Compared the prediction performance using chronologically cross-validation. & none & none & none & none \\ 
        Prykhodko and Prykhodko \cite{prykhodko2019multiple} & A multiple non-linear regression model to estimate the agile testing efforts for small web projects & Predict the effort (ML) & Not specified & work items & Using a multiple non-linear regression technique to predict the effort of a work item. & 40 small web projects (lack of detail about the dataset) & Evaluated the prediction accuracy with the actual effort and compared it with linear and non-linear regression models (in MMRE, confidence intervals, and prediction intervals). & moderate - report only relative accuracy units & none & none & none \\ 
        Ramessur and Nagowah \cite{ramessur2021predictive} & A predictive model to estimate effort in a sprint using machine learning techniques & Predict the effort (ML) & Sprint planning & sprints & Using six machine learning models to predict the effort of a sprint. & 2,100 simulated records & Compared the prediction performance among all models based on 10-folds random cross validation, and cross-datasets prediction. & high - using a generated dataset & none & high - using a generated dataset & high - using a generated dataset \\ 
        Rosa et al. \cite{rosa2021empirical} & Empirical effort and schedule estimation models for agile processes in the US DoD & Predict the effort (ML) & Project bidding & projects & Using initial function requirements, domain groups, and peak staff inputs to predict the size of a project during the bidding phase. & 36 agile and hybrid projects in US DoD & Compared the prediction accuracy with the final effort and schedule. & none & none & none & none \\ 
        Sakhrawi et al. \cite{sakhrawi2021support} & Support vector regression for enhancement effort prediction of Scrum projects from COSMIC functional size & Predict the effort (ML) & Maintenance phase & requirement changes & Using a Support Vector Regression model to estimate the COSMIC functional size of a software change. & 93 user stories from a company project & Compared the predicted effort with actual effort using 70$\backslash$\%-30$\backslash$\% hold-out strategy. Compared the prediction performance between the models trained using COSMIC sizing and Story Points. & none & none & none & none \\ 
        Sakhrawi et al. \cite{sakhrawi2022software} & Software enhancement effort estimation using stacking ensemble model within the scrum projects: a proposed web interface & Predict the effort (ML) & Not specified & software enhancements & Using a stacked ensemble models to predict the effort of a software enhancement. & Software enhancement dataset from an industry scrum project called. & Compared the prediction accuracy with the actual effort in MAE, MSE, and RMSE units. & none & none & none & none \\ 
        Sánchez et al. \cite{rodriguez2023effort} & Effort and Cost Estimation Using Decision Tree Techniques and Story Points in Agile Software Development & Predict the effort (ML) & Project planning & projects & Using a mix of decision trees (decision tree, random forest, and AdaBoost) to create an ensemble effort prediction model & 21 projects containing labeled completion time and costs in Pakistani rupees. & Performed a 10-fold cross-validation and compare the prediction performance with the other approaches from literature. & moderate - report only relative accuracy units & none & none & none \\ 
        Sánchez et al. \cite{sanchez2022software} & Software effort estimation for Agile Software Development using a strategy based on K-nearest neighbors algorithm & Predict the effort (ML) & Project planning & projects & Using a linear-based regression model and machine learning models to predict the effort. & 21 Agile projects from 6 software houses proposed by Ziauddin et al. & Compared the prediction accuracy with the actual effort. & none & none & none & none \\ 
        Satapathy et al. \cite{satapathy2014story} & Story Point Approach based Agile Software Effort Estimation using Various SVR Kernel Methods & Predict the effort (ML) & Not specified & projects & Using various Support Vector Regression models to estimate the project delivery time based on Story Points of work items and team velocity. & 21 projects retrieved from Ziauddin and Zia \cite{ziauddin2012effort} & Evaluated the performance of various Support Vector Regression models based on five-fold cross validation. & moderate - report only relative accuracy units & none & none & none \\ 
        Scott and Pfahl \cite{scott2018} & Using Developers’ Features to Estimate Story Points & Predict the effort (ML) & Sprint planning & work items & Using machine learning models to estimate the Story Points of issue reports based on the developer-related features. & 15,155 issue from 8 open-source projects & Compared the prediction performance with a model built based on text features. & none & none & none & low - open-source projects may adopt Agile differently \\ 
        Sharma and Chaudhary \cite{sharma2020linear} & Linear regression model for agile software development effort estimation & Predict the effort (ML) & Not specified & projects & Using multiple Linear Regression models to predict the project effort. & 21 projects retrieved from Ziauddin and Zia \cite{ziauddin2012effort} & Compared the predicted effort with actual effort. & none & none & none & none \\ 
        Sharma and Chaudhary \cite{sharma2022combined} & The combined model for software development effort estimation using polynomial regression for heterogeneous projects & Predict the effort (ML) & Project planning & projects & Using a polynomial regression technique that considering multiple project-related parameters to predict the effort of a project. & 21 projects from Python procedural projects, Zia et al.'s dataset agile projects, and a company dataset for object-oriented projects. & Compared the prediction accuracy with the actual effort in R-squared and MMRE. & none & none & none & none \\ 
        Tawosi et al. \cite{tawosi2022investigating} & Investigating the effectiveness of clustering for story point estimation & Predict the effort (ML) & Sprint planning & work items & Using a combination of LDA and hierarchical clustering to group issues based on their topic similarities to predict effort of work items & 26 open source projects with a total of 31,960 issues. & Compared the prediction accuracy in MAE with a random guessing. & none & none & none & none \\ 
        Vyas and Hemrajani \cite{hemrajanipredicting} & Predicting Effort Of Agile Software Projects Using Linear Regression, Ridge Regression And Logistic Regression & Predict the effort (ML) & Not specified & work items & Using Linear, Logistic, and Ridge regression techniques to predict the required man-hours of a user story. & 21 projects retrieved from Ziauddin and Zia \cite{ziauddin2012effort} & Compared the prediction performance of the models based on random cross validation (20$\backslash$\% for testing) & none & none & none & none \\ 
        Wińska et al. \cite{winska2021reducing} & Reducing the uncertainty of agile software development using a random forest classification algorithm & Predict the effort (ML) & Release planning & work items & Using complexity factors in effort classification for a work item. & 70k issues across four major releases from a bank & Applied the proposed solution in a commercial environment, and analyzed the impact of each complexity factor. & none & none & none & none \\ 
        Aslam et al. \cite{aslam2017risk} & Risk Aware and Quality Enriched Effort Estimation for Mobile Applications in Distributed Agile Software Development. & Predict the effort (parametric) & Not specified & projects & Using a parametric-based approach to considers size, complexity, quality, novelty, type, and risk to predict the effort of a project. & 128 generated user stories & Compared the prediction accuracy with the generated actual effort. & high - using a generated dataset, report only relative accuracy units, the definition of MER accuracy unit is not described  & none & high - using a generated dataset & high - using a generated dataset \\ 
        Basri et al. \cite{basri2016algorithmic} & An Algorithmic-Based Change Effort Estimation Model for Software Development & Predict the effort (parametric) & Not specified & requirement changes & Extending the COCOMO II estimation method to estimate the size of requirement changes. & 6 traditional and Agile software projects & Compared the predicted effort with actual effort. & moderate - report only relative accuracy units & none & none & none \\ 
        Butt et al. \cite{butt2023prediction} & Prediction based cost estimation technique in agile development & Predict the effort (parametric) & Not specified & projects & Using a structured approach to gather information of the projects from stakeholders, to be used as inputs for a parametric-based calculation method for project level estimation. & Two ongoing projects of different sizes and user stories. & Applied the proposed approach on the projects, followed by a developer survey to understand the developers' perception. & none & none & none & none \\ 
        Choudhari and Suman \cite{choudhari2012phase} & Phase wise Effort Estimation for Software Maintenance: An Extended SMEEM Model & Predict the effort (parametric) & Maintenance & maintenance phases & Using an algorithmic method to estimate the effort of a phase in software maintenance. & A case study with three Request for Changes & Used the method to predict the effort in a case study. & moderate - small dataset & none & none & none \\ 
        Kang et al. \cite{kang2010model} & Model-Based Dynamic Cost Estimation and Tracking Method for Agile Software Development & Predict the effort (parametric) & Sprint planning & sprints & Using the Kalman Filter to estimate and track the team velocity. & One case study consists of 32 user stories & Compared the performance of the model with the traditional methods through a case study. & moderate - small case study & none & none & none \\ 
        Parvez \cite{parvez2013efficiency} & Efficiency Factor and Risk Factor Based User Case Point Test Effort Estimation Model Compatible with Agile Software Development & Predict the effort (parametric) & Project bidding & projects & Adjusting the Use Case Points using algorithmic method based on efficiency and risk factors & 4 mobile and web applications & Compared the predicted effort with actual effort. & none & none & moderate - not report the accuracy using any well-known metrics & none \\ 
        Paz et al. \cite{paz2014approach} & An Approach for Effort Estimation in Incremental Software Development using Cosmic Function Points & Predict the effort (parametric) & Not specified & releases & Estimating the effort of an increment using the Cosmic Function Points and COCOMO's effort adjustment factor. & 2 student projects & Compared the predicted effort with actual effort. & high - conducted the experiment using student projects & none & none & high - students may not have sufficient experience in Agile software development \\ 
        Raslan and Darwish \cite{raslan2018enhanced} & An Enhanced Framework for Effort Estimation of Agile Projects & Predict the effort (parametric) & Project planning & projects & Using a combination of Story Points and COCOMO II methods to estimate the project effort. & 10 projects dataset extracted from COCOMONASA2 & Compared the predicted effort with actual effort. & moderate - old dataset, and report only relative accuracy units & none & none & none \\ 
        Rosa et al. \cite{rosa2017early} & Early Phase Cost Models for Agile Software Processes in the US DoD & Predict the effort (parametric) & Project bidding & projects & Constructing an equation that consider different variable combinations and investigate whether the variable should be considered in the estimation for project bidding. & 20 industrial projects & Used the three variations of the approach to predict the actual effort. & none & none & none & none \\ 
        Singal et al. \cite{singal2022integrating} & Integrating software effort estimation with risk management & Predict the effort (parametric) & Project planning & projects & Using a parametric-based (formula) that considers risk exposure with other factors to calculate an integrated effort for a software project. & Two industrial datasets from an Indian IT firm, one for waterfall (45 projects) and one for agile (30 projects) development. & Compared the prediction accuracy with CoCoMo II (for waterfall projects) and Story Point approach (for agile projects). & none & none & none & none \\ 
\hline
\end{longtable}
\end{scriptsize}
\end{landscape}

\begin{scriptsize}
\begin{landscape}

\begin{longtable}{p{.06\textwidth}p{.1\textwidth}p{.06\textwidth}p{.06\textwidth}p{.15\textwidth}p{.07\textwidth}p{.20\textwidth}p{.08\textwidth}p{.07\textwidth}p{.07\textwidth}p{.1\textwidth}}

\caption{Summary of the selected studies that proposed an approach to support the effort estimation practice (Section~\ref{sec:litrevpurposesupport}).}
\label{tab:table_litrev_studylist_rq2support}

\\

\textbf{Authors} & \textbf{Title} & \textbf{Planning level} & \textbf{Automated / Manual} & \textbf{Purpose and techniques} & \textbf{Dataset used / Participants} & \textbf{Evaluation method performed} & GRADE-Individual Study Bias & GRADE-imprecision & GRADE-Inconsistency & GRADE-indirectness  \\ \hline

Algarni and Magel \cite{algarni2019applying} & Applying software design metrics to developer story: A supervised machine learning analysis & Sprint planning & Manual & Proposing a developer story artifact to capture functional information and used it for the prediction approach. & 30 open source java systems & Used machine-learning models that were trained based on the developer stories to predict the effort. The prediction performance was measured using 10x10-folds cross-validation. & moderate - report only relative accuracy units & none & none & low - open-source projects may adopt Agile differently \\ 
Buglione and Abran \cite{buglione2013improving} & Improving the user story Agile technique using the INVEST criteria & Sprint planning & Manual & Proposing a process to apply the INVEST criteria to improve the quality of user stories. & One case study company & Evaluated the applicability of the INVEST process in an industrial context and record the lessons learned. & high - lack of detail on the application in industrial setting & none & none & none \\ 
Grapenthin et al. \cite{grapenthin2016supporting} & Supporting Feature Estimation with Risk and Effort Annotations & Sprint planning & Manual & Proposing a system to help the team annotate user stories with symbols that highlight risks and effort drivers for effort estimation. & 203 software features from 8 student teams & Conducted empirical analyses to examine whether the annotations could improve the team understanding of the project and increase the estimation accuracy & high - experiment conducted with students & none & none & high - student teams may not have sufficient software development experience \\ 
Hannay et al. \cite{hannay2019agile} & Agile Uncertainty Assessment for Benefit Points and Story Points & Not specified & Manual & Proposing a process to use a three-points estimation to illustrate uncertainty of the estimated effort & 8 epics & Used Monte-Carlo simulations to demonstrate the prediction of project effort based on the three-points estimates. & moderate - small dataset & none & none & none \\ 
Taibi et al. \cite{taibi2017operationalizing} & Operationalizing the experience factory for effort estimation in agile processes & Sprint planning & Automated & Proposing a system to help the team decide whether to collect a measure based on its Return on Invested Time. & 7 company projects & Compared the estimation performance between the projects that did not use the system and the projects that used the system. & none & none & none & none \\ 
Tanveer et al. \cite{tanveer2017utilizing} & Utilizing change impact analysis for effort estimation in agile development & Sprint planning & Automated & Proposing a framework to integrate effort estimation with change impact analysis to help the team estimate effort of software change. & 3 Agile teams from SAP SE & Evaluated the usefulness with Agile development teams and collected the feedback. & none & none & none & none \\ 
Usman et al. \cite{usman2018developing} & Developing and using checklists to improve software effort estimation: A multi-case study & Sprint planning & Manual & Proposing a checklist to help the team recall relevant factros of the tasks being estimated to improve the estimation accuracy. & Participants from 3 companies & Examined the estimation accuracy and the perceived usefulness of the proposed checklists. & none & none & none & none \\ 
Pasuksmit et al. \cite{pasuksmit2022story} & Story points changes in agile iterative development: An empirical study and a prediction approach & Sprint planning & Automated & Perform data analyses to study the change of Story Points. Using machine learning techniques to predict Story Points changes of a work item. & Seven open source projects with 19349 \ \wis{}. & Compared the classification accuracy with four baselines. & none & none & none & low - open-source projects may adopt Agile differently \\ 
Pasuksmit et al. \cite{pasuksmit2022towards} & Towards Reliable Agile Iterative Planning via Predicting Documentation Changes of Work Items & Sprint planning & Automated & Using machine learning and deep learning approaches to predict the documentation changes before effort estimation of a work item. & Seven open source projects with 17731 \ \wis{}. & Compared the prediction accuracy of the three predictive classifier with two baselines and performed a qualitative assessment on the reasons for the correctly predicted documentation changes & none & none & none & low - open-source projects may adopt Agile differently \\ 
\hline
\end{longtable}
\end{landscape}
\end{scriptsize}

\end{document}